\documentclass[modern]{aastex61}

\submitjournal{AJ}

\usepackage[latin1]{inputenc}
\usepackage[T1]{fontenc}
\usepackage[english]{babel}
\usepackage{amsfonts}
\usepackage{eucal}
\usepackage{amsthm}
\usepackage{amsbsy}
\usepackage{amssymb}
\usepackage{amsmath}
\usepackage{wasysym}
\usepackage{graphicx}
\usepackage{babel}
\usepackage{color}
\usepackage{makecell}
\usepackage{natbib}
\usepackage{fancyhdr}
\usepackage{booktabs}
\usepackage{float}
\usepackage{multicols}
\usepackage{multirow}
\usepackage{textcomp}
\usepackage[caption=false]{subfig}
\usepackage{epsfig}
\usepackage{mathrsfs}
\usepackage{natbib}
\usepackage{enumitem}
\usepackage{threeparttable}
\usepackage[rightcaption]{sidecap}
\usepackage{tablefootnote}

\begin{document}

\newcommand{\BetaPic}{$\beta$ Pictoris}

\title{Atmospheric characterization of  directly imaged exoplanets with JWST/MIRI.}

\correspondingauthor{Camilla Danielski}
\email{camilla.danielski@cea.fr}

\author[0000-0002-3729-2663]{Camilla Danielski}
\affil{AIM, CEA, CNRS, Universit\'e Paris-Saclay, Universit\'e Paris Diderot, Sorbonne Paris Cit\'e, F-91191 Gif-sur-Yvette, France}
\affil{GEPI, Observatoire de Paris, PSL Universit\'e, CNRS,  5 Place Jules Janssen, 92190 Meudon, France}
\affil{Institut d'Astrophysique de Paris, CNRS, UMR 7095, Sorbonne Universit\'e, 98 bis bd Arago, 75014 Paris, France}

\author[0000-0003-4061-2514]{Jean-Loup Baudino}
\affil{Department of Physics, University of Oxford, Oxford, UK}

\author{Pierre-Olivier Lagage}
\affil{AIM, CEA, CNRS, Universit\'e Paris-Saclay, Universit\'e Paris Diderot, Sorbonne Paris Cit\'e, F-91191 Gif-sur-Yvette, France}

\author{Anthony Boccaletti}
\affil{LESIA, Observatoire de Paris, Universit\'e PSL, CNRS, Sorbonne Universit\'e,  Univ. Paris Diderot, Sorbonne Paris Cit\'e, 5 place Jules Janssen, F-92195 Meudon, France}

\author{Ren\'e Gastaud}
\affil{AIM, CEA, CNRS, Universit\'e Paris-Saclay, Universit\'e Paris Diderot, Sorbonne Paris Cit\'e, F-91191 Gif-sur-Yvette, France}

\author{Alain Coulais}
\affil{LERMA, Observatoire de Paris, CNRS, F-75014, Paris, France}
\affil{AIM, CEA, CNRS, Universit\'e Paris-Saclay, Universit\'e Paris Diderot, Sorbonne Paris Cit\'e, F-91191 Gif-sur-Yvette, France}

\author[0000-0002-5433-5661]{Bruno B\'ezard}
\affil{LESIA, Observatoire de Paris, Universit\'e PSL, CNRS, Sorbonne Universit\'e,  Univ. Paris Diderot, Sorbonne Paris Cit\'e, 5 place Jules Janssen, F-92195 Meudon, France}



\begin{abstract}
The Mid-Infrared instrument (MIRI) on board the \textit{James Webb Space Telescope}  will perform the first ever characterization of young giant exoplanets observed by direct imaging in the 5-28 $\mu$m spectral range. This wavelength range is key for both determining the bolometric luminosity of the cool known exoplanets and for accessing the strongest ammonia bands. In conjunction with shorter wavelength observations, MIRI will enable a more accurate characterization of the exoplanetary atmospheric properties.

Here we consider a subsample of the currently known exoplanets detected by direct imaging and we discuss their detectability with MIRI, either using the coronagraphic or the spectroscopic modes. 
By using the Exo-REM atmosphere model we calculate the mid-infrared emission spectra of fourteen exoplanets, and we simulate MIRI coronagraphic or spectroscopic observations.
Specifically we analyze four coronagraphic observational setups, which depend on (i) the target-star and reference-star offset (0, 3, 14 mas), (ii) the wave-front-error (130, 204 nm rms), (iii) the telescope jitter amplitude  (1.6, 7 mas). 
We then determine the signal-to-noise and integration time values for the coronagraphic targets whose planet-to-star contrasts range from 3.9 to 10.1 mag. \\
We conclude that all the MIRI targets should be observable with different degrees of difficulty, which depends on the final in-flight instrument performances.

Furthermore, we test for detection of ammonia in the atmosphere of the coolest targets. Finally, we present the case of HR 8799 b to discuss what MIRI observations can bring to the knowledge of a planetary atmosphere, either alone or in combination with shorter wavelength observations.

\end{abstract}



%% to the KEY in th \bibitem in the reference list below. 

\section{Introduction} \label{sec:intro}

In the field of study of exoplanets, high-contrast imaging enables us to probe the outermost part of an exo-planetary system. 
%V1 : Planets in the outskirts ($d_{P}\geq$ 10 AU) neither receive sufficient stellar radiation nor reflect significant amounts of stellar light.  
%For this reason, and due to current instrumental capabilities in term of attainable contrast, direct imaging observations are limited to self-luminous 
%giant planets in young ($a~\leq$ 100 Myr) nearby ($d_{\star} \leq$ 100 pc) systems.\\
% V2: The flux coming from directly imaged exoplanets, i.e. of  nearby ($d_{\star} \leq$ 100 pc) young planets (<100 Myrs) in the outskirts of a planetary system ($d_{P}\geq$ 10 AU) is dominated by internal emission compared to stellar irradiation.\\
% * <dr.jean-loup.baudino@hotmail.com> 2018-03-29T13:00:23.824Z:
% 
% > Planets in the outskirts ($d_{P}\geq$ 10 AU) neither receive sufficient stellar radiation nor reflect significant amounts of stellar light.  
% > For this reason, and due to current instrumental capabilities in term of attainable contrast, direct imaging observations are limited to self-luminous 
% > giant planets
% 
% J'ai l'impression que quelque chose ne va pas avec ces deux phrases...
% Est ce que ça ne serait pas mieux quelque chose comme ça?
% 
% The flux coming from directly imaged exoplanets, i.e. of young planets (<100Myrs) in the in the outskirts ($d_{P}\geq$ 10 AU) is dominated by internal emission compared to stellar irradiation.
% 
% ^ <camilla.danielski@gmail.com> 2018-04-06T11:10:52.313Z.
Due to observational biases, the exoplanets detected so far by direct imaging are young objects (< 100 Myr),
orbiting  at large distances ($d_{P}\geq$ 10 AU) nearby stars ($d_{\star} \leq$ 100 pc).
Given that they reside in the outskirts of the planetary systems, the stellar irradiation they receive is negligible, causing their temperature to decrease in time. Consequently, they need to be young to be bright enough for being detected.\\
These young planets produce radiation from the heat of formation and gravitational contraction and they are therefore brighter at infrared wavelengths than 
their older equivalents. 
This particular aspect makes them optimal targets for spectroscopic studies with direct imaging from
which it is possible to derive important information about the planetary architecture, the atmospheric structure and dynamics, and about planetary formation.

In recent years new extreme adaptive optic cameras mounted on 8-m class telescopes, like the VLT/Spectro-Polarimetric High-contrast Exoplanet REsearch 
(SPHERE, \citealt{SPHERE}) or the Gemini Planet Imager (GPI, \citealt{GPI}), 
have enabled infrared detection and characterization of exoplanets at very small angular separation from their host stars, reaching contrasts as large as  
$5\cdot10^{-7}$ at 0.5$\arcsec$ \citep{Vigan2015, Mesa2017}.
%Conversely, although this new generation of instruments is 
%providing excellent constraints on exoplanetary physical parameters (such as luminosity, temperature, mass and age), these extreme adaptive optics cameras still yield
%significant uncertainties due to the limited spectral window in which they operates ($\Delta \lambda$ =1 - 2.3 $\mu$m). 
This new generation of instruments enables spectroscopic observations, which are needed to characterize the atmosphere of the exoplanets (such as their atmospheric structure, dynamics, and molecular content).  However, the limited spectral window in which they operate ($\Delta \lambda$ = 1 - 2.3 $\mu$m) is a serious shortcoming to study this kind of exoplanets, especially the coolest ones. 
Only planets warmer than 1200 K have a large enough thermal emission in this wavelength range.
Cooler planets like, for instance, HD 95086 b ($T_\mathrm{eff} \sim$ 1050 K), present a very low flux in the Y-band ($\Delta \lambda$ = 0.96 - 1.08 $\mu$m), J-band ($\Delta \lambda$ = 1.11 - 1.33 $\mu$m) and H-band ($\Delta \lambda$ =  1.48 - 1.78 $\mu$m), making the detection difficult (e.g. \citealt{Chauvin2018}). Though, at longer wavelengths, like the L$^\prime$-band ($\Delta \lambda$ = 3.49 - 4.11  $\mu$m) where the planetary flux is higher, the detection is clear (e.g. \citealt{Rameau2013}).\\
In addition, observations restricted to the 1 - 2.3 $\mu$m spectral range do not allow to sample the strongest features of molecules %expected to be present in the atmosphere of these \textcolor{red}{cool} planets (for instance 
like CH$_4$  ($\sim$3.3 $\mu$m), 
CO$_2$ ($\sim$ 4.3 $\mu$m), 
PH$_3$ ($\sim$ 4-5 $\mu$m), 
CO ($\sim$5 $\mu$m), 
NH$_3$ ($\sim$10.65 $\mu$m), 
C$_2$H$_2$  and HCN (both $\sim$ 14 $\mu$m, visible only in a case of planets with a C/O ratio larger than 1).\\
The situation will dramatically change with the launch of the \textit{James Webb Space Telescope} (JWST) in mid-2020.
%A natural answer to this issue is to extend the planetary characterization to a larger spectral band, solution that will indeed soon be possible with the advent of  
%the \textit{James Webb Space Telescope} (JWST). 

Among JWST instruments, the Mid-Infrared Instrument (MIRI, \citealt{Rieke2015,Wright2015} and references therein) will be pivotal to the 
characterization of gas-giant exoplanets. MIRI will overcome the limited sensitivity of the largest ground-based 
observatories (e.g. the Large Binocular Telescope Interferometer, \citealt{Hinz2009})
extending planetary characterization to the mid-infrared thermal regime, 
where objects are too faint to be detected from the ground.
MIRI covers a wavelength range from 5-28 $\mu$m and it combines imaging \citep{Bouchet2015}, coronagraphy \citep{CoronoBoccaletti}, 
low  resolution spectroscopy (LRS, \citealt{Kendrew2015}) 
and medium resolution spectroscopy with integral-field unit (\citealt{Wells2015}); additional information about MIRI can be found at \url{https://jwst-docs.stsci.edu/display/JTI/Mid-Infrared+Instrument\%2C+MIRI}.\\
In this manuscript we will mostly focus on simulated exoplanet observations with the coronagraphic and low resolution spectrometer observational modes.

MIRI coronagraphic imaging incorporates one Lyot mask (30\arcsec x 30\arcsec) at $\lambda$ = 23 $\mu$m and three four-quadrant phase masks (4QPM) at $\lambda$ = 10.65, 11.40, 15.50 $\mu$m, which cover a field of view of 24\arcsec x 24\arcsec  \citep{Rouan2000}. The 4QPM transparent masks confer phase differences in diverse parts of the focal plane and make the light interfering more destructively than with a normal Lyot mask. 
This concept allows to reach an inner working angle (IWA, separation at which the throughput of an off-axis object achieves 50$\%$)  of $\sim\lambda/D$. It is the first time that such coronagraphs are used in a space-based instrument.\\
Among its various capabilities MIRI coronagraphic mode	
was specifically conceived to detect the ammonia feature
(located at $\lambda$ = 10.65 $\mu$m) and measure its intensity \citep{CoronoBoccaletti}.
The first 4QPM filter is centered on the ammonia absorption band, while the second one is strategically placed beside the first one to give the level of the continuum.
Coronagraphic observations, in combination with shorter wavelength observations, will also allow us to
better constrain the planetary parameters (such as the effective temperature, bolometric luminosity, chemical equilibrium or non-equilibrium chemistry and gravity). In such a way, 
MIRI will support comprehensive modeling of the atmospheric properties of an exoplanet (e.g. \citealt{Bonnefoy2013}).

In this manuscript 
%we update the description of the MIRI coronagraph performances presented in \citealt{CoronoBoccaletti} and 
we present the exoplanetary science that can be done with the 4QPM coronagraph and low-resolution spectrometer observational modes. 
For the coronagraphic mode, in particular, we present the degree of difficulty of various sources' observations, according to possible observational conditions and on-orbit performance.\\
In $\S$~\ref{sec:targets} we discuss the selection of targets studied in this manuscript and how 
we have chosen the type of observation, i.e. coronagraphic versus spectroscopic.  
In $\S$~\ref{sec:models} we introduce the Exo-REM planetary models and PHOENIX stellar models used for our analysis. 
$\S$~\ref{sec:coronosimu} presents our MIRI coronagraphic mode simulations.
%, while $\S$~\ref{sec:LRSsimu} presents the LRS ones.
In $\S$~\ref{sec:discussion} we elaborate on the results, discussing
both sources' detectability via coronagraphic or spectroscopic observations, and the significance of ammonia detection in the atmosphere of the "cold" planets belonging our target list.
We discuss in $\S$~\ref{sec:MIRINIR} the science constraints that MIRI can put when working in synergy with NIR observatories and we present our conclusions in $\S$~\ref{sec:conclusion}.

\newpage
\section{Targets}
\label{sec:targets}
\begin{figure}[t]
\centering
%\framebox(200,200){}
\includegraphics[width=0.9\linewidth]{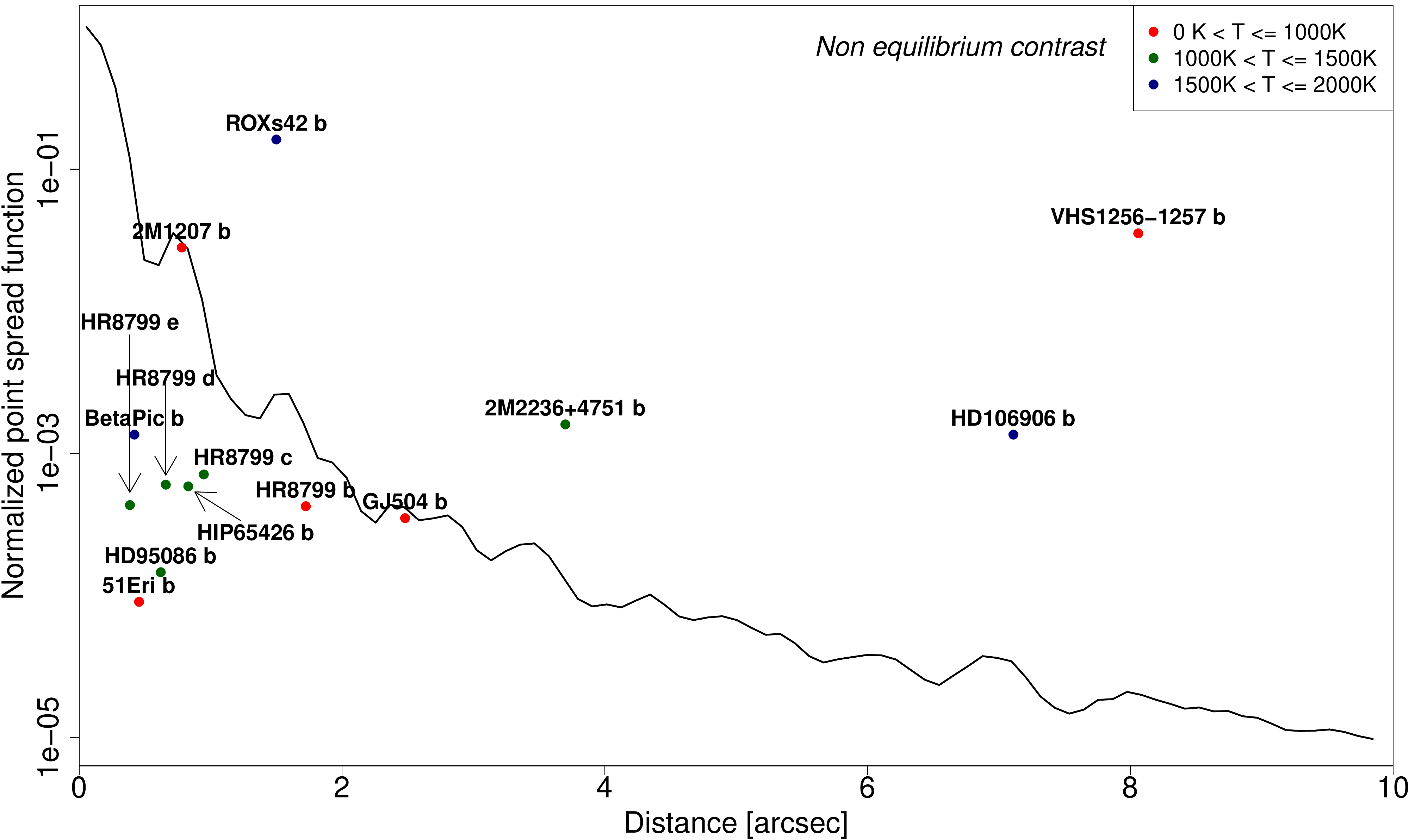}
\caption{Non coronagraphic point spread function versus the distance to the peak signal, normalized to 1. The 1D curve has been measured from a simulated image of a MIRI observation with a filter at 11.3$\mu$m (using the WebbPSF software). The signal has been averaged over an annulus of 1 pixel width (i.e. 0.11$\arcsec$). Dots indicate the planet-to-star  contrast of the targets under consideration. Colors correspond to the temperature of the target as indicated in the top legend.
Those planetary systems whose planet-to-star contrast lies above the curve will be observed with the spectroscopic mode. Note that 2M1207 b contrast takes into account the host star and disk flux.}
\label{fig:PO_PSFcontrast}
\end{figure}

For our study we selected from the Extrasolar
Planets Encyclopaedia\footnote{\url{http://exoplanet.eu}} a subsample of known directly imaged exoplanets with a planetary mass $M_{P}<$ 13 M$_{J}$, effective temperature $T_P \leq$ 2000 K, and located at angular distance d > $\lambda_{C}/D$  as well as $d < 10\arcsec$ from the host star, where $\lambda_{C}$ is the central wavelength of each coronagraphic filter 
(see Tab. \ref{tab:filters}) and where $D$ is the telescope mirror diameter.
 Table \ref{tab:targets} summarizes the list of the selected targets with their mean parameters. For each target we also indicated which MIRI mode we used to simulate the observations: coronagraphic (C) or spectroscopic (S).
The choice between the type of observations depends on the contrast and the angular distance between the exoplanet and its host star. To make the selection we have used the JWST simulated Point Spread Function (PSF) simulation tool at 11.30 $\mu$m i.e. the WebbPSF\footnote{\url{https://jwst.stsci.edu/science-planning/proposal-planning-toolbox/psf-simulation-tool-webbpsf}} software to determine the PSF as a function of the angular distance from the star itself.
Fig. \ref{fig:PO_PSFcontrast} shows that for four exoplanets (i.e. VHS 1256-1257 b, HD 106906 b, 2M2236+4751 b and ROXs 42B b) the contrasts (which is measured using the Exo-REM model, see \S \ref{sec:models}) and angular distances are such that coronagraphic observations are not required. The stellar PSF contribution to the signal is indeed much lower than the planet contribution at the angular distance of the planet. \\
For two exoplanets (i.e. 2M1207 b and GJ 504 b), the stellar contribution at the planet location is comparable to the planet signal itself, hence coronagraphic observations, as well as spectroscopic observations, can be considered. 
For all the sources whose contrast is well below the stellar PSF signal, coronagraphic observations are really beneficial in terms of signal-to-noise ratio.

\begin{deluxetable*}{cccccccc}[t]
\tablecaption{Coronagraphic modes of MIRI \label{tab:filters}}
\tablecolumns{8}
\tablewidth{0pt}
\tablehead{
\colhead{Filter} & \colhead{Coronagraph} & \colhead{Mean filter } &\colhead{Stop transmission} & \colhead{$\lambda_{C}$} &\colhead{Bandwidth\tablenotemark{*}$_\mathrm{(50\%)}$} & \colhead{IWA} & \colhead{Rejection\tablenotemark{$\dagger$}}\\
\colhead{} & \colhead{} & \colhead{\small{transmission [$\%$]}} & \colhead{\small{[$\%$]}} & \colhead{\small{[$\mu$m]}} &\colhead{[$\mu$m]} &\colhead{[arcsec]} & \colhead{[on-axis]}
} 
\startdata
F1065C  & 4QPM1 & 72 & 62 & 10.575 &  0.558 & 0.33 & 304 \\
F1140C & 4QPM2 & 78 & 62 & 11.30 & 0.537 & 0.36 & 293 \\
F1550C & 4QPM3 &68  & 62 &  15.50 & 0.731 & 0.49 & 334 \\
F2300C & Lyot spot & 71 &72  & 22.75 & 4.504 &  2.16 & 918 \\
\enddata
\tablenotetext{*}{Width at half maximum filter transmission while \cite{CoronoBoccaletti} is quoting the width at 10$\%$ of the maximum.}
\tablenotetext{\dagger}{Values are marginally different from \cite{CoronoBoccaletti} due to different simulations parameters.}
\end{deluxetable*}

\begin{deluxetable*}{ l | c | c | c | c | c | c | c | c | c }[t!]
\fontsize{9}{13}\selectfont
 \tablecaption{Table of analyzed targets. For each target we report here the planetary effective temperature ($T_{P}$), the planetary radius ($R_{P}$), surface gravity ($\log$(g)), angular separation (sep), stellar-to-planet contrast for non equilibrium models, filter F1550C ($C_{\rm FC3}^{neq}$), the stellar 2MASS $K_s$ magnitude ($K_s$), the stellar effective temperature ($T_{\rm eff}$), the system distance ($d$) and the type of observations analyzed (OBS: C for coronograph, S for spectroscopy). \label{tab:targets}}
\tablecolumns{10}
\tablehead{
\colhead{Name} & \colhead{$T_{P}$~[K]} &  \colhead{$R_{P}$~$[R_{\rm J}]$} & \colhead{$\log$(g) [cgs] }& \colhead{sep [\arcsec] }& 
\colhead{$C_{\rm FC3}^{neq}$[mag] }& \colhead{$K_s$ } & \colhead{$T_\mathrm{eff}$[K]} & \colhead{$d$ [pc]} & \colhead{OBS}
}
\startdata
	2M1207 b &  1000\tnote{$a$}  &  1.5&  4  & 0.78 & 3.88\tablenotemark{$w$}& 11.945 & 2500 & 52.4 &C, S\\
    2M2236+4751 b & 1050\tablenotemark{$m$} & 1\tablenotemark{$\dagger$} & 4.5\tablenotemark{$\dagger$} & 3.7 & 6.87 & 9.148 & 4000\tablenotemark{$\dagger$} & 63 & S \\
	51 Eri b&   700 &  1. &   3.5 & 0.45 & 10.07 & 4.537 & 7400 & 29.4& C\\ 
	\BetaPic~ b &  1700  &  1.65 &  3  & 0.42\tnote{$\textasteriskcentered$} & 7.23 & 3.48\tablenotemark{$\ddagger$} & 8000& 19.3&C\\  
    GJ 504 b &   544  & 0.96 &  3.9 & 2.48\tablenotemark{$m$}& 8.83 & 4.033 & 6234& 17.56 &C, S \\ 
    %GU-Psc b &  1050  &  1.35  & 4.5   & 42 & -- & 9.34 & 3300\tablenotemark{$m$} & 48& S\\ 
	HD 106906 b &  1950  &  1\tablenotemark{$\dagger$}   &  4\tablenotemark{$\dagger$}  & 7.11 & 7.15 & 6.683 & 6516& 92 &S\\ 
	HD 95086 b & 1050\tablenotemark{$m$}   &  1.3 & 3.3   &  0.6 & 9.58 & 6.789& 7550& 90.4&C \\  
 	HIP 65426 b& 1300 & 1.5 & 4.5 & 0.83 & 8.09 & 6.771 & 8840 & 111.4 & C\\
    HR 8799 b & 950  &  0.96  & 4.8  &  1.7241 & 8.19 & 5.24 & 7430& 39.4 &C\\
	HR 8799 c&   1150 &  1.07 &  5.4  & 0.9481 &  7.71 & - & -& -&C\\ 
    HR 8799 d &    1150 &  1.14 &  5.4 &  0.6587 & 7.88 & - & -& -&C \\  
 	HR 8799 e &  1200 & 1.06  & 5.2   & 0.3855 & 8.19 & -  & - & - & C \\ 
%	PSO-J318.5338 b &  1127 & 1.46 & 4.01 & 3.86  & {\color{red} POL}& NA & NA &  24.6  &S\\ 
	%ROSS-458 (AB) c &  670  & 1.07   &   4.35  & 102 &  -- & 5.578 &  3500 &  11.7 &S\\ 
	ROXs 42B b&  1975\tnote{m}  &  2.5  & 3.6   & 1.5 & 2.02 & 8.671 & 2200 &  135 & C, S \\ 
 	VHS 1256-1257 b&   880 & 1\tablenotemark{$\dagger$} &  4.24  & 8.06 &  3.49 &  10.044 & 2620 & 12.7 &S\\ 
	%WD-0806-661B b & 325\tnote{m} & {\color{red}POL? }  &  4.01 &  130 &  -- & 13.781 & 330 & 19.2 &S\\ 	
\enddata
\tablerefs{
 2M1207 b: \cite{2Mdiscovery} ;
 2M2236+4751 b: \cite{2M2236discovery}
 51 Eri b: \cite{51Eridiscovery};
  \BetaPic ~ b: \cite{Betapicdiscovery}; 
  GJ 504 b: \cite{GJ504discovery};
 %GU-Psc b: \cite{GUPscdiscovery};
  HD 106906 b: \cite{HD106discovery};
  HD 95086 b: \cite{HD95discovery};
  HIP 65426 b: \cite{Chauvin2017};
  HR8799 b,c,d,e: \citet{HRbcddiscovery, HRediscovery}, 
  %PSO-J318.5338 b: \citet{PSOJ318discovery,PSO318parameters};
  %ROSS-458 (AB) c: \citet{Ross458discovery, Ross458parameters};
  ROXs 42B b: \cite{ROXs42discovery};
  VHS 1256-1257 b : \cite{VHS1256discovery};
  %WD-0806-661B/GJ-3483B b: \cite{WDdiscovery}
 }%
  \begin{tablenotes}
 %\item {\color{red}($a$)\small{\cite{Barman11}},   ($b$) \small{ SOMETHING},   ($c$) \small{ SOMETHING}, ($\textasteriskcentered$) \small{max distance}, ($bau$) \small{mean value}, ($\dagger$)  \small{assumed}}
\item($\dagger$) \small{assumed}; 
\item ($\ddagger$) \small{magnitude in $K$ band.};
\item($m$) \small{mean value of a given range};
\item($w$) \small{WISE W3 ($\lambda \sim 12\mu$m, \citealt{WISE}) contrast};\\
 Note that HR 8799 system's parameters are retrieved from the fit on the available data, non-equilibrium case, see \S~\ref{sec:HR8799models}. 
    \end{tablenotes}
\end{deluxetable*}

Concerning GJ 504 we decided to keep the companion GJ 504 b among our targets despite the uncertainty on the age estimation of its host star GJ 504 (\citealt{FuhrmannChini15}, \citealt{Dorazi2017}) which could 
classify the companion as a brown-dwarf with mass $M \approx$ 30 - 40 M$_{\rm J}$. 
The decision was driven by the fact that the planetary very low temperature, which allows for ammonia studies, and large angular separation, make it a good test case for MIRI coronagraphy observations. Furthermore, the planet is very much documented 
by ground-based observations and it provides an interesting comparison (\citealt{Bonnefoy2018}, submitted).

Similarly, the mass of VHS 1256-1257 b is contentious due to a controversy about the distance of the system itself.
In the discovery paper \citep{VHS1256discovery} the authors report a parallactic distance of d = 12.7 $\pm$ 1.0 pc, while a more recent study \citep{Stone2016} reports a spectrophotometric distance of d = 17.2 $\pm$ 2.6 pc. A farther system would imply the mass of the planetary companion to be $M \approx$ 35 M$_{\rm J}$, locating it in the brown dwarfs class. 
%Till new astrometric data will address this disagreement, we thought it was worth presenting the case of VHS 1256-1257 b as exoplanet with a very low planet-to-star contrast.

\section{Models}
\label{sec:models}
To compute the synthetic spectra of planets we have used the Exoplanet Radiative-convective Equilibrium Model (Exo-REM), developed by \citealt{Baudino2015, Baudino2017} and tailored for directly imaged exoplanets. 
For each target, the parameters required to generate a model (effective planetary temperature $T_P$, surface gravity $\log$(g), radius $R_P$ and distance $d$ to Earth) were
taken from the literature.\\
A different approach has been used for HR 8799 system. Given its scientific interest we produced for each companion the best set of models  (one at the equilibrium and one at non-equilibrium) fitting the available near-infrared observations (see \S$\ref{sec:HR8799models}$).  

~\\For the stellar spectrum models  we used the BHAC15 PHOENIX spectra \citep{Baraffe2015}.

\subsection*{HR 8799 planetary models}
\label{sec:HR8799models}

\begin{table}[t]
\centering
\caption{Physical parameters of the planetary companions in the HR 8799 system
where the best fit (i.e. minimal $\chi^2$, non-equilibrium model) was performed on data by \cite{Bonnefoy2017}, within 2 $\sigma$.
Best fit parameters with equilibrium chemistry ($k_\mathrm{zz}$= 0) are also shown in the bottom panel. 
All cases are with clouds and solar metallicity. Uncertainties for the equilibrium chemistry are not reported as no agreement with the data was found within 2 $\sigma$.
}
\begin{tabular}{l|c|c|c}
\hline 
\hline 
& $T_P$ [K] & $\log(g)$  [cgs]& $k_\mathrm{zz} ~ [cm^2 s^{-1}$ \\ 
\hline 
HR 8799~b & 950$ \substack{+100 \\ -0}$ & 4.8$\substack{+0.1 \\ -0.1}$ & 10$^8$ \\ 
HR 8799~c & 1150$\substack{+50 \\ -0}$ & 5.4$\substack{+0 \\ -0.8}$ & 10$^8$ \\  
HR 8799~d & 1150$\substack{+50 \\ -150}$ & 5.4$\substack{+0 \\ -0.6}$ & 10$^8$  \\ 
HR 8799~e & 1200$\substack{+0 \\ -300}$ & 5.2$\substack{+0.2 \\ -1.4}$ & 10$^8$ \\ 
\hline 
\hline 
HR 8799~b,~c,~e & 1200 & 5.2 & 0 \\ 
HR 8799~d & 1200 & 5.4 & 0 \\ 
\hline 
\end{tabular} 
\label{tab:hrbcdemodelfitneqeq}
\end{table}

We performed a $\chi^2$ analysis \citep[following the one performed in][]{Baudino2015}, 
comparing four grids of models to the observational data of the HR 8799 
system from \cite{Bonnefoy2016}. 
Each grid corresponds to a combination of set of clouds and a type of atmospheric chemistry. 
More precisely we considered a case with silicate and iron clouds (with a mean particle 
radius of 30 $\mu$m and $\tau_\mathrm{ref}$ = 0.5) 
and a case with no clouds at all. For the atmospheric chemistry we considered both equilibrium and non-equilibrium chemistry with an eddy mixing coefficient  $k_\mathrm{zz}$ = 
10$^8$~cm$^2$~s$^{-1}$. 
Refer to  \citealt{Baudino2017} for a description of the non-equilibrium chemistry formalism.\\
Grids were generated with Exo-REM spanning a temperature $T_P$ range 
from 400~K to 1200~K (by step of 50~K), a surface gravity $\log_{10}(g)$ range from  3.0 to 5.4 
(by step of 0.2, where $g$ is in cgs units) and a metallicity ($z$) range from -0.2 dex to +1.4 dex (by step of 0.1).

\begin{figure}[t!]
\begin{center}
\begin{tabular}{cc}
%~~~~~~~a) HR 8799~b&~~~~~~~b) HR 8799~c\\
\includegraphics[width=0.48\linewidth, trim={0 1.3cm .5cm 1cm},clip]{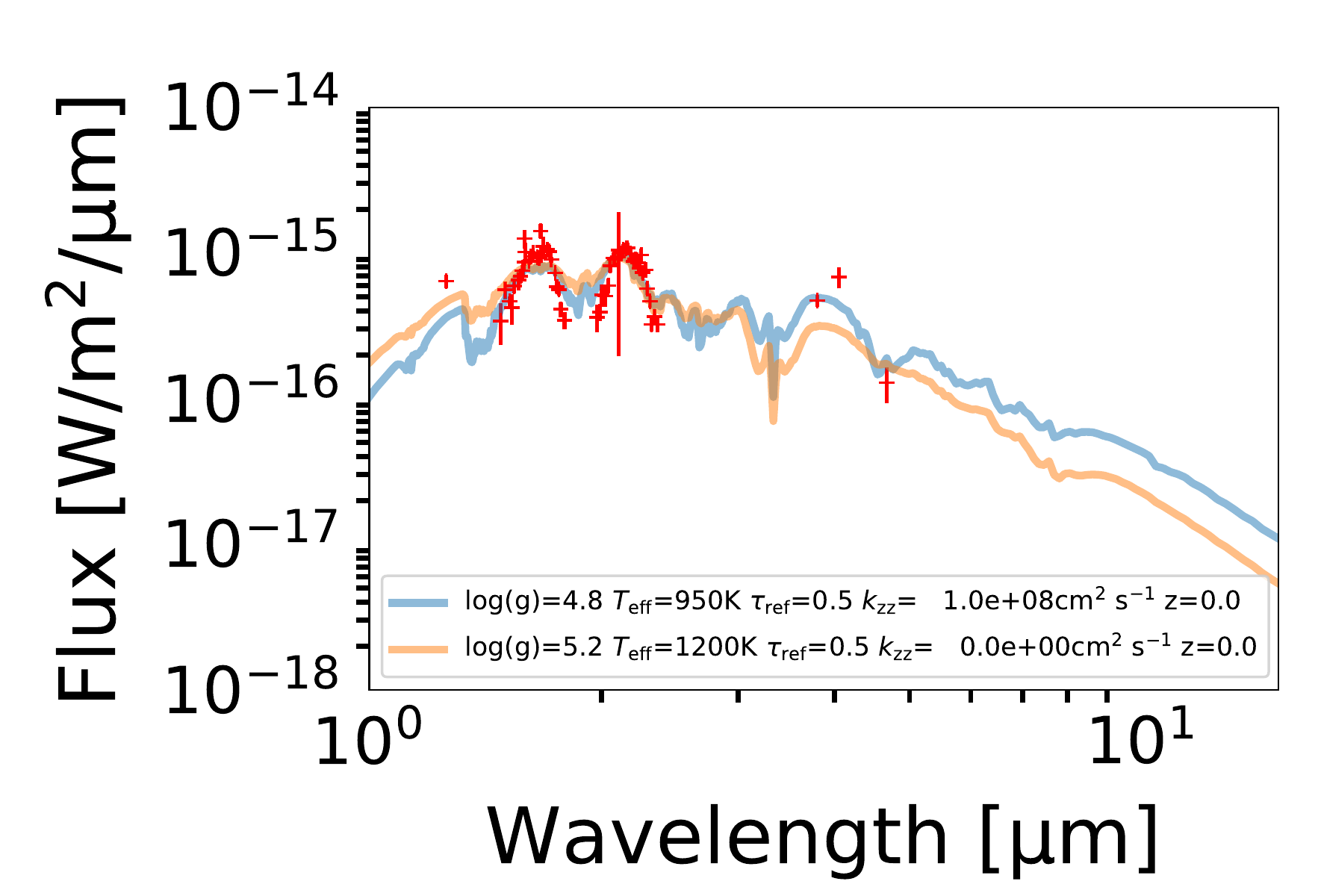}&
~~\includegraphics[width=0.47\linewidth, trim={2.6cm 1.5cm .5cm 1cm},clip]{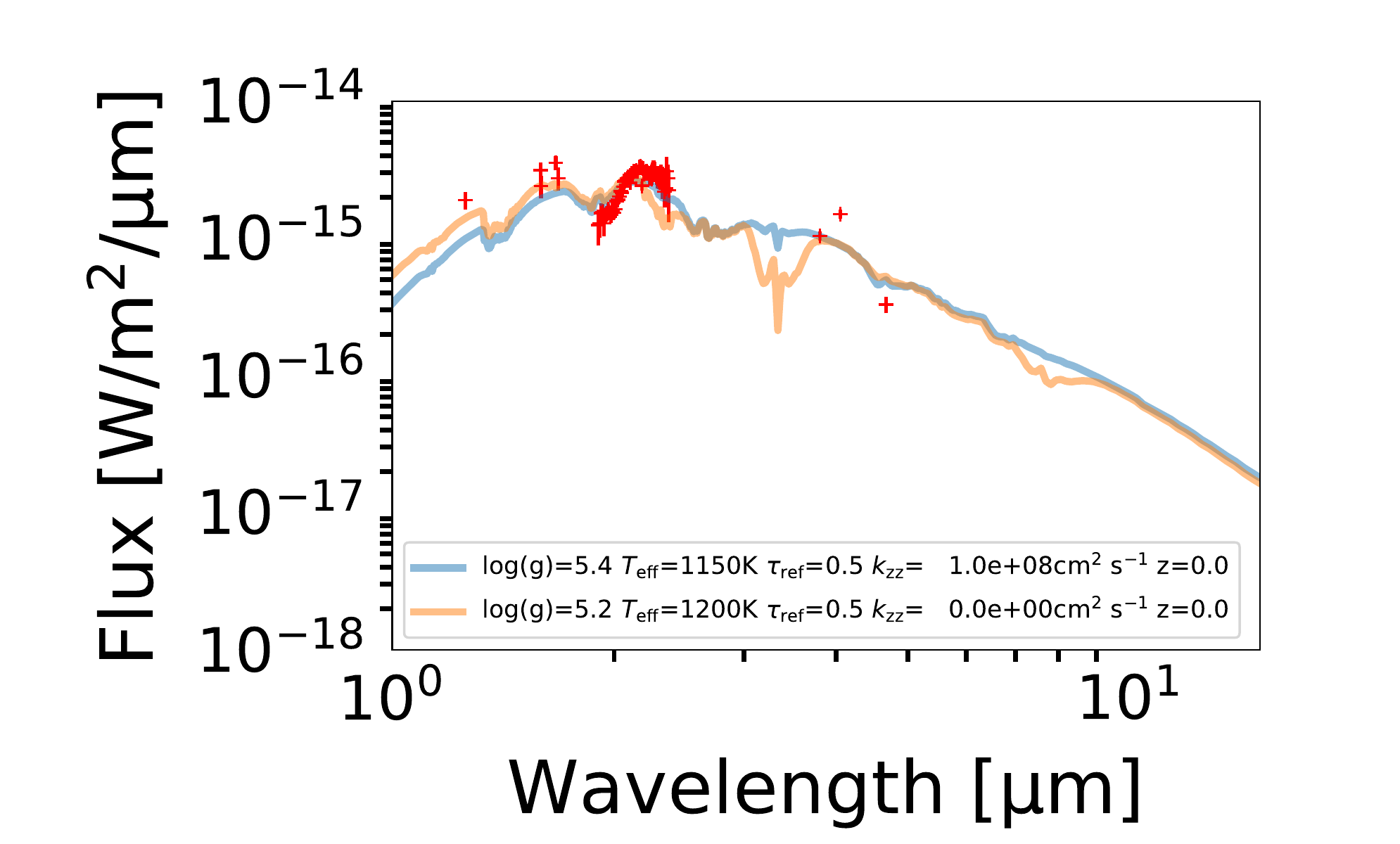}\\
%~~~~~c) HR 8799~d&~~~~~~~d) HR 8799~e\\
\includegraphics[width=0.48\linewidth, trim={0 .2cm 2cm 1cm},clip]{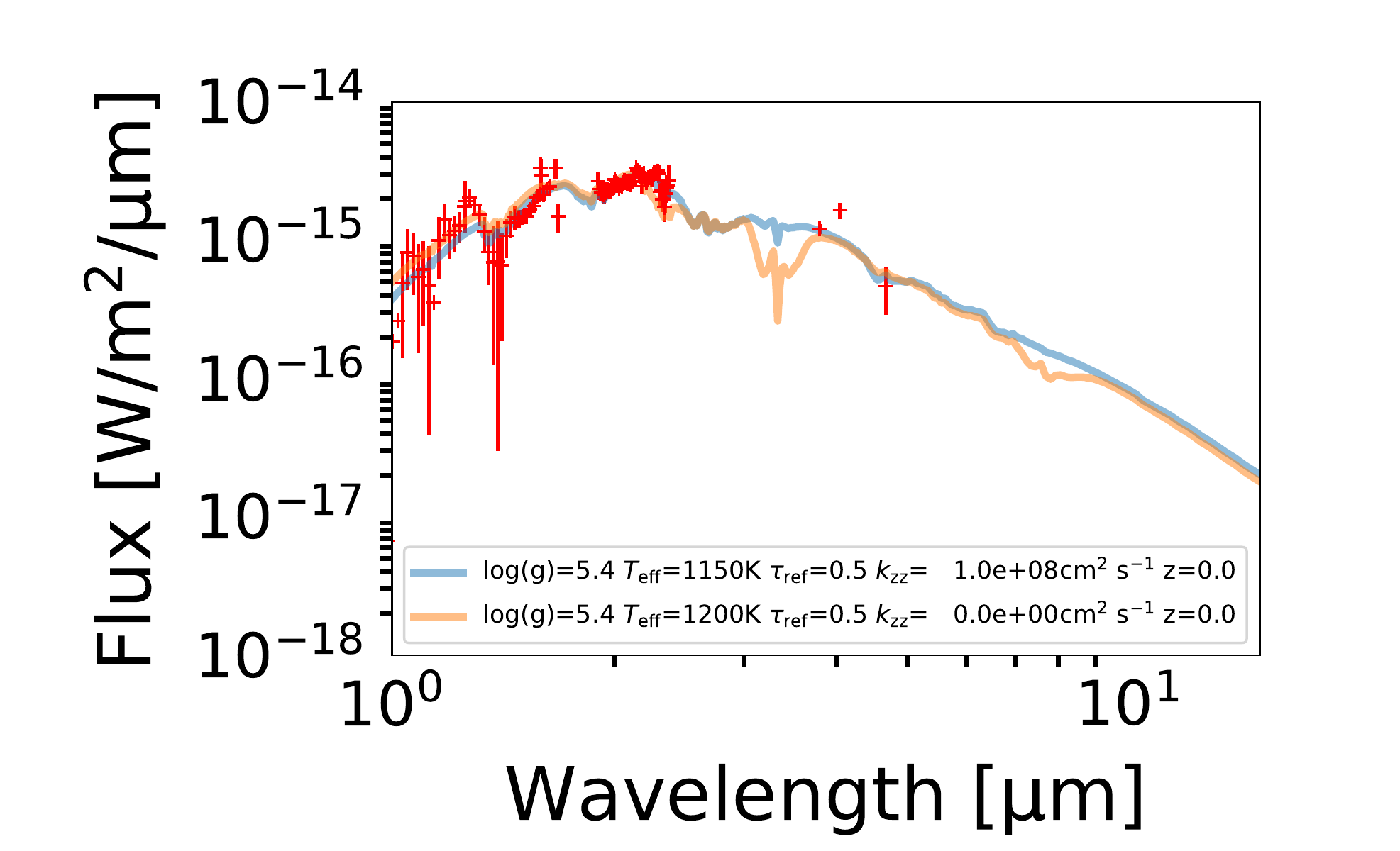}&
\includegraphics[width=0.47\linewidth, trim={2.6cm .2cm .5cm 1cm},clip]{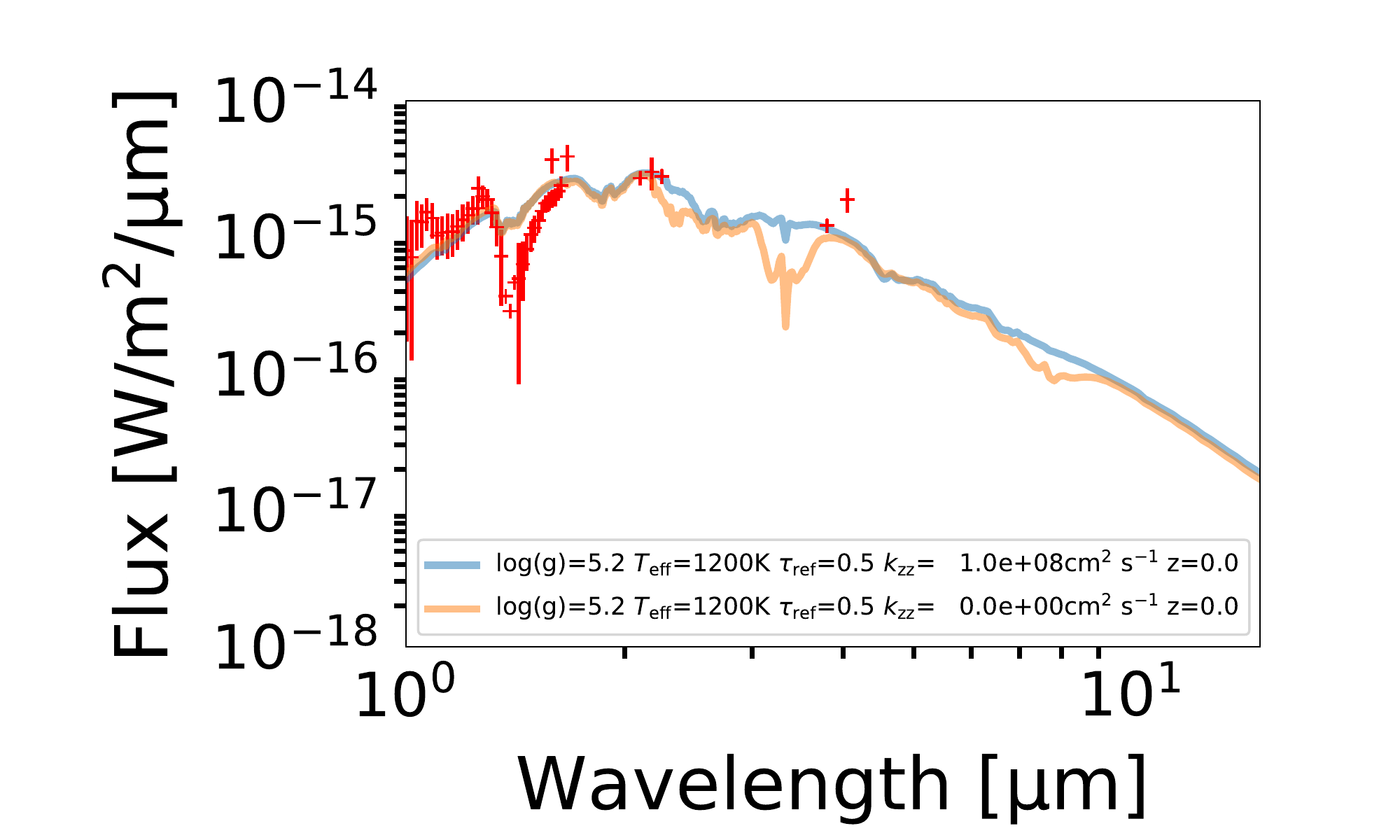}\\
%trim={0 0 1.5cm 0}
\end{tabular} 
\end{center}
\caption{Exo-REM synthetic spectra (equilibrium \textit{orange} and non-equilibrium \textit{light-blue}) with the best $\chi^2$ compared to HR 8799 system observations (\textit{red}, \citealt{Bonnefoy2016}. From top left, clockwise: HR 8799 b, HR 8799 c, HR 8799 e, HR 8799 d.}
\label{fig:plothrbcdemodelfitneqeq}
\end{figure}

Tab.~\ref{tab:hrbcdemodelfitneqeq} shows the parameters of the best fit on the data \citep{Bonnefoy2016} for each planet in the HR 8799 system. 
The best model is always for solar metallicity ($z$=0), with clouds, 
non-equilibrium chemistry and at less than 2 $\sigma$ from the observations (less than1 $\sigma$ for HR 8799 e). We also kept the best cases with equilibrium 
chemistry for comparison purposes, though we note that all cases deviate from the observations at more than 2 $\sigma$. Figure \ref{fig:plothrbcdemodelfitneqeq} shows
the corresponding spectra.

Note that, since the publication by \cite{Bonnefoy2016}, the Exo-REM model has been updated
\citep[see Appendix B by][]{Baudino2017} and the grids used in the present paper are different from the ones used in the original analysis. The result of the characterization is hence slightly different.

\newpage

\section{Coronagraphic simulations}
\label{sec:simulations}

\subsection{Coronagraphic observations simulations}
\label{sec:coronosimu}
In this section, we will focus on describing the coronagraphic observations simulation process; for a thorough description of both MIRI coronagraphs 
and  target acquisition process refer to \citealt{CoronoBoccaletti}.

The simulation of a science image is a two-step process that consists in simulating the diffraction patterns of all objects in the system under study (on-axis pattern for the star as well as off axis patterns for planets, see $\S$~\ref{sec:PSFsimu}) 
and in creating the science image itself, which also includes sources of noise (see $\S$~\ref{sec:scienceimage}). 
This procedure is applied to both the observed planetary system  %(target star and planet/s)
and a reference star (hereafter called \textquotedblleft reference\textquotedblright ~for simplicity).  
Note that, for contrast maximization reasons,  the acquisition of a reference image is  
necessary because it will be subtracted from the science target image.
For a best result, the magnitude and spectrum of the reference should be identical to the ones of the target star.

\begin{deluxetable*}{c c c c c c}[t!]
\tablecaption{The different cases analyzed and their respective instrumental configuration for the observations. The value of (x,y) position of star and reference are 
 relative to the center of the coronagraphic mask.}
\tablecolumns{7}
\tablewidth{0pt}
\tablehead{
\colhead{Case} & \colhead{WFE rms} &  \colhead{Jitter} & \colhead{Star $(x,y)$} & \colhead{Reference $(x,y)$} & \colhead{Star-Reference} \\
 & [nm] &  \colhead{amplitude [mas]} & \colhead{ [mas]} & \colhead{[mas]}& \colhead{Offset [mas] }
}
\startdata
	$k_A$  &   130 & 1.6  & 0 ; 0 & 0 ; 0 & 0 \\  
	$k_B$  &   130 & 1.6  & 0 ; 0 & +2.12 ; +2.12 & 3  \\ 
	$k_C$  &   204 & 7  & 0 ; 0 & +2.12 ;  +2.12 & 3\\ 
	$k_D$  &   204 & 7  & -4.95 ; -4.95  & +4.95 ; +4.95 & 14  \\ 
	$k_P$  &  204 & 0 &  0 ; 0 & 0 ; 0 & 0\\
\enddata
\tablecomments{For case $k_P$ (photon noise) no jitter and no stellar offset have been included and its WFE = 204 nm has been chosen for conservative reasons. 
The only difference from the other cases is within the PSFs generation step, the science image generation is  the same as the other cases (see $\S$~\ref{sec:PSFsimu} 
and $\S$ \ref{sec:scienceimage}
for more details).}
\label{tab:cases}
\end{deluxetable*}

\subsubsection{Coronagraphic PSFs and coronagraphic image simulation} 
\label{sec:PSFsimu}
\begin{figure}[t!]
\centering
\includegraphics[clip,trim=7.5cm 0.5cm 5.5cm 1.2cm,width=1.05\linewidth]{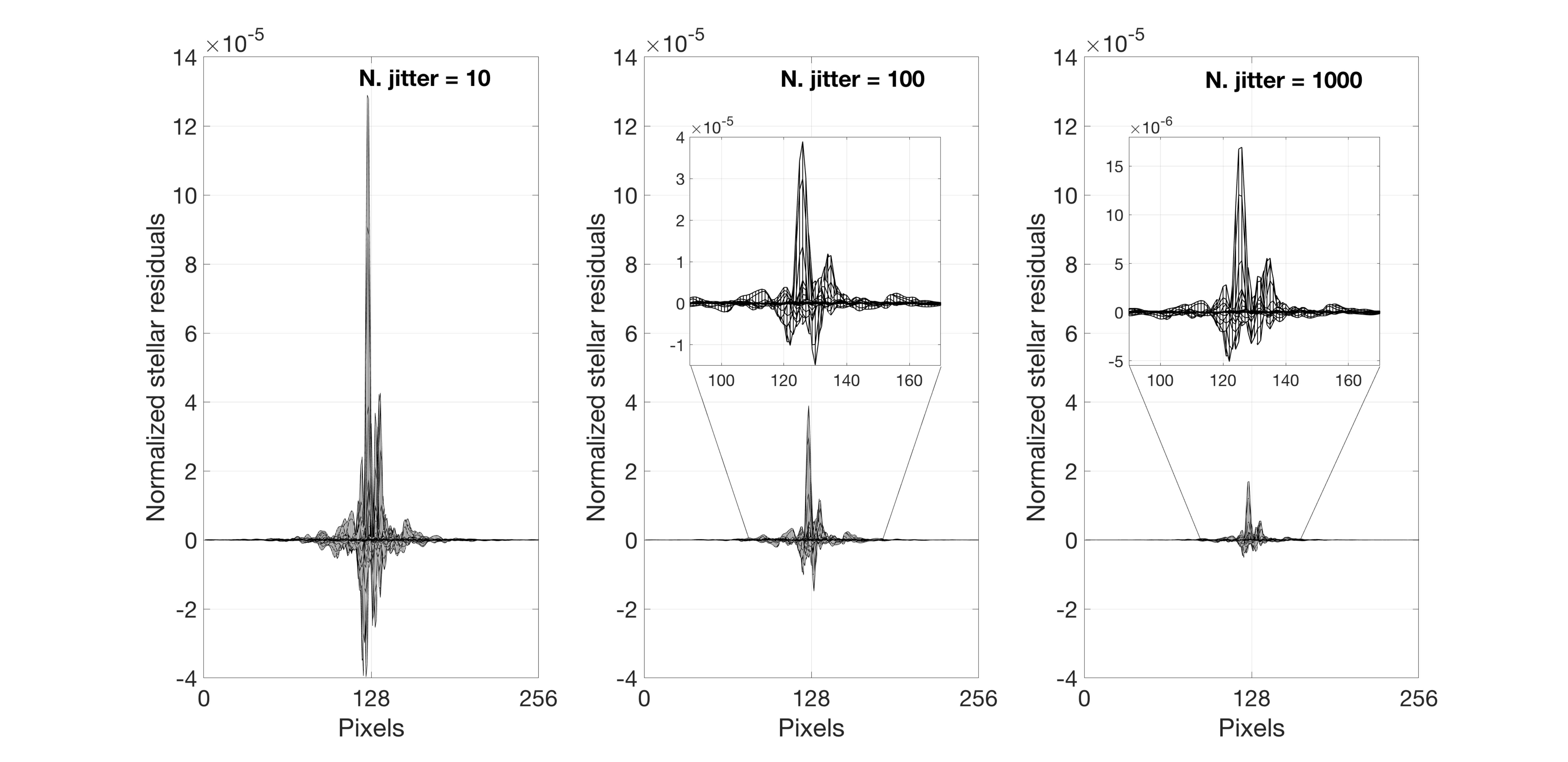}
 \caption{F1065C $k_A$ normalized coronagraphic residuals (i.e. given by the star - reference coronagraphic images subtraction) in a case of 10 (\textit{left}), 100 (\textit{center}) and 1000 (\textit{right}). For the central and right panel zoom-in 
 of the speckles are shown to better appreciate the structure and intensity values. The images show a vertical cut on the intensity of the speckles, where the X axis 
represents the pixels in the X direction of the image. Units corresponds to the number of pixels as seen in the X axis of Fig. \ref{fig:stellarpsf}; pixel 128 is the central pixel. Telescope jitter was applied in both X and Y axis, with a jitter amplitude $jamp$ = 1.6 mas (see \S \ref{sec:PSFsimu} for more details on the jitter simulations).}
\label{fig:psfsjitter}
\end{figure}

To build the diffraction patterns we followed the principle of operation of the 4QPM coronagraph (Fig.1 by \citealt{CoronoBoccaletti}).
Note that, in order to encompass different observational settings, we considered five specific cases $k$ to account for variations of the 
wavefront error (WFE), for different amplitudes of the telescope jitter and for different offsets between the target star and the reference star.
Table \ref{tab:cases} shows the configuration for these cases: an optimistic one ($k_A$), a pessimistic one ($k_D$) and two intermediate cases ($k_B$ and $k_C$). 
Each case represents a different configuration of the following values:
for the wave front error we used WFE  $\sim$130 nm and  WFE $\sim$204 nm root mean square (rms) (also used in \citealt{CoronoBoccaletti}), 
while for the amplitude of the jitter ($jamp$) of the pointing we used a minimum value of 1.6 mas (E. Nalan, private communication, July 8, 2015) and a maximum 
value of 7 mas (1 $\sigma$ dispersion value). \\
We modeled the telescope jitter by changing the pointing step by step for a discrete number of iterations. As currently not enough information is available to foresee the jitter performances in space, we set the number of iterations to be what we thought to be a realistic number i.e. N = 1000.
Figure \ref{fig:psfsjitter} compares the 
intensity of the normalized stellar residuals (i.e. after subtracting the normalized reference's coronagraphic image from the target's normalized coronagraphic image)  for the number of jitter steps N = 10, 100, 1000 for $k_{A}$ and filter F1065C. The highest the number of jitter steps, the faintest the speckle residuals.
More specifically, the telescope jitter was applied on both X and Y axes and each jitter step followed the normal distribution $\sim$ $\mathcal{N}(0,jamp^2)$ where $jamp$ = 1.6 mas or 7 mas, depending on the case $k$ we are in. For each of the thousand frames we supposed the telescope to be steady, hence no smearing effect was included during the integration. The effect of this approximation (i.e. no smearing) is negligible when averaging over a thousand frames. We note that a jitter amplitude $jamp$ = 1.6 mas corresponds to a spatial movement on the detector of 1.45 $\cdot10^{-2}$ pixel, while for a $jamp$ = 7 mas the movement corresponds to 6.3 $\cdot10^{-2}$ pixel\footnote{
An offset in the focal plane corresponds to a shift of phase in the coronagraphic pupil, meaning that, even if the jitter amplitude is small, it has an impact on the final image due to the high sensitivity of the phase-mask coronagraph to phase changes.}. 
\\ 
Jitter realizations were applied differently for target star and reference, meaning that their coronagraphic PSFs differ. Accordingly, when no telescope jitter is included and when target star and reference star have the same offset (e.g. in case $k_P$), both stellar coronographic PSFs are identical.
Concerning the offset between the target star and the reference, we considered one offset of 0 mas, one of 3 mas where the star is perfectly centered behind the coronagraph,  and one of 14 mas  where both stars are displaced in opposite direction on the coronagraph center, each one at a distance of 1$\sigma$ jitter amplitude value from the center.

The fifth case ($k_P$) represents an ideal case where no jitter and no star-reference offset were added and where, for conservative reasons, we used WFE $\sim$ 204 nm rms.\\

\begin{figure}[!t]
   \begin{minipage}{0.49\textwidth}
     \centering\includegraphics[clip,trim=1.cm .cm .0cm 1.35cm,width=1.1\linewidth]{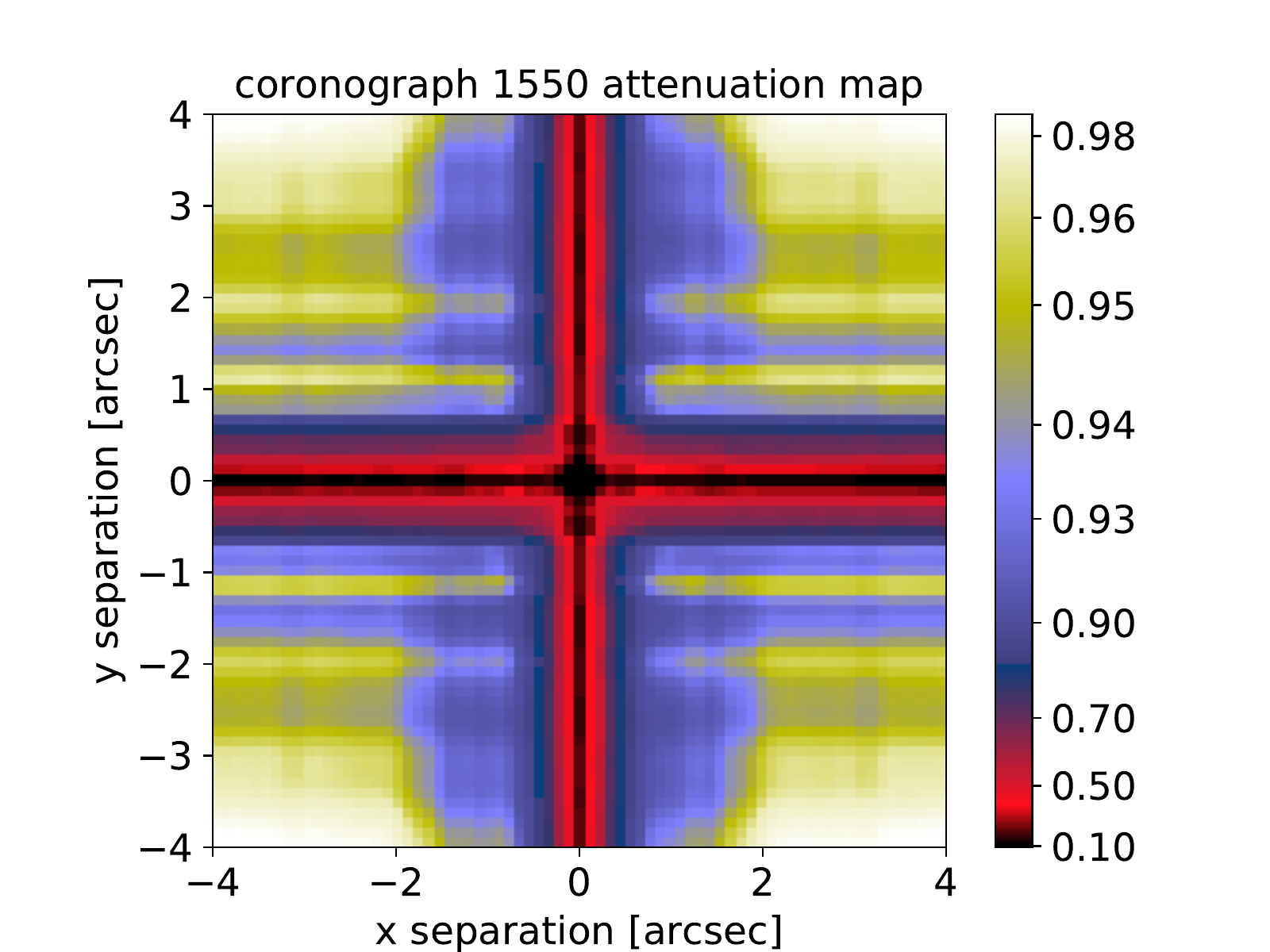}
     \caption{Coronagraphic F1550C transmission map within $\pm$4\arcsec ~~from the 
center of coronagraph mask.\\
~\\}
   \label{fig:attenuationMap}
  \end{minipage}\hfill
   \begin {minipage}{0.48\textwidth}
     \centering
        \includegraphics[clip,trim=.2cm .0cm .3cm .3cm,width=1\linewidth]{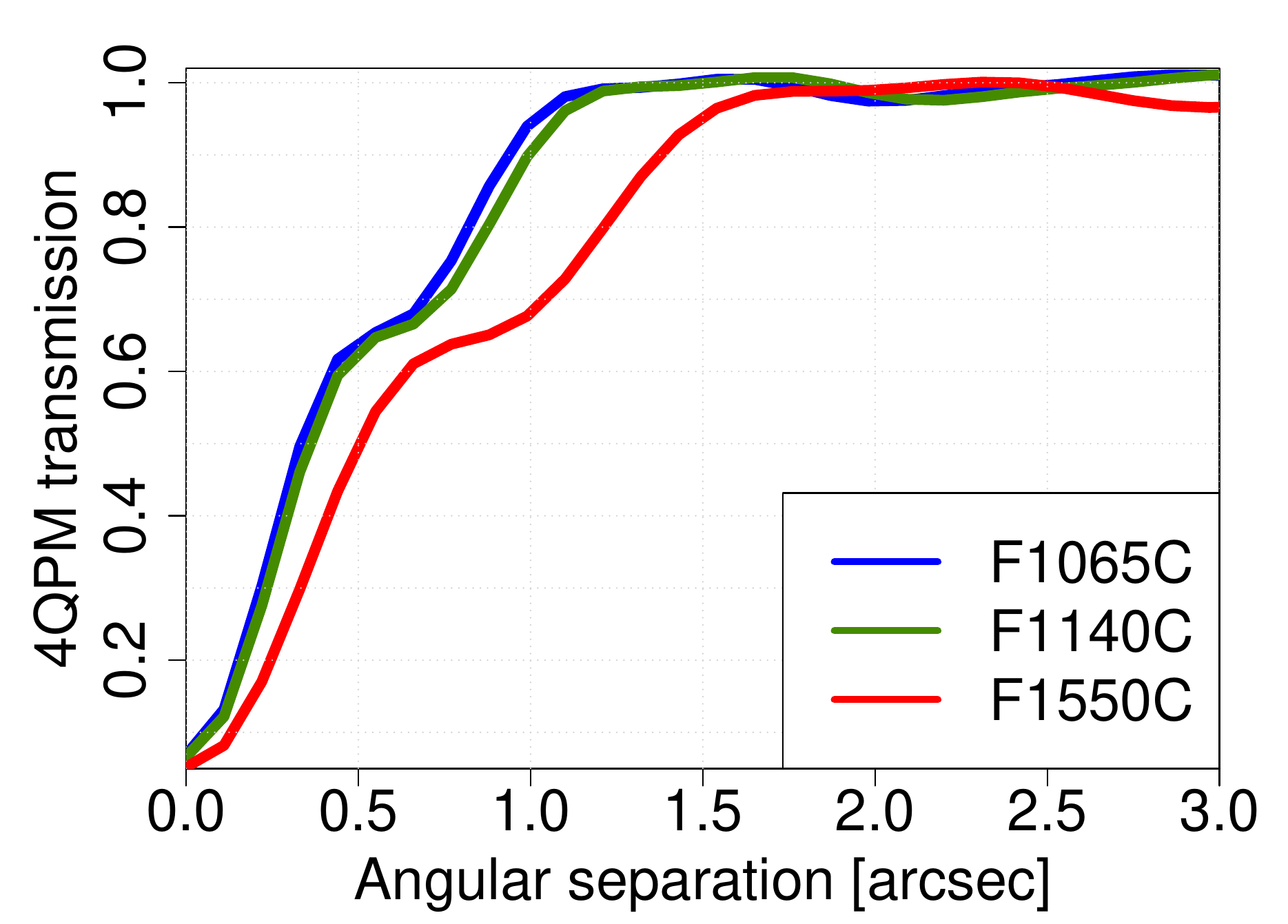} 	
     \caption{Radial transmission of the MIRI 4QPM coronagraph filters. Transmission is measured as a function of the distance from the center of the 4QPM mask, at 45$\degr$, where pixel size is 0.11\arcsec x 0.11\arcsec.\\
     ~\\
     ~\\}
     \label{fig:coronotrasmission}
  \end{minipage}
\end{figure}

 \begin{figure}[b!]
 \centering
 \subfloat[Normalized flux as a function of the angular separation between the planet and the star, for an exoplanet located at PA = 30$^\circ$, 45$^\circ$, 60$^\circ$. Values in the X-axis reports the angular distance relative to the center of the coronagraphic mask along these specific PAs. Black full dots on the positive X-axis side correspond to the dots at PA = 30$^\circ$, 45$^\circ$, 60$^\circ$ of Fig. \ref{fig:vsPA}, while those on the negative side to those at PA = 210$^\circ$, 225$^\circ$, 240$^\circ$ (Fig. \ref{fig:vsPA}). ] 
{\includegraphics[width=0.49\linewidth]{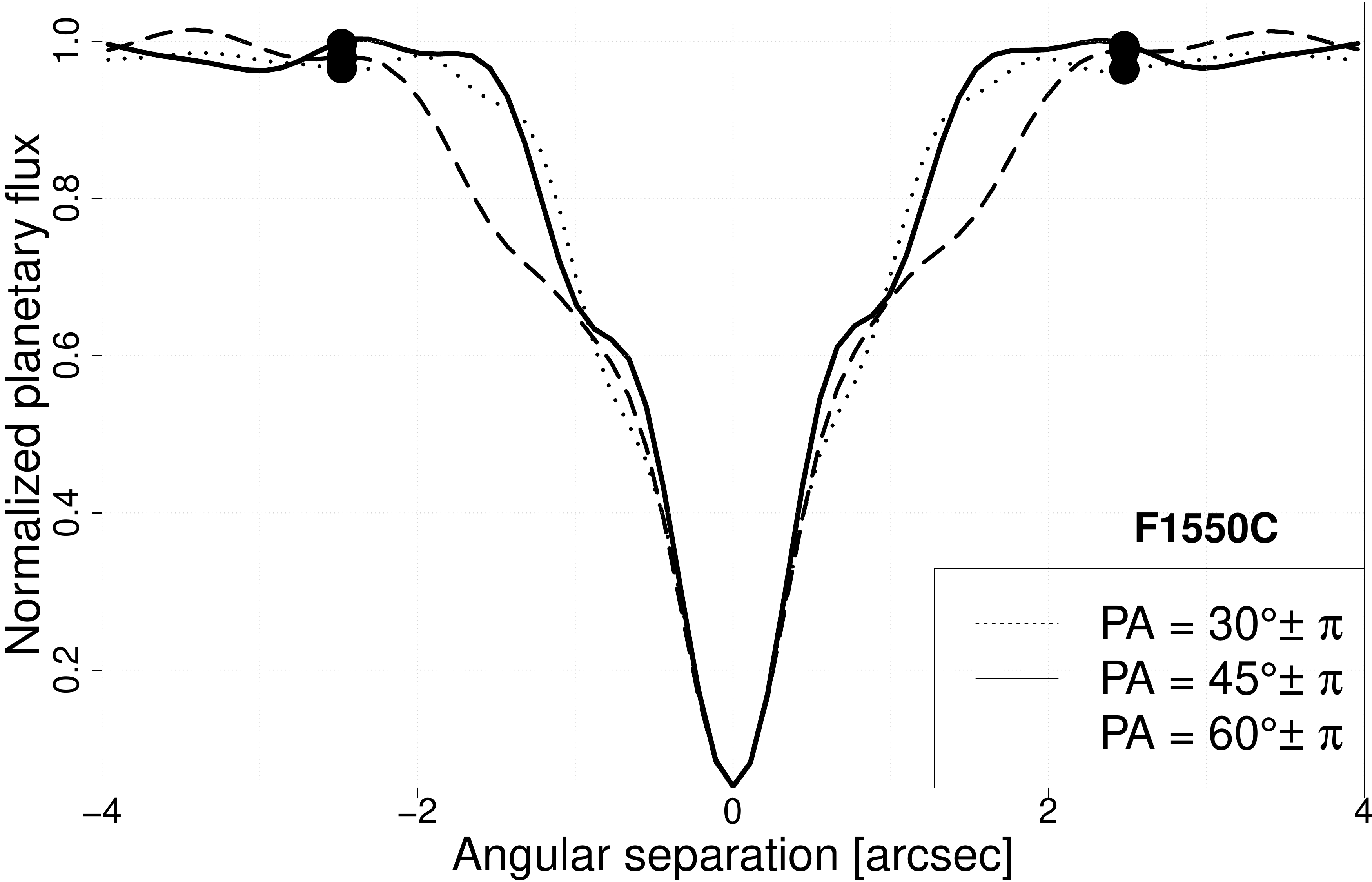}\label{fig:vsSEP}} \quad \quad 
\subfloat[Normalized flux as a function of the PA, measured for a fixed angular separation ($sep$ = 2.48\arcsec). Each black full dot shows the planetary flux value measured for a planet when located at the position angle labeled in the outermost ring. Radial scale corresponds to the flux values marked on each gray circle.]{\includegraphics[clip,trim=0cm 0.cm 0.cm 0cm,width=0.38\linewidth]{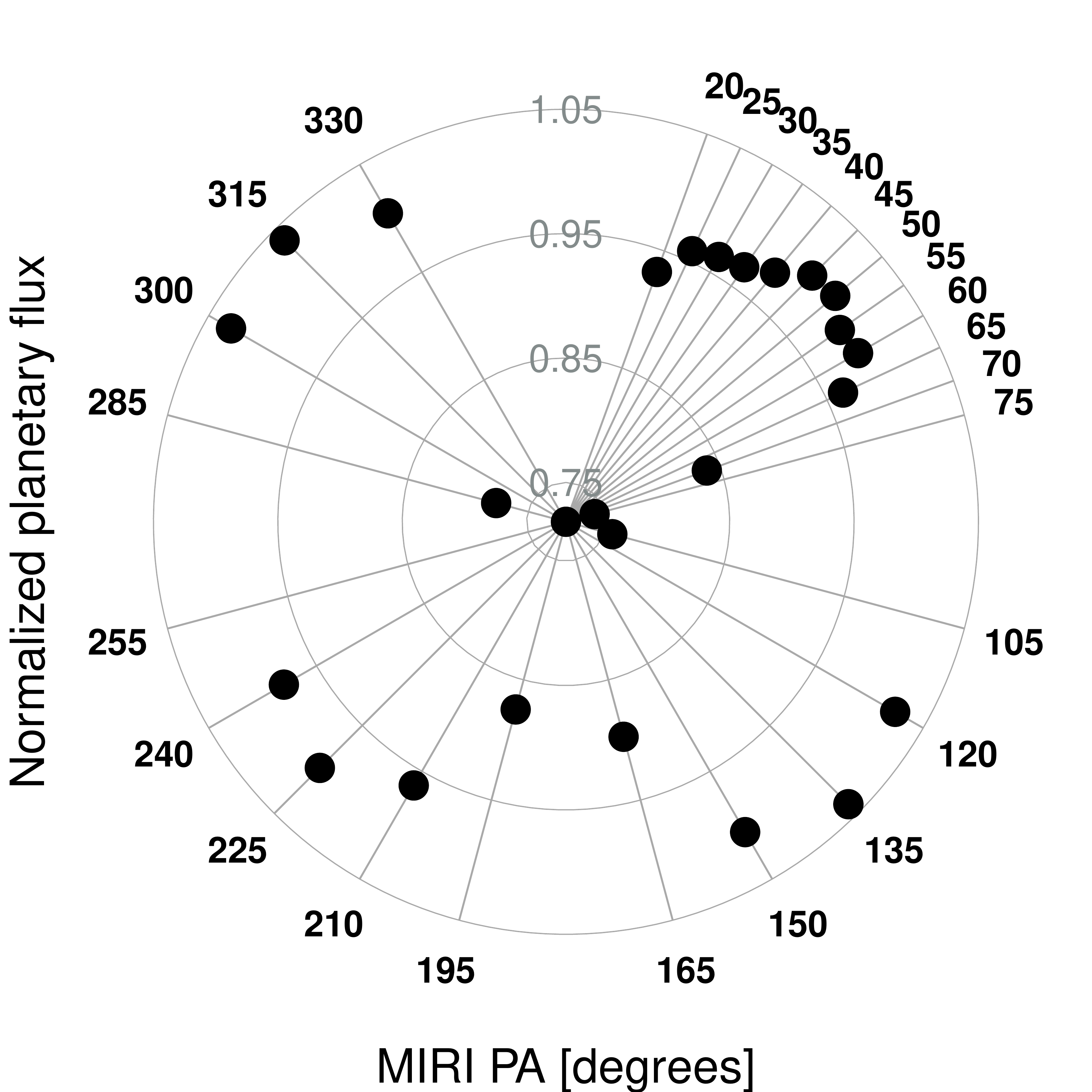}\label{fig:vsPA}}\\
~\\

\caption{Normalized planetary flux as a function of the angular separation (\textit{left}) and the MIRI position angle PA (\textit{right}) for filter F1550C (IWA $\sim$ 0.49\arcsec). 
Flux at 315\textdegree ~and 135\textdegree ~is larger than those at 45\textdegree ~or 225\textdegree ~mainly because of the stellar pupil shift applied (see Tab. \ref{tab:intrumentparams}).
For some PA the normalized flux is larger than 1 (but within 3\% deviation) because the PSF used for the normalization was itself produced at 45\textdegree (see \S \ref{sec:PSFsimu}), and hence presenting a lower flux than at 315\textdegree, asymmetry due to the stellar shift mentioned above.\\
~\\}
\label{fig:vsPAandSEP}
\end{figure}

%\begin{SCfigure}[0.9][b!]\label{fig:stellarpsf}
\begin{figure}[b]
\centering
\includegraphics[clip,trim=0cm 0cm 0.cm 1.35cm,width=0.6\linewidth]{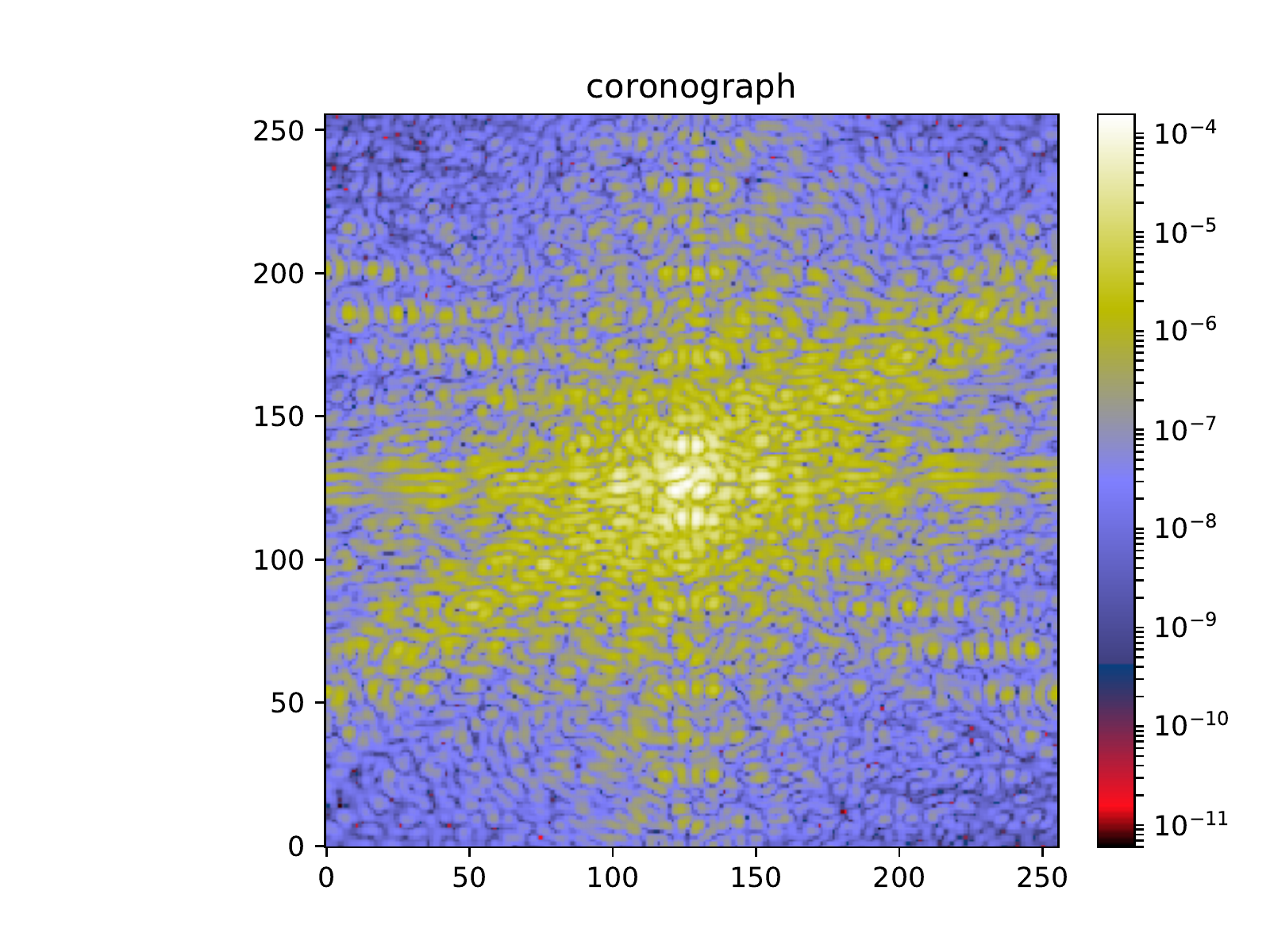}
\caption{Coronagraphic image of the target star (F1065C, $k_A$, N jitter = 1000) normalized to the PSF of the target star as seen on the imager at 45$^\circ$, far from the center so that the coronagraphic effect is negligible. The color scale indicates the contrast with respect to the image maximum.}
\label{fig:stellarpsf}
%\end{SCfigure}
\end{figure}

To choose the best position angle (PA) for the science image simulations, we studied the
effect of the coronagraphic transmission as a function of the MIRI PA\footnote{The MIRI position angle has an offset angle from the JWST V3 axis of 4.45$^{\circ}$.}, hereinafter referred to as \textquotedblleft PA\textquotedblright~for simplicity.
 
To map the coronagraphic transmission we
produced a set of coronographic images (with no jitter) within $\pm 4\arcsec$ from the center of the coronagraphic mask. For each image, generated using a different combination of planetary $sep$ 
(-4$\arcsec$ -- 4$\arcsec$) and PA (0$^{\circ}$ - 360$^{\circ}$), we measured the total flux.
This process was performed for the three 4QPM filters. 
We show in Figure \ref{fig:attenuationMap} the coronagraphic transmission map measured for filter F1550C.
We note that the coronagraphic transmission, as a function of the position of the source, scales with wavelength. For instance the radial transmission at wavelength 15.5 micron, is the radial attenuation at 10.65 expanded by multiplying the separation angle by 15.5/10.65. We show in Figure
\ref{fig:coronotrasmission} the transmission at PA = 45$^{\circ}$ for the three 4QPM filters.\\
~
Figure \ref{fig:vsSEP} shows how the total flux in filter F1550C varies as a function of the angular separation (for a fixed PA), while Figure \ref{fig:vsPA} shows how the total flux varies depending on the PA (for a fixed $sep$). \\
Consequently, in order to account for the minimal coronagraphic attenuation, we suggest to set the planetary companions along the diagonals of the detector, i.e., PA = 45$^{\circ}$, 135$^{\circ}$, 225$^{\circ}$, 315$^{\circ}$. For our simulations we set the PA of all planetary companions at PA = 45\textdegree.
The only exception is the HR 8799 system, which needed a different configuration in order to simultaneously image all four planets, without having any of them hidden by the coronagraph axes (i.e. where the coronographic attenuation is maximal and the centroid measurement error is the largest; \citealt{Lajoie2014SPIE}). 
We did not vary the choice of PA over the filters to be coherent in our results.

After having set the planetary PA we generated for each 4QPM filter (Tab. \ref{tab:filters}), 
for each element in the coronagraph target list (Tab. \ref{tab:targets}) and for each case (Tab. \ref{tab:cases}), one coronagraphic image for the star, 
one coronagraphic image for the reference, and one coronographic image for each planet in the system. Finally, for each planetary system we produced a total of 15 sets (i.e. 3 filters, 5 cases)  of 3 coronagraphic images (i.e. the trio star, planet, reference).\\
These images (an example is given in Fig. \ref{fig:stellarpsf}) were normalized to the PSF of the target star as seen on the imager at 45$^\circ$, far from the center so that the coronagraphic effect is negligible.

\begin{deluxetable*}{l l}[t!]
\tablecaption{Values of MIRI and telescope parameters utilized in the simulations. }
\tablecolumns{2}
\tablewidth{0pt}
\tablehead{
\multicolumn{2}{c}{System parameters values} 
}
\startdata
	Mirror diameter \dotfill & 6.57 m \\ 
    Primary mirror area\dotfill  & 25 m$^{2}$ \\
    Amplitude of telescope defocus \dotfill & 2 mm\\
	Telescope transmission  \dotfill & 0.9216 \\
	MIRI transmission \dotfill & 0.9272 \\
    Pupil rotation\tablenotemark{*}\dotfill  & 0.5$^{\circ}$\\
    Star pupil shift\tablenotemark{*} \dotfill &   3$\%$   \\  
	Reference pupil shift\tablenotemark{**} \dotfill & 0$\%$ \\
	Read out noise amplitude \dotfill  & 14 e$^{-}$ \\
    Detector quantum efficiency \dotfill  & $\geq 60 \%$ \\
	Detector subarray dimension \dotfill & 256 x 256 pxl  \\
    Pixel field of view \dotfill & 0.11\arcsec x 0.11\arcsec\\
	Flat field error measurement \dotfill & 1e-3 \\
    Detector saturation \dotfill & 250000 e$^{-}$ \\
    Limit $K_s$ magnitude for coronagraph saturation \dotfill & $\sim$ ~-1.9 \\
\enddata
\centering 
\tablecomments{~(*) relative to the telescope pupil; (**) relative to the star pupil\\.}
\label{tab:intrumentparams}
\end{deluxetable*}

\subsubsection{Science image creation} 
\label{sec:scienceimage}
To produce the science image we multiply the normalized coronagraphic image of each object by its respective spectrum, 
integrated over the MIRI bands (see $\S$~\ref{sec:models} for more details on the modelling). 
Note that the spectral flux in the coronagraphic image, and consequently its photon noise, is a function of the planetary system distance, the telescope mirror area, both telescope and optics transmissions, the MIRI detector quantum efficiency, and of the integration time $t_{\rm int}$. 
Table \ref{tab:intrumentparams} summarizes the values of these and more parameters, specific of both MIRI and the telescope, that were used during the simulations.

From the resulting three images (star, reference and planet) we merged stellar target and planetary image by adding them together. 
In the case of a system with multiple planets, we produce one image for each planet and we combined them together with the target star image.

Thereafter we added to both planetary system (star + planet) and 
reference images both  sky and telescope background (\citealt{Glasse2015}, Fig.1) and photon noise.
At the detector level we added readout noise (\citealt{Ressler2015}) and 
flat field error measurement (Tab. \ref{tab:intrumentparams}).
Note that, for $k_P$, the only sources of noise are Poisson noise, background, and detector noise.\\

\begin{figure}[t!]
\centering
\includegraphics[clip,trim=1cm 0cm 1cm 1.3cm,width=0.49\linewidth]{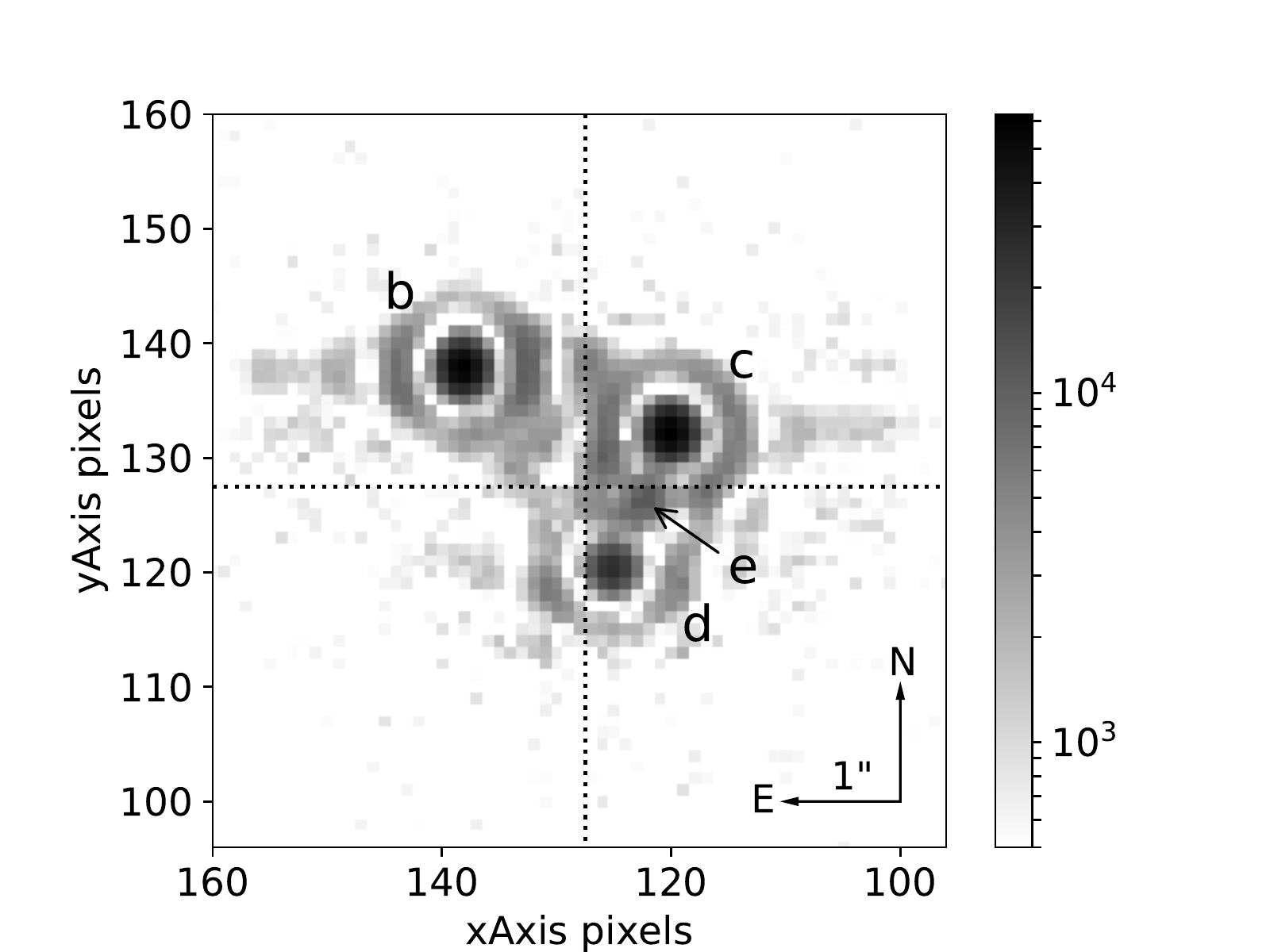}
 \includegraphics[clip,trim=1cm 0cm 1cm 1.3cm,width=0.49\linewidth]{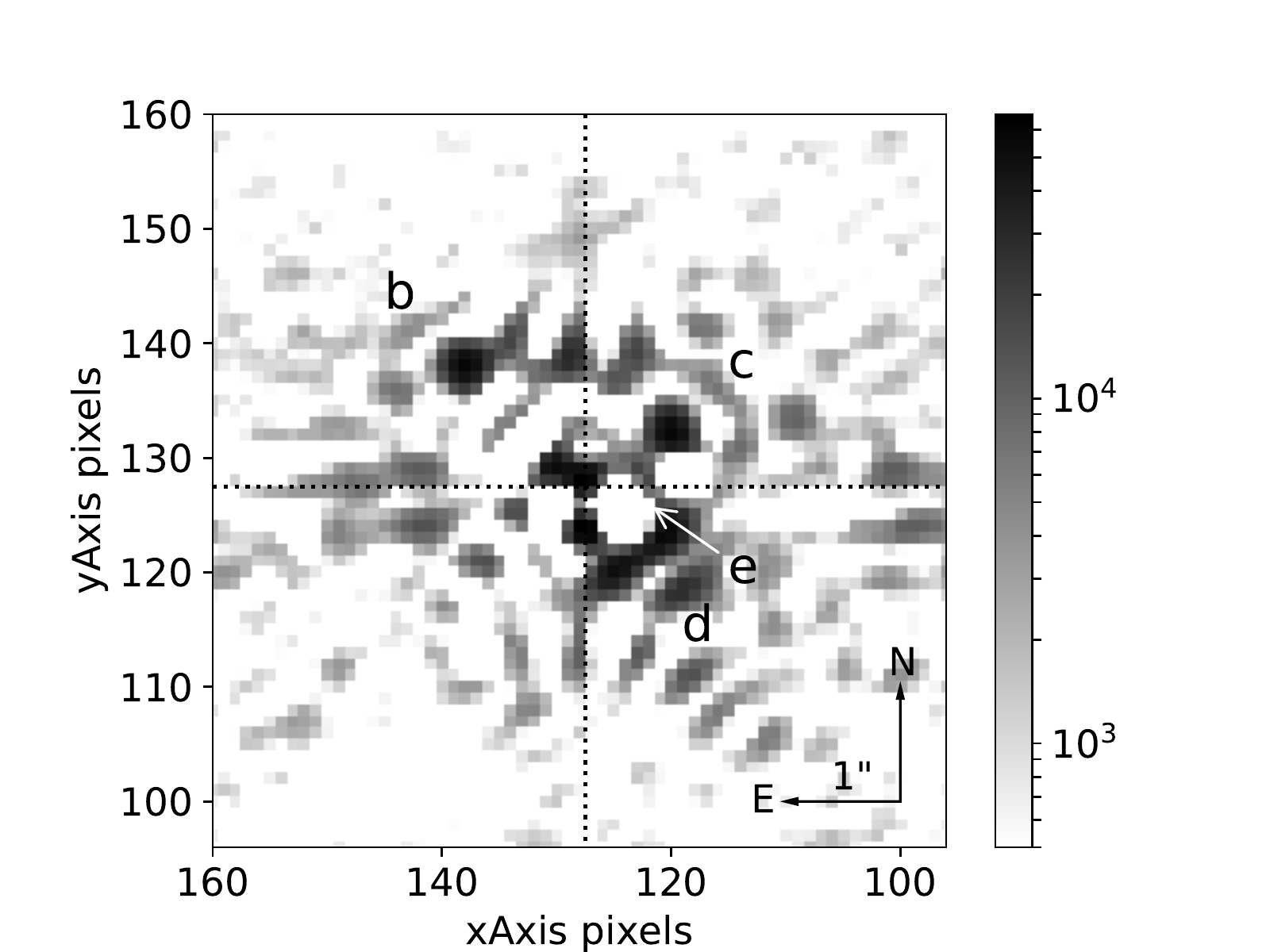}
 \caption{Zoom-in of the HR 8799 system final scientific image ($k_A$, \textit{left} and $k_D$, \textit{right}, equilibrium case, filter F1065C, $t_{\rm int}$  = 1800 s) after the reference image subtraction. The color scale unit is electrons, pixel size is 0.11\arcsec} 
\label{fig:diffimage}
\end{figure}

Finally we obtained the science image by subtracting the reference image from the planetary system image.
Figure \ref{fig:diffimage} shows the science image of the  
HR 8799 system, for cases $k_A$ and $k_D$, both at F1065C, in the atmospheric equilibrium case.\\
We remark that, throughout the whole science image simulation process, both star and reference images share the same WFE, telescope defocus amplitude, and telescope jitter amplitude (but not the same jitter realizations, which differ for the two objects). Variations in the WFE rms and/or in the jitter amplitude values between star and reference images are not taken into account in this study.

%% ________________________________________ PHOTOM UNCERTAINTIES _______________________________________
\subsection{Coronagraphic SNR and photometric uncertainties}
\label{sec:SNRandUnc}
To account for different noise realizations, and to estimate the photometric uncertainties, we 
generated  for each target, filter, case ($k_{\rm A,B,C,D,P})$ and for different 
integration times, a cube of n = 500 scientific images (see $\S$~\ref{sec:scienceimage}).
Images within a cube were produced using the same set of three normalized coronagraphic images: one for the target star, one for the planet and  one for the reference star (see \ref{sec:PSFsimu}), meaning that science images in the cube differ only in random noise but not in WFE and telescope jitter.\\
We then performed aperture photometry on each image to measure the planetary flux $F$ by using a mask of fixed aperture centered on the planet's centroid, which was measured by knowing the exact position of the planet on the FOV. 
The radius of the photometric mask used  was r = 2.5 pixels for filters F1065C and F1140C,
and r = 3.5 pixels for F1550C to account for the larger planetary PSF.
  
For each cube of case $k$ (Tab. \ref{tab:cases}),  filter $\lambda$ (Tab. \ref{tab:filters}) and integration time $t_\mathrm{int}$ (i.e. 600 s, 1200 s, 1800 s), we measured the photometric uncertainty 
$(\sigma_{F})_{k, \lambda, t_{\rm int}} =  \sqrt{\frac{\sum\limits_{i =1}^{\mathrm{n}}
(F_{i} - \bar{F})_{k, \lambda, t_{\rm int}}^2}{\mathrm{n}}}$ %_{k, \lambda, T_{int}}$
and signal-to-noise ratio 
(SNR)$_{k, \lambda, t_{\rm int}}$ = ($\bar{F}/\sigma_{F})_{k, \lambda, t_{\rm int}}$ corresponding to each image in the cube. $\bar{F}_{k, \lambda, t_{\rm int}}$ is the mean number of electrons measured over each cube $k$ and $i$ are the images in the cube.

%% ________________________________________ CONTRAST PART _______________________________________

\subsection{About speckles and planetary detection}
\label{sec:contrast}
To validate a planetary detection it is first necessary to quantify the effect of stellar residuals' noise, 
and compare it to the planet-to-star contrast of each target, while taking into account the coronagraphic attenuation, which has effect up to an angular distance of about 3$\arcsec$ from the center (see Fig. \ref{fig:coronotrasmission}).

\begin{figure}[!p]
\centering
\includegraphics[clip, trim=0.cm 0.cm 0.3cm 2.1cm, width= 0.94\linewidth]{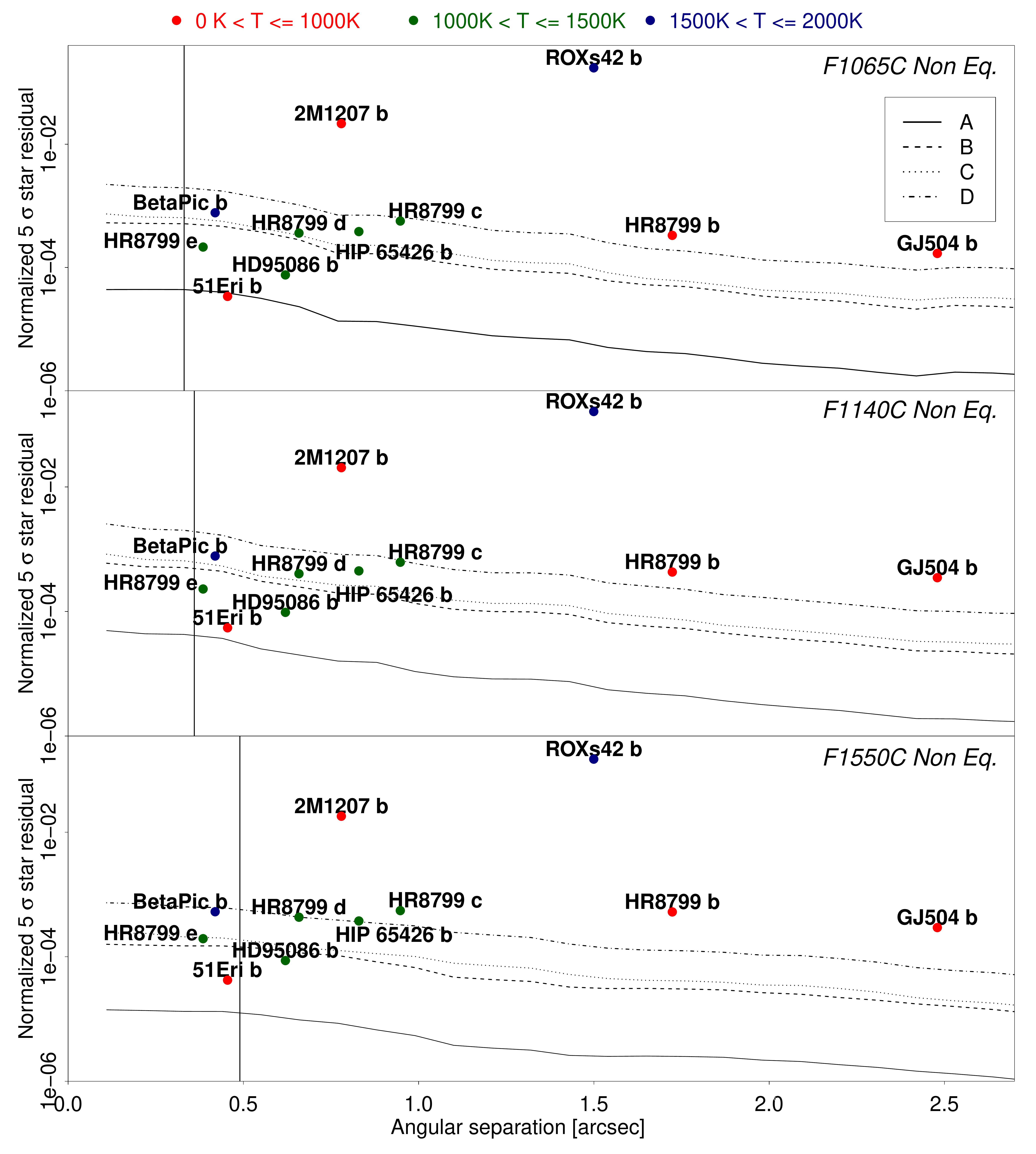}
\caption{Log-normalized 5 $\sigma$ contrast curves for the four $k$ cases under study (i.e. A, B, C, D, Tab. \ref{tab:cases}) and 4QPM filters, as a function of the distance from the coronograph center. 
The values reported were obtained using differential coronographic images (see \S~\ref{sec:contrast}), they are hence a function of WFE, telescope jitter realizations and stellar offset. No random noise is included.
Dots represent the planet-to-star contrast (non-equilibrium case) of each target, attenuated by the coronagraph. The coronagraphic attenuation varies with the distance from the center of the mask, following the coronagraph radial transmission curve (Fig. \ref{fig:coronotrasmission}).
Colors correspond to the temperature of the target as indicated in the top legend.
When a planet is lying above a curve it means that we can detect it at least at 5 $\sigma$
while, when it is below, the detection is lower than 5 $\sigma$. 
Black vertical lines indicate the IWA value of each filter.} 
\label{fig:contrast}
\end{figure}

To do so we built for each $k$ case a differential coronagraphic image by subtracting 
the normalized coronagraphic image of the reference star from the normalized coronographic image of the target star. 
We then measured the standard deviation of the stellar residuals over various annular areas of five pixels width and increasing radius (of one pixel step), all centered on the 4QPM center. For filter F1550C the annular width was set to seven pixels to cover the size of the PSF. 
We note that, by working on the normalized images (see \S~\ref{sec:PSFsimu}) and not on the science images, we quantified only the intensity of the residual speckle noise (i.e. there are no other sources of noise apart from the stellar residual flux due to the offset between star and reference offset, and the telescope jitter).
Figure \ref{fig:contrast} shows the 5 $\sigma$ stellar residual flux (henceforth labeled as \textit{contrast curves}) for $k$ cases  A, B, C, D for each MIRI coronagraphic filter, as a function of the distance from the center of the differential coronagraphic image.
Planets lying above a contrast curve are those that can be detected at 5 $\sigma$ at least.\\
For each case $k$ and filter $\lambda$, we measured the signal-to-noise ratio (SNR$_\mathcal{S})_{k,\lambda}$ of the planet-to-star contrast (henceforth referred to as \textit{planetary contrast}) over the 
speckle residuals' noise $\mathcal{S}$  by dividing the planetary contrast of each planet by the value of the contrast curve at the respective planetary separation. 
We provide in Appendix \ref{app:A} planetary contrast values and in Appendix \ref{app:B} contrast curves values for every case $k$ and 4QPM filter.

\begin{figure}[!b]
\centering
\includegraphics[clip,trim=0.2cm 0cm 0.4cm .2cm, width=0.45\linewidth]{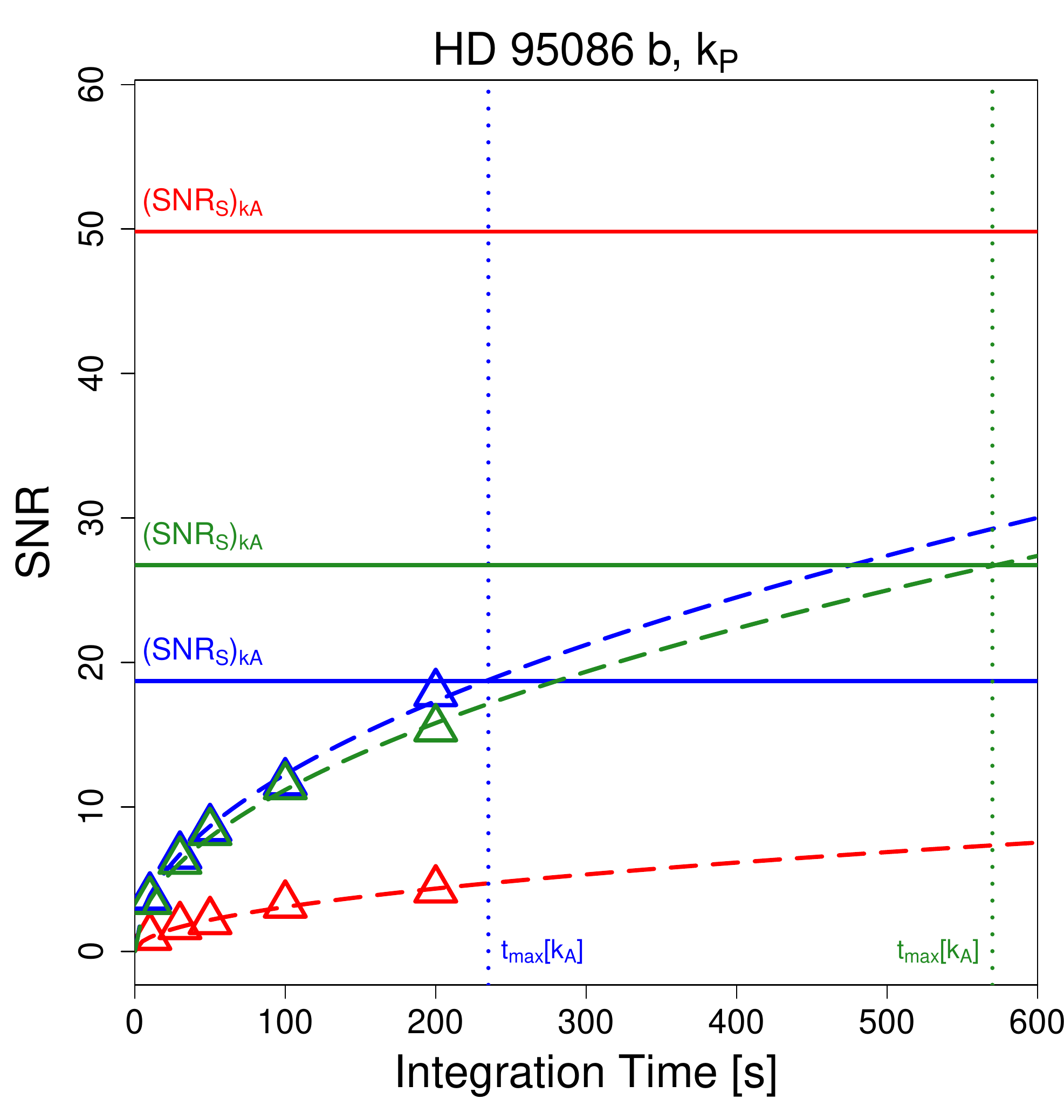}\quad
\includegraphics[clip,trim=0.2cm 0cm 0.4cm .2cm, width=0.45\linewidth]{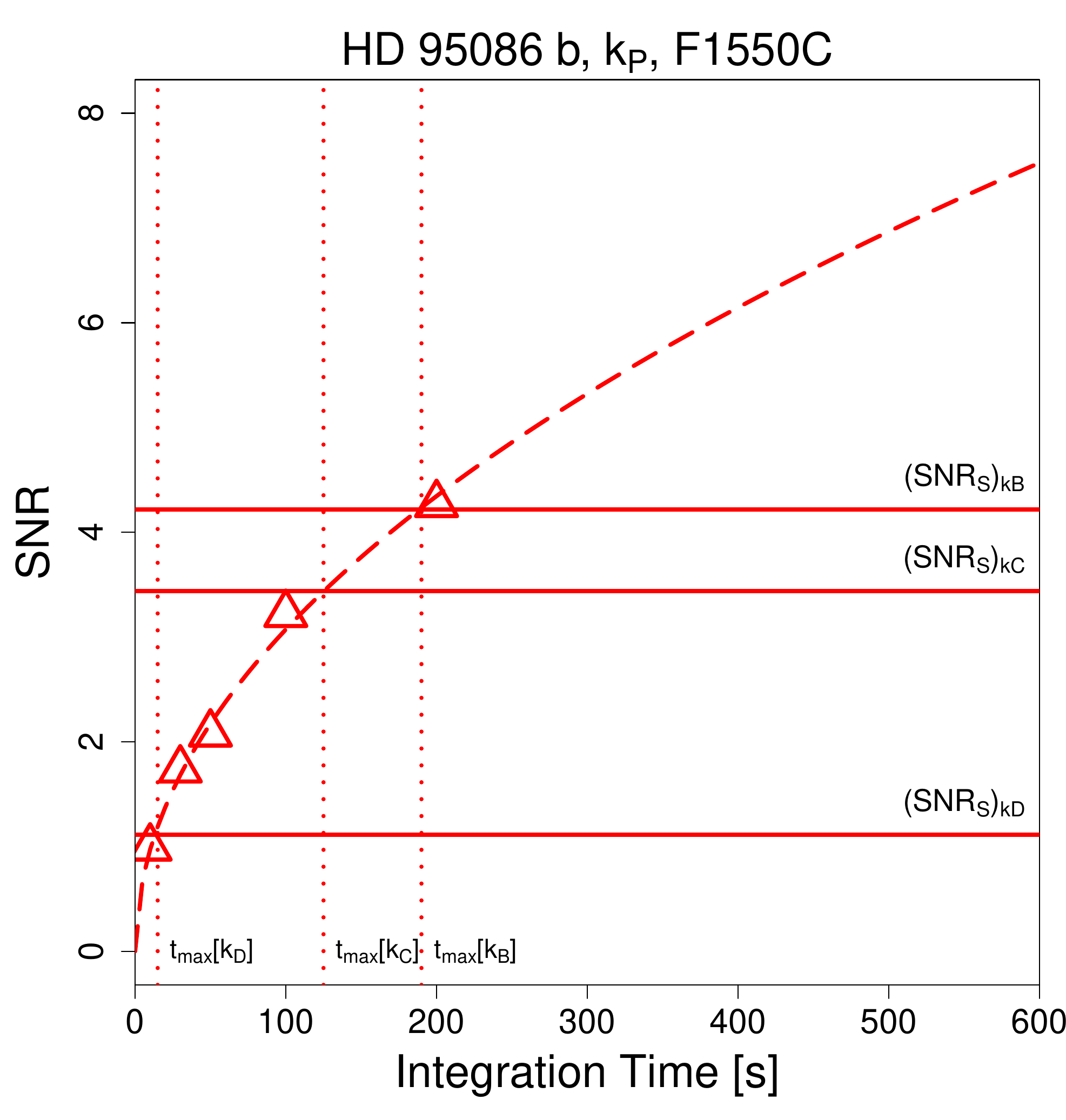}
\caption{\textit{Left: }SNR$_{kP}$ of HD 95086 b as a function of integration 
time $t_{\rm int}$. Colors represent values relative to filter F1065C (\textit{blue}),  F1140C (\textit{green}) and F1550C (\textit{red}). Triangles represent 
the SNR measured in $\S$ \ref{sec:SNRandUnc} for chemical-equilibrium model, for integration times of $t_\mathrm{int}$ = 10, 30, 50, 100, 200 s. 
Dashed lines are the fit to the simulated SNR values ($\sqrt{t_{\rm int}}$), while solid lines represent the  signal-to-noise relative to the speckle intensity of case $k_A$, (SNR$_\mathcal{S}$)$_{kA}$. The higher the (SNR$_\mathcal{S}$)$_{kA}$, the longer it is possible to integrate and to improve the planetary signal.
The integration time $t_{\rm max}$, defined as the time when SNR$_{kP}$ = (SNR$_\mathcal{S}$)$_{kA}$, is marked by vertical dotted lines and it is reported in Table \ref{tab:tmax}. Note that red{$t_{\rm max} $ = > 7200 s} for filter F1550C is outside the plot.
\textit{Right:} Same as the left panel, but only for filter F1550C. Here 
solid lines represent the signal-to-noise relative to the speckle intensity of case $k_B$: (SNR$_\mathcal{S}$)$_{kB}$, $k_C$: (SNR$_\mathcal{S}$)$_{kC}$, and $k_D$: (SNR$_\mathcal{S}$)$_{kD}$.
Note how $t_{\rm max}$ scales, depending on the observational case ((SNR$_\mathcal{S}$)$_{kA}$ does not appear due to a different y-axis scale, which was employed for clarity reasons). 
}
\label{fig:tmax}
\end{figure}

%% ________________________________________ TMAX ______________________________________
The intensity of stellar residuals, likewise planetary signal, is proportional to the square root of the integration time, we hence 
measured the optimized integration time ($t_{\rm max})_{k,\lambda}$ beyond which the planetary signal would not better stand out of speckle residuals noise. \\
($t_{\rm max})_{k,\lambda}$ is a function of both filter $\lambda$ and case $k$ as the stellar residuals' intensity varies for each case $k$.  For the following we focused on speckle residuals of the optimistic case $k_A$ (and therefore we use the notation $t_{\rm max}$ for simplicity).
To compute $t_{\rm max}$  we fitted the function $\sqrt{t_{\rm int}}$ 
to the (SNR)$_{kP, \lambda}$ values, measured for different integration times (see $\S$ \ref{sec:SNRandUnc}) for case $k_P$ and we calculated the time where (SNR)$_{kP, \lambda}$ =  (SNR$_\mathcal{S})_{kA,\lambda}$ (Figure \ref{fig:tmax}, left panel).  
Table \ref{tab:tmax} reports values of $t_{\rm max}$ with their corresponding signal-to-noise values.
Note that we utilized the (SNR)$_{kP, \lambda}$ values of case $k_P$ because this specific case presents no speckle residuals, we could hence extrapolate the pure planetary signal when fitting the function $\sqrt{t_{\rm int}}$. For the same concept it is possible to use the values reported in the table to infer the SNR measured at an integration time different from the one reported, by fitting the same function.

As mentioned above the signal-to-noise ratio is proportional to the square root of the integration time, such as (SNR)$_{kP, \lambda}  = \gamma \sqrt{(t_{\rm max})_{k,\lambda}}$ where $\gamma$ is a coefficient that depends on the exoplanet analyzed, in term of planetary contrast and angular distance values. Values of $(t_{\rm max})_{k,\lambda}$ for cases $k$ other than $k_A$ can therefore be deduced from the $k_A$ case using the formula: 
~
\begin{equation} 
(t_{\rm max})_{k,\lambda} = t_{\rm max} \left(\frac{({\rm SNR}_\mathcal{S})_{k,\lambda}}{({\rm SNR}_\mathcal{S})_{kA,\lambda}}\right)^2 
\end{equation}
~
Using the planetary contrast values (Appendix \ref{app:A}) and the contrast curves values (Appendix  \ref{app:B}), it is hence possible to obtain the $({\rm SNR}_\mathcal{S})_{k,\lambda}$ and, consequently, $(t_{\rm max})_{k,\lambda}$ for each case $k$.
Figure \ref{fig:tmax}, (right panel) shows an example of how the maximal integration times scales as a function of the observational cases $k$.\\

Throughout this work we used the telescope background values provided by \cite{Glasse2015}, though we tested for a higher background, to leave a margin of error. We simulated the case k$_P$ of HIP 65426 b (i.e. faintest coronagraphic target star) for filter F1550C (i.e. with highest background among the 4QPM filters) non-equilibrium chemistry, increasing background values by \cite{Glasse2015} of 50$\%$.
We found that  $t_{\rm max}$ > 7200 s also for the increased background case. More specifically:
at $t_{\rm int}$  = 1200 s we measured a  SNR$_{\rm +50\%}$ = 10 (previously  SNR = 13), and 
 at $t_{\rm int}$  = 2400 s we measured a SNR$_{\rm +50\%}$ = 15 (previously  SNR = 18.)   \\

%% _______________________________________ LRS SIMU _______________________________________

\begin{deluxetable*}{l l | r | r || r | r || r | r  }[!ht]
\tablecaption{Measured integration time $t_{\rm max}$ beyond which planetary signal would not better stand out of case $k_A$ speckle noise. $t_{\rm max}$ value corresponds to the integration time when the signal-to-noise of case $k_P$ (i.e. where Poisson noise and background are the only sources of noise:
no jitter and no offset between target star and reference are included) is equal to the signal-to-noise of the planetary contrast versus the speckle residuals of case $k_A$, hence when SNR$_{kP}$ = (SNR$_\mathcal{S})_{kA}$. 
The equilibrium and non equilibrium cases are represented by the \textquotedblleft eq\textquotedblright~ and  \textquotedblleft neq\textquotedblright~ strings, respectively.}.
\tablecolumns{5}
\tablewidth{0pt}
\tablehead{
\colhead{Planet} & \colhead{} &  \multicolumn{2}{c}{F1065C} & \multicolumn{2}{c}{F1140C} &\multicolumn{2}{c}{F1550C} \\
\colhead{}  & \colhead{} &  \colhead{(SNR$_\mathcal{S})_{kA}$} & \colhead{$t_{\rm max}$[s]} &  \colhead{(SNR$_\mathcal{S})_{kA}$ } & \colhead{$t_{\rm max}$ [s]} & \colhead{(SNR$_\mathcal{S})_{kA}$} & \colhead{$t_{\rm max}$[s]} 
}
\startdata
\hline
{\BetaPic~ b}  & eq & 98 & 15 & 102  & 20 & 198 & 700 \\
			   & neq & 95 & 15 & 102 & 20 & 198 & 680 \\ 	
\hline
{51 Eri b }   & eq & 2 & 25 & 7 & 40 & 15 & 1020 \\
			   & neq & 4 & 25 &  8 & 40 & 16 & 995 \\	
\hline
{GJ 504 b } & eq & 152 & 4180 & 570 & > 7200 & 227&  > 7200 \\
		    & neq & 447 & 5185 & 749 & > 7200 & 245 & > 7200  \\ 		
\hline
{HD 95086 b} & eq & 19 & 235 & 27 & 570 & 26 & > 7200 \\ 
			 & neq & 14 & 210 & 22 & 565 & 22 & > 7200 \\ 		
\hline
{2M1207 b} & eq & 625 & > 7200 & 528 & > 7200 & 144 &  > 7200\\
		   & neq & 542 & > 7200 & 430 & > 7200 & 123 & > 7200 \\ 	
\hline
{ROXs42 b} 	& eq & 820 & > 7200 & 663 & > 7200 & 196 &  > 7200\\
			& neq & 820 & > 7200 & 684 & > 7200 & 196 &  > 7200\\ 	
\hline
{HIP 65426 b} & eq & 162 & > 7200 & 147  & > 7200 & 34 & > 7200 \\
			  & neq & 129 & > 7200 & 124 & > 7200 & 30 & > 7200 \\ 	
\hline
{HR 8799 b} & eq & 771 & >7200 & 682 & > 7200 & 296 & > 7200 \\
			& neq & 402 & 6120 & 442 & > 7200 & 203  & > 7200 \\ 	
\hline
{HR 8799 c} & eq & 293 & 950 & 289 & 1995 & 206 & > 7200  \\
		   & neq & 235 & 855 & 245 & 1805 & 192 & > 7200  \\ 	
\hline
{HR 8799 d} & eq & 99 & 1070 & 124 & 2110 & 101  & > 7200 \\
		   & neq & 78 & 1000 & 100 & 1815  & 92 &  > 7200 \\ 	
\hline
{HR 8799 e} & eq & 31 & 990 & 33 & 1150 & -- & --  \\
		   & neq & 27 & 950 & 29 & 1095 & -- & -- \\ 	               
\hline
\enddata
\label{tab:tmax}
\tablecomments{For those cases where $t_{\rm max}$ > 7200 s, we report the SNR measured at exactly 7200 s.}
\end{deluxetable*}

%% ________________________________________ DISCUSSION PART _______________________________________

\section{Discussion }
\label{sec:discussion}

We note that in our analysis no advanced post-processing techniques  (e.g. PCA, \citealt{Choquet2016}) have been applied:
results are obtained by directly using the science image obtained after the subtraction of the reference image from the target image and by knowing the exact position of the planet on the image.
The use of the currently known modern post-processing techniques will improve the planetary SNR for cases $k_B$, $k_C$ and $k_D$ reaching up to the $k_A$ level, where optimal stellar subtraction is achieved (even though there are both jitter effects and some residual stellar photon noise, because such noise is different for target star and reference).
\noindent As an example of the improvement we can have with modern post-processing techniques we refer to the work by \cite{Soummer2014SPIE} which presents a comparison between the contrast achieved with a classical reference subtraction and with the small-grid dither technique.
At 1\arcsec ~and F1140C, while the classical subtraction yields a 5 $\sigma$ contrast of the order of $\sim$ 10$^{-3.4}$ (i.e. in between our 5 $\sigma$  $k_C$ and $k_D$ cases), the 9-point dithers (dither step = 10 mas) returns a 5 $\sigma$ contrast of $\sim$ 10$^{-5.6}$, meaning an improvement factor of $\approx$ 150.  
~
We note that the value of the 9-point dither contrast is smaller than the one achieved in our $k_A$ case. Though, in their simulations the telescope jitter, which is the limiting contrast factor in our case $k_A$, was not taken into account. We can thus conclude that case $k_A$ provides a good limit of what can be achieved with small-grid dither technique, meaning, for instance, that our $k_D$ contrast could be improved at levels very close to the $k_A$ one. 

We remind the reader that $k_P$ is not realistic as it does not include the telescope jitter noise.  

\subsection{Planets' detectability}
\label{sec:planetsdetectability}
As explained in \S~\ref{sec:contrast} a planetary detection is feasible when the planetary contrast, attenuated by the coronagraph, is larger than the speckle contrast; this dimension, called (SNR$_\mathcal{S})_{k,\lambda}$, has been quantified for each planet with respect to all the cases $k$ of speckles' residuals and filters $\lambda$ (Fig. \ref{fig:contrast}), and needs to be  larger than 5 to assure a planetary detection. Table \ref{tab:tmax}  reports the values of (SNR$_\mathcal{S})_{kA}$ for all the planets observed in coronagraphic mode.
We remind the reader that contrast curves have been estimated by computing the standard deviation of the speckle residuals over various annular areas around the 4QPM center (see \S~\ref{sec:contrast}). 
While some sources appear easily detectable due to their intrinsic luminosity (e.g. 2M1207 b, ROXs42 b), for others the prospect of a detection strongly varies with the case $k$ in which the telescope is observing.

51 Eri b, HD 95086 b and HR 8799 e have relatively small (SNR$_\mathcal{S})_{kA}$ and 
their detection is more difficult, even with the coronagraph. \\
We note that the parameter that plays a major effect in a coronagraphic observation is the offset between the target star and the reference. Hence, for sources with low SNR$_\mathcal{S}$, specific coronagraphic target acquisition and operations, such as the small grid dither concept, in conjunction with sophisticated image analysis algorithms for optimizing the PSF subtraction (e.g. LOCI: \citealt{Lafreniere2007}; KLIP: \citealt{Soummer2012}), are necessary \citep[and references therein]{Soummer2014SPIE,Lajoie2016SPIE}. 

In our analysis 51 Eri b cannot be detected at 5 $\sigma$ in filter F1065C in the optimistic case observation ($k_A$) because it is buried in the speckle residuals.
%its detection in filter F1140C is borderline while it is not doable in filter F1550C as the planet is buried below speckle residuals noise. 
Note that beside detectability issues, 
it is possible that the planet is moving towards the star \citep{deRosa2015}, meaning that there is the possibility that 
it will not be observable by MIRI coronagraph in 2020, after JWST launch, being within the IWA. 

The planet HD 95086 b is detectable at 5 $\sigma$ only in case $k_A$ for all filters (both chemical states).

The planet \BetaPic~ b can be detected at 5 $\sigma$ in filter F1065C in all cases but case $k_D$. 
We note that a 5 $\sigma$ detection is possible at F1550C for cases $k_A, k_B, k_C$ despite the planet being inside $\lambda/D$ ((SNR$_\mathcal{S})_{kA}$ = 198,  Fig. \ref{fig:contrast}, Tab. \ref{tab:tmax}). 
This means that it is possible to push MIRI observations also for those planets with angular distance 
smaller than the IWA (here 0.42$\arcsec$ compared to $\lambda/D$ = 0.49$\arcsec$). 
By design, the 4QPM does not provide an abrupt cut at the IWA level, but instead a smooth transition which may allow the detection of planets within the IWA.
However, the planet image is significantly non-linearly affected by the coronagraphic mask  and so such a detection inside the IWA cannot be made in practice at any separations. This remains to be quantified yet as it is out of scope of this manuscript.

Concerning the HR 8799 system the only planet detectable in any condition is HR 8799 b.  HR 8799 c detection in $k_D$ is borderline for F1065C and F1140C both equilibrium and non-equilibrium cases. 
HR 8799 d can be detected at 5 $\sigma$ for all the cases but case $k_D$ in filters F1065C and F1140C (equilibrium case). For the non-equilibrium case the detection is difficult also in F1065C, $k_C$.  In F1550C, case $k_D$, the detection is borderline for both chemical states. 
HR 8799 e is only visible in filters F1065C and F1140C, case $k_A$. With only a classical reference subtraction technique the planet is not directly detectable in filter F1550C as the 
IWA is too large and also because, at this wavelength, its PSF is highly contaminated/covered by the larger PSFs of the other planets.

GJ 504 b is always detectable at 5 $\sigma$ apart from $k_D$, F1065C equilibrium case.

For the HIP 65426 b simulation we used the Exo-REM model at $T_P$ = 1300 K which is the 
minimal $T_P$ estimated by 
\citealt{Chauvin2017}, meaning that we measure the inferior limit of  (SNR$_\mathcal{S})_{kA}$ 
and that larger signal-to-noise could 
be achieved if the planet is warmer than 1300 K ((SNR$_\mathcal{S})_{kA}$ = 34 for F1550C and integration time $t_{\rm int}$ = 7200 s for the equilibrium case). A 5 $\sigma$ planetary detection is not possible only for case $k_D$ in filters F1065C and F1140C (both equilibrium and non-equilibrium cases). For F1550C the detection becomes challenging also for $k_D$, both chemical states. 

We note that these simulations were performed with a specific angular position (i.e., 45$^{\circ}$), in order to maximize the planetary flux. The results are the outcome of a optical path with specific parameters (such as the amplitude of the telescope defocus, the stellar pupil shift, the pupil rotation, etc., see Tab. \ref{tab:intrumentparams}) which can vary when the telescope will be in orbit.

%% ________________________________________ AMMONIA _______________________________________
\subsection{$NH_3$ in an exoplanetary atmosphere.}
\label{sec:ammonia}

\begin{figure}[t!]
 \centering
 \includegraphics[clip,trim=0.cm 0cm 2.cm 1.cm,width=0.9\linewidth]{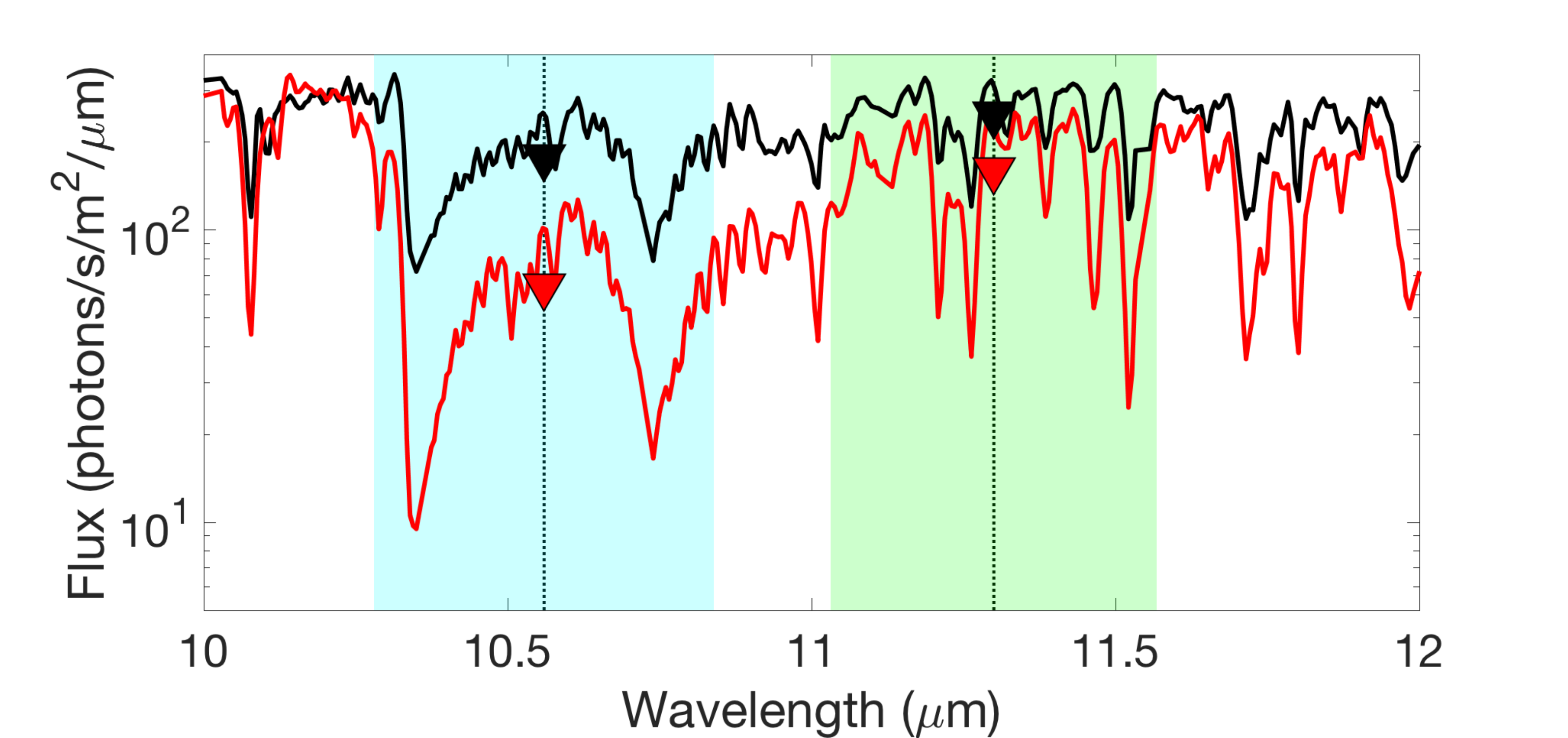} 
 \caption{Exo-REM models for GJ 504 b  equilibrium case (\textit{red}) and non-equilibrium case  (\textit{black}). Note that, given the temperature of $T_P = 554$ K , the equilibrium model has a deeper NH$_3$ feature than the non-equilibrium model.  Black and red triangles mark the integrated flux in each filter. Shaded areas mark the F1065C (\textit{light blue}) and F1140C (\textit{green}) filters' range, while dashed vertical lines mark the filters' central wavelength.}
 \label{fig:NH3models}
\end{figure}

The ammonia absorption band (centered at $\lambda = 10.65~\mu$m) is visible in the synthetic spectra of 
planets whose temperature is $T_P \lesssim$  1200 K and it can be a useful index of both planetary temperature and atmospheric chemical equilibrium state.
While for the temperature index the general rule is that the cooler the planet, the stronger the NH$_{3}$ band,  for the chemical equilibrium index it is important to distinguish two regimes: one with temperature $T_P$ < 1000 K and one  with 1000 K < $T_P \lesssim$ 1200 K.
This difference is due to the vertical profile of NH$_3$ at thermochemical equilibrium, which increases with height in the lower atmosphere, reaches a minimum, and then increases with height at higher levels. This behavior results from a competition between decreasing temperature (that favors NH$_3$) and decreasing pressure (that favors N$_2$) with height. For objects with relatively low $T_P$ (GJ 504 b and 51 Eri b), the NH$_3$ equilibrium abundance at the quench level is smaller than that in the emitting region (aka. photosphere), leading to a stronger absorption feature in the equilibrium case. The situation is opposite in the second regime (e.g. HR 8799 b), as the NH$_3$ abundance profile at equilibrium decreases between the quench level and the emitting region.
Figure \ref{fig:NH3models} shows an example for the case of GJ 504 b, whose temperature put the planet in the first regime ($T_P$ = 544 K).

In our coronagraphic target list 2M1207 b, 51 Eri b, GJ 504 b, HD 95086 b, HR 8799 b, c, d, e  have such temperatures to potentially host ammonia in their atmosphere. 
In the event that this molecule is present and abundant in the atmosphere, we would observe an increase of flux from filter F1065C to filter F1140C. On the other hand, for a relatively low abundance of NH$_3$,
it was thought to use the combination of F1550C and F1140C observations, to retrieve the ammonia line and its abundance, by fitting to the observations a black body curve.

We applied this method in our analysis in order to validate a possible MIRI/4QPM detection of NH$_{3}$ in the atmosphere of the above mentioned exoplanets.
We produced a black body (BB) curve at the planetary temperature $T_P$
of reference (Tab. \ref{tab:targets}), we then integrated this BB curve on the 4QPM coronagraphic filters, obtaining three BB data points centered on the 4QPM filters' central wavelengths $\lambda_C$. 
We then corrected each of these three BB data points for the respective coronagraphic attenuation and filter transmission values (these corrections are necessary because they are wavelength dependent).
Therefore we minimized the flux offset between the two BB points at $ \lambda_2$ = 11.40 $\mu$m and $ \lambda_3$ = 15.50 $\mu$m  and their respective simulated ones.
The uncertainty on this minimization is given by $\sigma_M = \sqrt{\sigma^2_{\lambda_2}+\sigma^2_{\lambda_3}}$ where $\sigma^2_{\lambda_2}$ and $\sigma^2_{\lambda_3}$ are the observed variance at $\lambda_2$ and $\lambda_3$, respectively.\\
We then compared the simulated data point with the black body point at $\lambda_1$: in the case where the simulated flux is dimmer and not consistent with the equivalent BB one, it means that we are detecting an absorption in the planetary spectra, in this specific case the NH$_{3}$ one. 
The significance $\sigma_\mathrm{NH3}$ of this detection has been measured as:

\begin{equation}
\sigma_\mathrm{NH3} = \frac{(F_\mathrm{BB})_{\lambda_1} - (F_\mathrm{obs})_{\lambda_1}}{\sigma_\mathrm{tot}}
\end{equation}

\noindent where 
($F_\mathrm{BB})_{\lambda_1}$ is the BB point calculated at $\lambda_1$, 
$\sigma_\mathrm{tot} = \sqrt{\sigma^2_M + \sigma^2_{\lambda_1}}$, and where $(F_\mathrm{obs})_{\lambda_1}$ and $\sigma^2_{\lambda_1}$ are the observed flux and variance at $\lambda_1$, respectively.  
~\\
We note that we are assuming no error on the black body curve, though this error 
will have to be accounted for when performing the same analysis on real data. 

\begin{SCtable}
\centering
\begin{tabular}{l c c }
\hline \hline
Planet & eq & neq\\
\hline
    2M1207 b & 3 $\sigma$  & 4 $\sigma$  \\  
    GJ 504 b &  98 $\sigma$  &  66 $\sigma$ \\ 
    HR 8799 b &  16 $\sigma$ & 30 $\sigma$ \\ 
    HR 8799 d &  -- & 3 $\sigma$  \\ 
    \hline
\end{tabular} 
\caption{Significance of ammonia detection for case $k_A$ when the planet has SNR$_\mathcal{S}$ > 5 in all 4QPM filters at $t_{\rm int}$ = 1800 s for both chemical state of equilibrium (\textquotedblleft eq\textquotedblright) and non-equilibrium (\textquotedblleft neq\textquotedblright) obtained using a black-body approximation}. 
\label{tab:NH3significance}
\end{SCtable}

\begin{figure}[t!]\centering
\includegraphics[clip,trim=0.5cm 0cm 2cm 0.cm,width=0.5\linewidth]{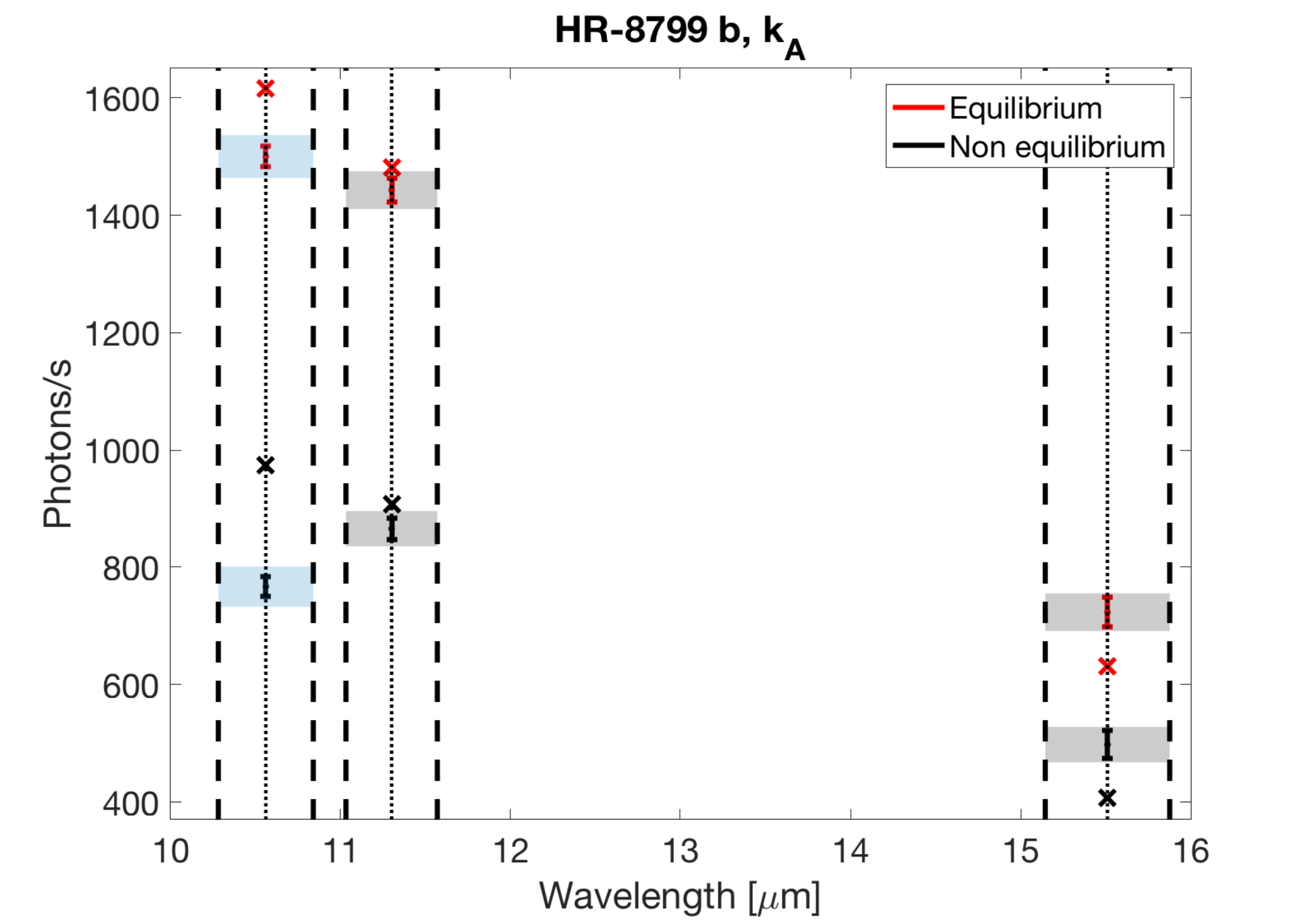}\includegraphics[clip,trim=0.5cm 0cm 2cm 0.cm,width=0.5\linewidth]{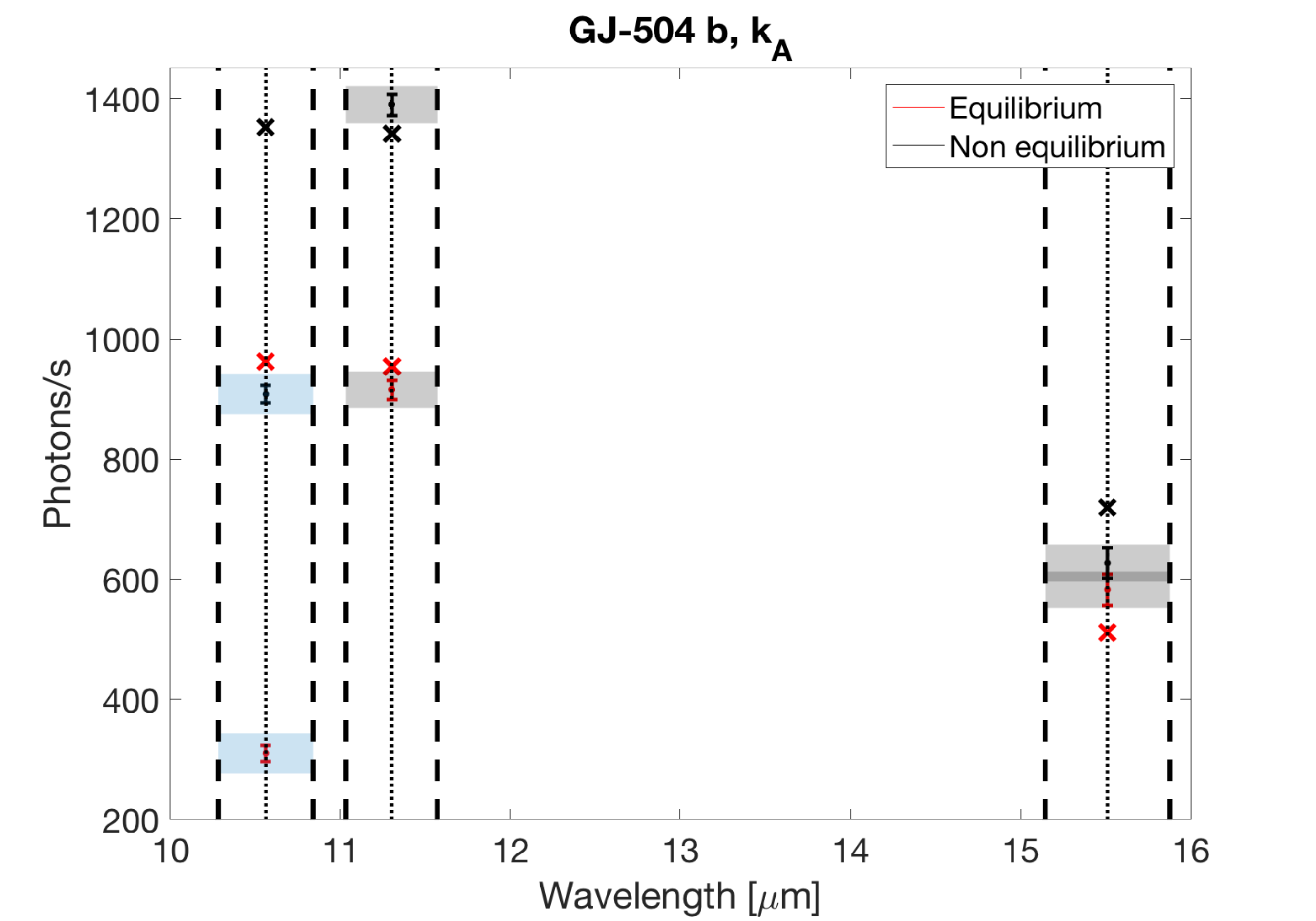}
     \caption{Case $k_A$ simulated flux (\textit{errorbars}) vs blackbody flux (\textit{crosses}) of HR 8799 b (\textit{left}) and GJ 504 b (\textit{right}). 
     The 5 $\sigma$ errorbars are plotted for both equilibrium (\textit{red}) and non equilibrium (\textit{black}) cases. Shaded areas correspond to $5~\sigma_M$ (\textit{grey}) and 5 $\sigma_\mathrm{tot}$ (\textit{blue}). 
     Dashed vertical lines mark the filters wavelength range, while dotted vertical lines mark filters' central wavelength.} 
     \label{fig:HR8799_GJ504_NH3}
\end{figure}

For an integration time of $t_{\rm int}$ = 1800 s,  our analysis shows a clear detection (i.e. $\sigma_\mathrm{NH3}$ > 5) of ammonia in the atmosphere of HR 8799 b and GJ 504 b in both chemical states. For 2M1207 b (both chemical states) and  HR 8799 d (only non-equilibrium) the significance is 3 $ \leq \sigma_\mathrm{NH3}$  < 5. Table  \ref{tab:NH3significance}
reports the significance of the ammonia detection for those planets whose (SNR$_\mathcal{S})_{k_A}$ > 5 in all the three coronagraphic filters.

Figure \ref{fig:HR8799_GJ504_NH3} shows our results for HR 8799 b and GJ 504 b, $k_A$. We observe that for the HR 8799 b case the black body point and the observed point at F1550C (both chemical states) are not consistent with each other. The reason for this lies in the fact that the Exo-REM model and the BB curve at temperature $T_P$ differ. It therefore follows that it not always possible to approximate (to the first-order) a planetary spectrum to a black body curve. This approach risks to give imprecise results such as under/over-estimation of both ammonia detection significance and its absolute abundance. 
For this reason we strongly suggest the use of radiative transfer models possibly coupled with \textit{inverse retrievals} techniques (e.g. \citealt{Irwin2008, Line2013, Waldmann2015taurex, Rocchetto2016}) to objectively determine the absolute gas abundances .\\
A specific study presenting the use of retrievals methods on MIRI data will be presented in a forthcoming paper.

\subsection{Other molecules}

\begin{figure}[t!]
\centering
\includegraphics[clip,trim=0cm 0cm 0cm 0.cm,width=0.7\linewidth]{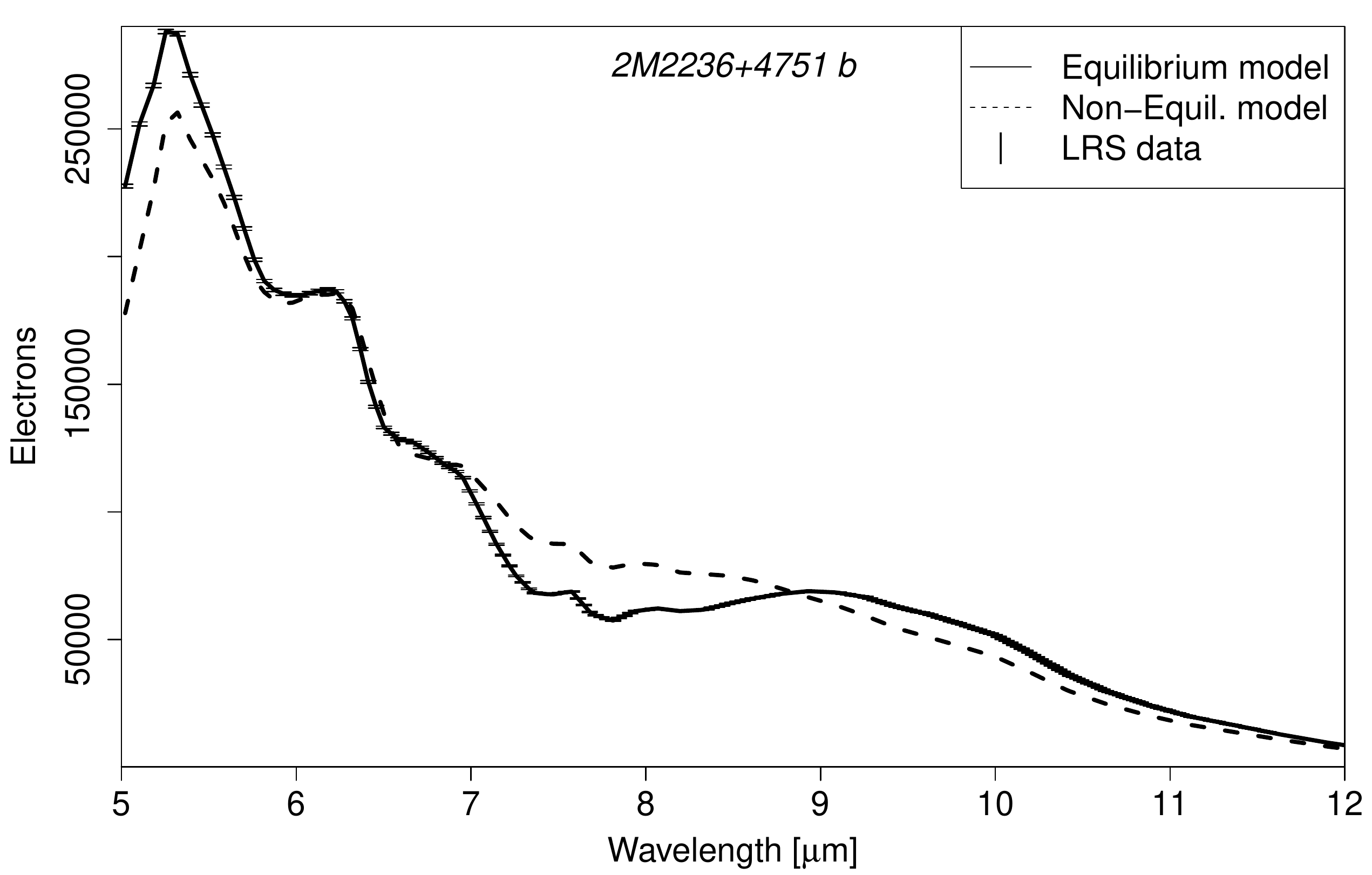}
\caption{Extracted spectrum of 2M2236+4751 b (equilibrium case) after one hour of simulated observation with the MIRI Low Resolution Spectroscopic mode (using the JWST ETC); overplotted the 3 $\sigma$ noise. The ETC sources of noise are: photon noise (source + background), detector noise, dark current noise and flat field errors; detector drifts effect are not taken into account.
Synthetic  models (see legend) have been plotted for comparison. }
\label{fig:LRS}
\end{figure}

Several molecules other than NH$_{3}$ have features in the MIRI wavelength range (i.e. CH$_4$, PH$_3$, CO$_2$, H$_2$O; e.g. Fig. 18 by \citealt{Baudino2017} concerning the spectrum of VHS 1256-1257 b).  MIRI spectroscopic observations, when possible, can give access to those lines. 
A high signal-to-noise ratio close or higher than 100 can be achieved in one hour (see Tab. \ref{tab:LRSsnr}). We show in Fig. \ref{fig:LRS} the extracted LRS spectrum of 2M2236+4751 b for one hour integration. Such observations will allow us to distinguish between equilibrium or non-equilibrium chemistry in the atmosphere of an exoplanet.

\begin{table}[h!]
\centering
\caption{Signal over noise ratio for various sources observed with the Low Resolution Spectroscopic mode of MIRI as calculated with the JWST  exposure time calculator (ETC)$^{a}$. Effects not taken into account, such as on how well the star signal at the position of the planet can be removed (especially for 2M1207 b and GJ 504 b) or calibration precision, will limit the SNR achievable to a value lower than that quoted here.}
\begin{tabular}{l c c c}
\\
\hline\hline
Sources & SNR at 5 $\mu$m &  SNR at 8 $\mu$m & SNR at 11 $\mu$m\\
\hline
   2M1207 b &	292 &	130	&45\\
   2M2236+4751 b	&230&	58&	25\\
   GJ 504 b	&290&	42	&29\\
   HD 106906 b	&323	&134 &	35\\
   ROXs42B b	&417&	198	&55\\
   VHS 1256-1257 b	&880&	351	&266\\   
\hline
\end{tabular}
\label{tab:LRSsnr}
\tablecomments{~(a)~\url{https://jwst.etc.stsci.edu/}}
\end{table}

\begin{figure}[b!]\centering
\includegraphics[clip,trim=0.8cm 0.5cm 0cm 0.cm,width=0.95\linewidth]{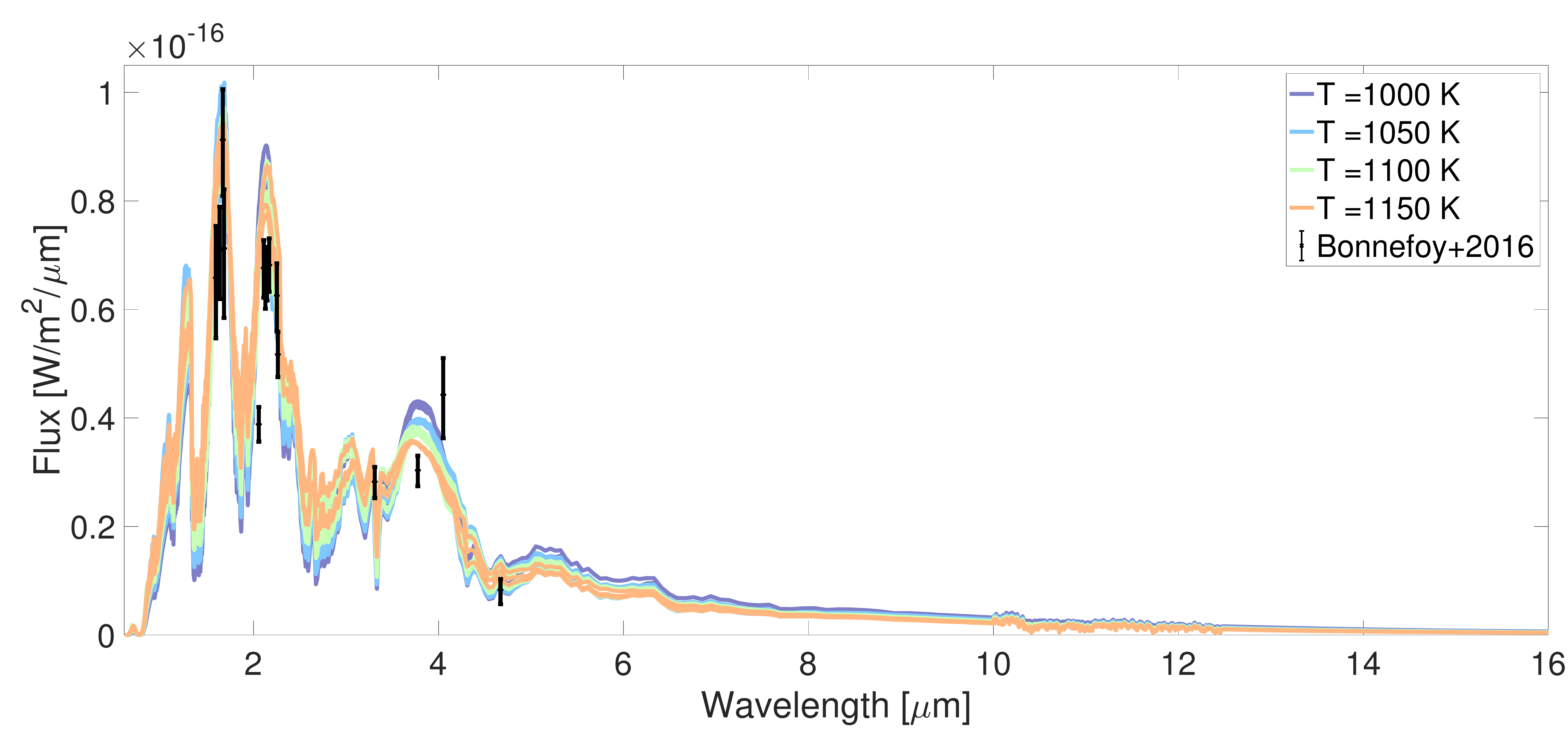}\\
\includegraphics[clip,trim=2.2cm 0cm 0.cm 0.cm,width=0.57\linewidth]{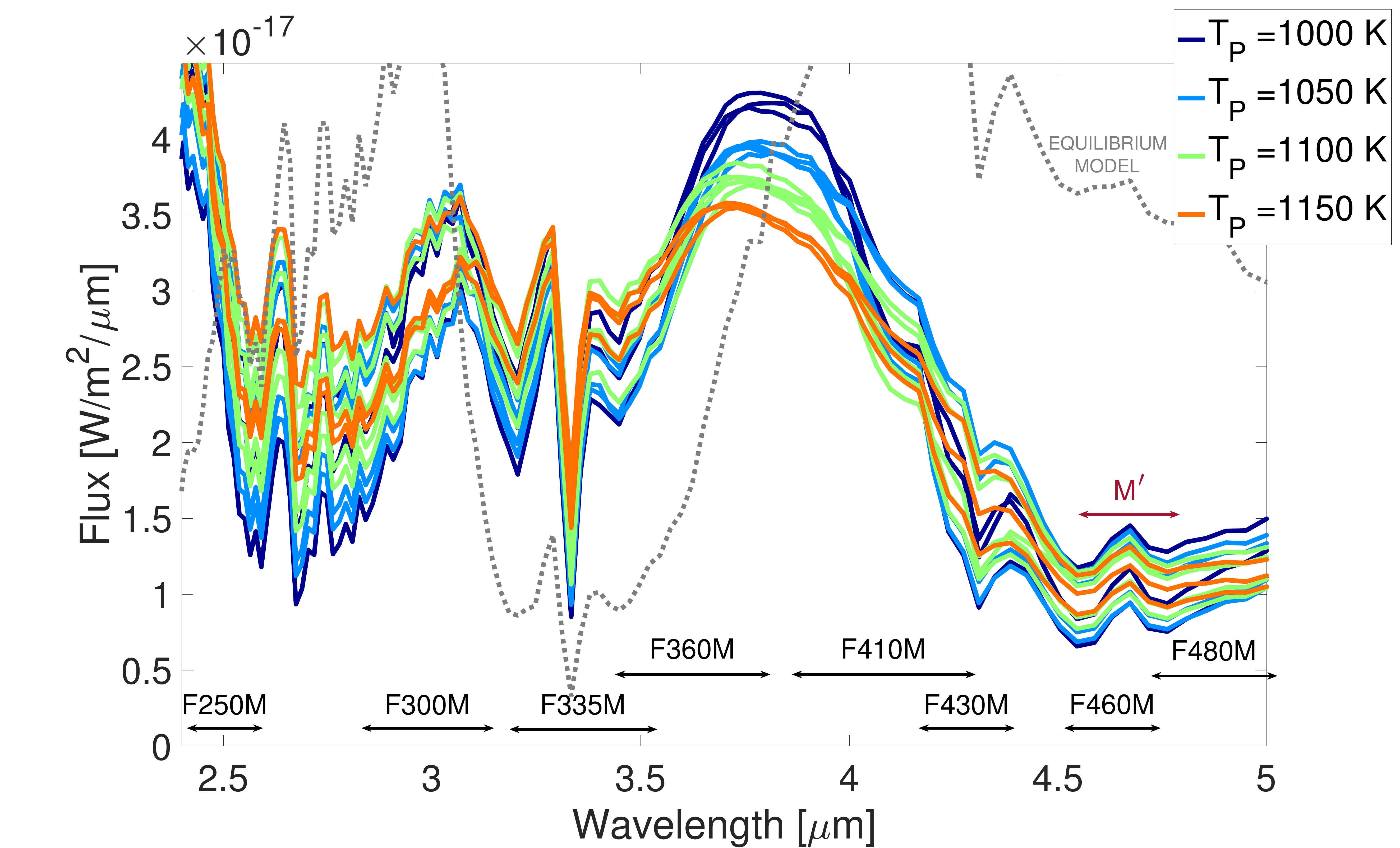}
\includegraphics[clip,trim=0.8cm 0.9cm 2.5cm 0.cm,width=0.4\linewidth]{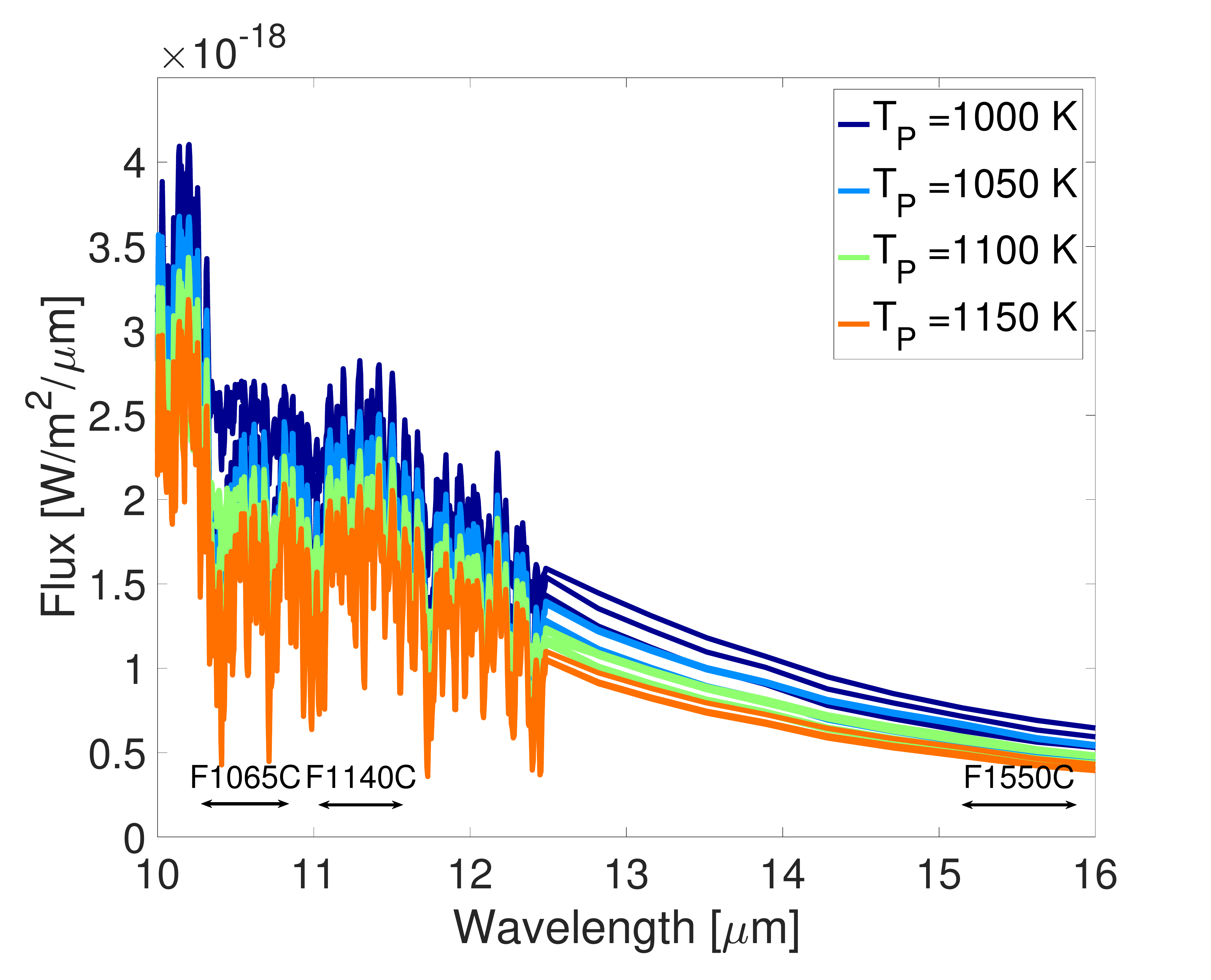}
\caption{HR 8799 b non-equilibrium synthetic models (measured at Earth) fitting the available photometric NIR observations by \cite{Bonnefoy2016}(\textit{top}) as seen in the NIRCam long wavelength channel (\textit{bottom left}) and MIRI coronagraphic filters (\textit{bottom right}). 
In the bottom left plot we show the available medium NIRCam filters allowed for coronagraphic imaging (\textit{black}) and Keck/NIRC2 M$^\prime$ filter (\textit{red}). Dotted line shows the equilibrium model ($T_P$ = 1200 K, $\log(g)$ = 5.2) used in the simulations.
In the bottom right plot we mark the 4QPM filters spectral range. We note that in the F1065C and F1140C wavelength range all models are at high resolution to prevent the under-sampling of the ammonia feature in the observations simulation.}
  \label{fig:familymodels}
\end{figure}

%% _______________________________________ SYNERGIES WITH NIR  _______________________________________

\subsection{Synergies between MIRI and NIR instruments}
\label{sec:MIRINIR}

As previously mentioned one of the known problems of planetary atmospheric characterization is the 
limited wavelength coverage in which current instruments work.
Measurements limited to a narrow spectral range yield significant uncertainties which do not 
allow to break degeneracy between planetary parameters.

This is where MIRI, conjointly with NIRCam\footnote{\url{https://jwst-docs.stsci.edu/display/JTI/NIRCam+Overview}, the JWST near-IR camera with coronagraphic capabilities},  plays a key role: by extending planetary 
observations to the mid-infrared it will be possible to 
constrain the atmospheric properties with higher precision.

We report here HR 8799 b as a test case to show the performances of MIRI/4QPM in term of 
accurate modeling of the exoplanetary atmosphere features.\\
\noindent We generated a family of Exo-REM models ($\Delta \lambda$ = 0 - 28 $\mu$m) fitting the planetary photometric observations reported in \cite{Bonnefoy2016} to have a group of models reproducing the observations at 1$\sigma$ (Fig. \ref{fig:familymodels}, top panel). 
We only selected photometric data in order to analyze a representative case. \\

The fitting models resulted to be only non-equilibrium chemistry ones spanning temperatures from 1000 to 1150 K, surface gravity values $\log(g)=$ 4.4 - 4.8 and metallicity values z = 0 and +0.5 dex. Note that both HR 8799 b equilibrium and non-equilibrium models, previously used in the science image simulations, are not part of this family of models because of different sets of data used for the minimization process (see \S~\ref{sec:HR8799models}).

On a general level, by using the JWST/NIRCam F460M or/and F480M filters, and/or the  M$^{\prime}$-band filter in the Adaptive Optic Keck/NIRC2 near-infrared narrow-field camera \citep{McLeanSprayberry}, it will be possible to distinguish between chemical equilibrium and non-equilibrium state of the atmosphere of a directly imaged exoplanet. 
In this spectral range (4-5~$\mu$m) there are overlapping features of at least three 
molecules, PH$_3$, CO and 
CO$_2$, whose abundances are impacted by non-equilibrium chemistry \cite[Sec.~6.4]{Baudino2017}.
Figure \ref{fig:familymodels} bottom left panel shows the different emission spectra in this particular wavelength range, in the cases whether the atmosphere is in chemical equilibrium or not.

\begin{figure}[p]\centering
\includegraphics[clip,trim=0.cm 5cm 0.5cm 3cm,width=1\linewidth]{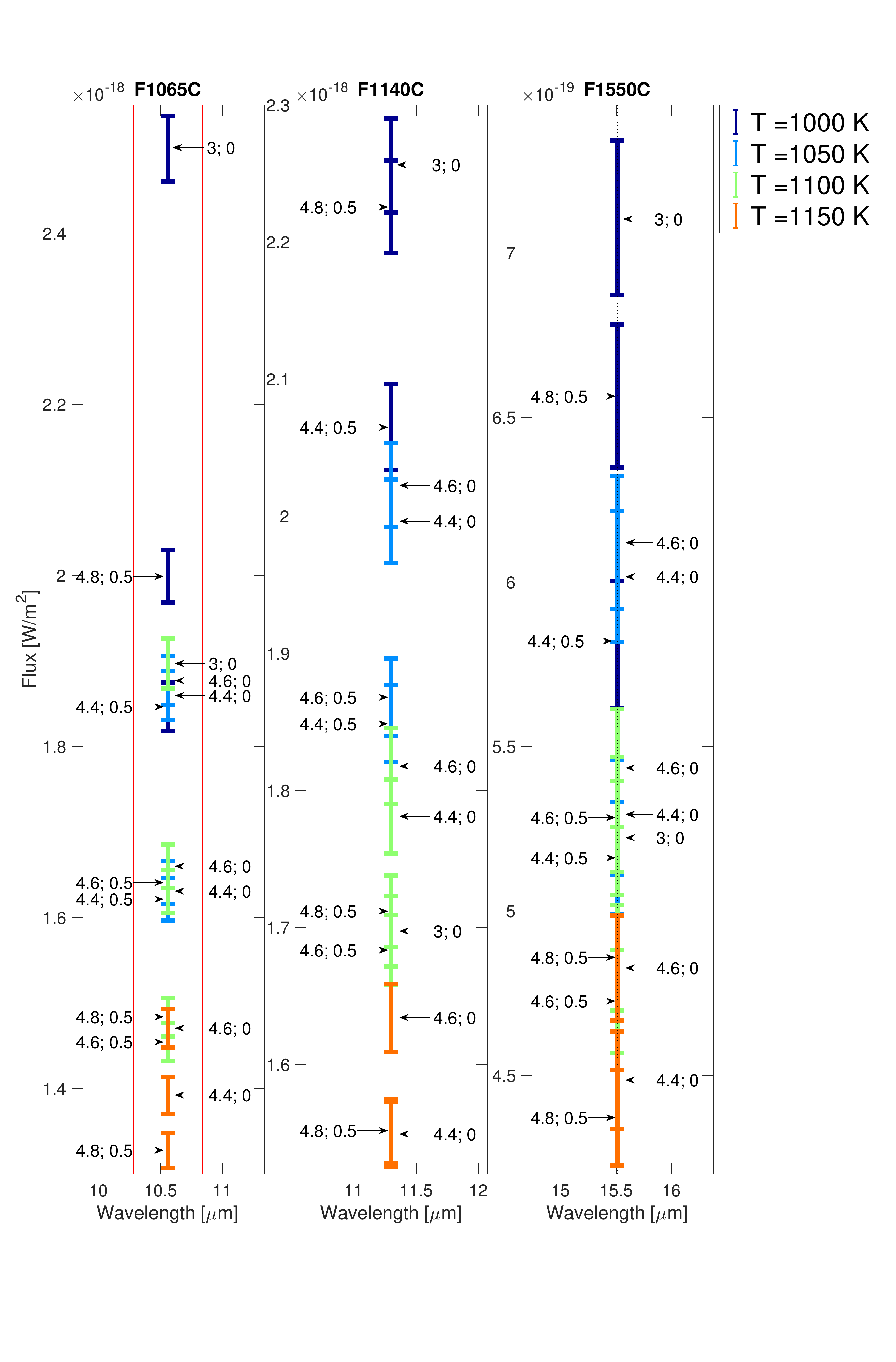}
\caption{HR 8799 b synthetic models integrated over the 4QPM filters with 5 $\sigma$ error. Uncertainties are the ones calculated for case $k_P$ at $t_{\rm max}$ = 4000 s. Colors mark the effective temperature of the models, while arrows correspond to the equivalent surface gravity [cgs] followed by the metallicity information.}
  \label{fig:errorbars}
\end{figure}

When knowing the chemical state we can move to the mid-infrared, in the MIRI bands, to retrieve the effective temperature of the planet. 
To test the precision achievable on this parameter with MIRI we integrated each synthetic model over the 4QPM filters and we added the corresponding uncertainties on the flux measured 
in \S~\ref{sec:SNRandUnc} at $t_{\rm max}$ = 4000 s for the photon noise case $k_P$. 
Figure \ref{fig:errorbars} shows the 5 $\sigma$ resulting measurements for each filter, with the corresponding temperature, surface gravity and metallicity. 
We find that, using filter F1140C, it is possible to disentangle models at better than $\Delta$T = 100 K apart from each other. Though, by combining the data-points at F1140C and F1550C it will be possible to better constrain the temperature and the bolometric luminosity of the planet due to the high SNR in these filters (Tab. \ref{tab:tmax}, Fig. \ref{fig:familymodels} bottom right). 
This said we notice that various degeneracies between temperature, surface gravity and metallicity persist. A clear example can be found in Fig. \ref{fig:errorbars}  around $F_\mathrm{obs} = 1.9\cdot10^{-18}$ W~m$^{-2}$ at F1065C or between
$F_\mathrm{obs} = 5 \cdot10^{-19}$  W~m$^{-2}$ and $F_\mathrm{obs} = 6\cdot10^{-19}$ W~m$^{-2}$  at F1550C, where various temperature, surface gravity and metallicity values are consistent with one another at the 5 $\sigma$ level.\\
One way to start breaking these degeneracies is to get spectroscopic observations in the 3.4 - 4.1 $\mu$m wavelength range (i.e. the L$^{\prime}$ band) to determine the planetary temperature. Even though this will not be sufficient for breaking all the degeneracies, it will allow us to discern more data-points in the mid-infrared. Having various observations that homogeneously cover a wide spectral range is key for mapping the planetary spectrum.
Furthermore, given the increased complexity in the interpretation of the spectrum due to high accuracy data, it is necessary to maximize the objectivity of the analysis from the start by not assuming anything about the atmospheres composition and structure. For this reason we recommend again the use of inverse atmospheric
retrieval modeling for objectively interpreting the data.

%---------------------------------- CONCLUSIONS ------------------------------------------------
\section{Conclusions}
\label{sec:conclusion}

We discussed here the simulated Mid-Infrared instrument (MIRI) coronagraphic observations for a set of known directly imaged exoplanets, whose emission spectrum has been calculated using the Exoplanet Radiative-convective Equilibrium Model.
The sample of analyzed objects presented effective temperatures ranging from 544 K to 1975 K, planet-to-star contrasts from 3.88 mag to 10.07 (measured at 15.50 $\mu$m), and angular separations from 0.42\arcsec~ to 8.06\arcsec. We provide planetary contrasts for each target and MIRI coronagraphic filter.

\begin{enumerate}[label=(\arabic{enumi}) ,wide, labelindent=0pt,labelwidth=! ] 

\item We studied the effect of the coronagraphic transmission as a function of the MIRI angular position (PA) of the planet, providing a coronagraphic transmission map over an area of 8$\arcsec$ x 8$\arcsec$ centered on the coronagraphic mask. 
When planning MIRI observations, if the observational window allows it, we suggest to set the planet at either one of the the MIRI PA = 45$^{\circ}$, 135$^{\circ}$, 225$^{\circ}$, 315$^{\circ}$. 

\item We examined the detectability of each target as a function of the four-quadrant phase mask (4QPM) filters and various observing telescope conditions i.e. by taking into consideration variations of: 
~
\begin{itemize}[noitemsep]
\item[] (i) offsets between the target star and the reference star: either 0 mas, 3 mas or 14 mas.
\item[] (ii) wave-front-error (WFE): either 130 nm rms or 204 nm rms.
\item[] (iii) jitter amplitude: either 1.6 mas or 7 mas.
\end{itemize}
~
\noindent More specifically we analyzed four combinations of these parameters, resulting in one optimistic observation, a pessimistic one and two intermediate observations.
For each combination and each 4QPM filter we provided the corresponding contrast curves.

\item We analyzed a fifth observational case whose only source of noise is Poisson noise. For this 
specific case we measured the signal-to-noise (SNR) and integration times for each coronagraphic target.
For an integration time $t_{\rm int}$ = 35 s the SNR spans values from 3 to 37 depending on the planetary contrast level (the highest or lowest in our target list, respectively).\\
We tested for the effect of a 50$\%$ higher background, at 15.50 $\mu$m, by studying the case of HIP 65426 b, whose star is the faintest among our targets. We found small differences: The SNR$_{50\%}$ measured  SNR$_{50\%}$ = 15 at t$_{\rm int}$ = 2400 s, compared to a SNR = 18, for the same integration time.

\item We showed that planetary detection strongly depends on the JWST in-flight performances and that a perfect stellar removal is highly necessary to obtain the best science results. For this reason we overall recommend the use of small grid dither concept while observing a target, in conjunction with sophisticated image-analysis algorithms for optimizing the PSF subtraction. 

\item We note that ammonia is a useful indicator of both planetary effective temperature and atmospheric chemical equilibrium state. We tested for the detectability of this molecule in the atmosphere of the coolest planets of our target list, predicting a possible NH$_3$ detection in HR 8799 b, d, GJ 504 b and 2M1207 b. The significance of this detection ranges from 3 $\sigma$ to 98 $\sigma$, depending on the planetary contrast and effective temperature. 

\item We have shown that several exoplanets detected by direct imaging can be observed using MIRI spectroscopy mode. Such observations will bring information not only on NH$_3$ but also on CH$_4$, H$_2$O, CO$_2$, PH$_3$ where very high SNR (>> 100) can be obtained.

\item MIRI, together with NIRCam, will provide strong constraints on the spectral characterization of young giant planets in wide orbits, pinning down the effective temperature and bolometric luminosity to an unprecedented accuracy. \\
The use of retrieval modeling techniques is advisable to maximize the objectivity of the analysis to infer the properties of exoplanetary atmospheres, including molecular abundances and temperature profiles.
A specific study presenting the use of retrievals methods on MIRI data will be presented in a forthcoming paper.\\

\end{enumerate}

~\\
~\\
\acknowledgments
C.D. acknowledges support from the LabEx P2IO, the French ANR contract 05-BLAN-NT09-573739,  the  Centre National d'etudes Spatiales (CNES) post-doctoral funding project and the P2IO LabEx (ANR-10-LABX-0038) in the framework $\ll$Investissements d'Avenir$\gg$(ANR-11-IDEX-0003-01).
J. L. B. acknowledges the support of the UK Science and Technology Facilities Council. 
All simulations presented in this work were performed in GDL (\url{https://github.com/gnudatalanguage/gdl/}). The authors would like to thank the anonymous referee whose comments and suggestions helped to improve and clarify this paper.

\newpage
\bibliography{Danielski_atmospheric_characterization_MIRI_R2} 


\newpage
\begin{appendix}
\section{PLANETARY CONTRAST}
\label{app:A}
\begin{deluxetable*}{l l | r  | r | r }[!ht]
\tablecaption{Planetary contrast of the coronagraphic targets measured in the three 4QPM filters using Exo-REM atmospheric model.
The equilibrium and non equilibrium cases are represented by the \textquotedblleft eq\textquotedblright~ and  \textquotedblleft neq\textquotedblright~ strings, respectively.}.
\tablecolumns{5}
\tablewidth{0pt}
\tablehead{
\colhead{Planet} & \colhead{} & \colhead{F1065C} & \colhead{F1140C} & \colhead{F1550C}
}
\startdata
\hline
{\BetaPic~ b}  & eq & 7.19 & 7.17& 7.23  \\
			   & neq & 7.22	& 7.17&  7.23 \\
\hline
{51 Eri b }   & eq & 11.32& 10.29 & 10.16  \\
			   & neq & 10.66 &10.11 &10.07  \\
\hline
{GJ 504 b } & eq & 10.60 & 9.01 & 8.91 \\
		    & neq &  9.43 & 8.64 & 8.83\\ 		
\hline
{HD 95086 b} & eq &  9.59 & 9.38 & 9.38 \\ 
			 & neq & 9.87 & 9.59 & 9.58 \\ 	
\hline
{2M1207 b} & eq & 2.19&  2.12&  2.0 \\ 
		   & neq & 3.88 & 3.88 & 3.88  \\  	
\hline
{ROXs42 b} 	& eq & 1.91 & 1.98 & 2.02 \\ 
			& neq &  1.91 & 1.98 & 2.02 \\  	
\hline
{HIP 65426 b}   & eq & 8.05 & 7.88 & 7.95   \\ 
			& neq &  8.32 & 8.08 & 8.09  \\ 
\hline
{HR 8799 b} & eq &  7.96 & 7.86 & 7.776  \\ 
			  & neq &  8.70  &8.43 & 8.19 \\ 	      
\hline
{HR 8799 c} & eq &  7.78 & 7.69 & 7.6 \\ 
			   & neq & 8.02 & 7.87 & 7.71 \\ 	
\hline
{HR 8799 d} & eq &   7.93 & 7.82 & 7.78  \\ 
			& neq &  8.19 & 8.05 & 7.88 \\ 
\hline
{HR 8799 e} & eq & 8.33 & 8.24 & 8.15 \\ 
			& neq &  8.54 & 8.41 & 8.19  \\                
\hline
\enddata
\label{app:planContrast}
\end{deluxetable*}

\newpage
\section{CONTRAST CURVES}
\label{app:B}
\subsection{F1065C}
\begin{longtable}{ccccc}
\caption{F1065C 1 $\sigma$ contrast curves as a function of the angular distance (Ang.Dist) for observational cases $k_A$, $k_B$, $k_C$, $k_D$.}\\
\hline\hline
Ang. Dist.[$\arcsec$]& $k_A$ & $k_B$ & $k_C$ & $k_D$  \\
\hline
\endfirsthead
\caption{continued.}\\
\hline\hline
Ang.Dist.[$\arcsec$]& $k_A$ & $k_B$ & $k_C$ & $k_D$ \\
\hline
\endhead
\hline
\endfoot

 0.11 & 8.7e-06 & 1.1e-04 & 1.5e-04 & 4.5e-04\\
 0.22 & 8.8e-06 & 1.0e-04 & 1.3e-04 & 4.0e-04\\
 0.33 & 8.8e-06 & 1.0e-04 & 1.3e-04 & 3.9e-04\\
 0.44 & 7.9e-06 & 9.3e-05 & 1.1e-04 & 3.5e-04\\
 0.55 & 6.3e-06 & 7.5e-05 & 8.7e-05 & 2.7e-04\\
 0.66 & 4.6e-06 & 5.6e-05 & 6.7e-05 & 2.1e-04\\
 0.77 & 2.7e-06 & 3.4e-05 & 4.6e-05 & 1.4e-04\\
 0.88 & 2.6e-06 & 3.3e-05 & 4.6e-05 & 1.4e-04\\
 0.99 & 2.2e-06 & 2.8e-05 & 4.0e-05 & 1.2e-04\\
 1.10 & 1.9e-06 & 2.3e-05 & 3.3e-05 & 1.0e-04\\
 1.21 & 1.6e-06 & 1.9e-05 & 2.6e-05 & 8.0e-05\\
 1.32 & 1.4e-06 & 1.7e-05 & 2.4e-05 & 7.3e-05\\
 1.43 & 1.3e-06 & 1.6e-05 & 2.3e-05 & 7.0e-05\\
 1.54 & 1.0e-06 & 1.2e-05 & 1.6e-05 & 5.0e-05\\
 1.65 & 8.6e-07 & 1.0e-05 & 1.3e-05 & 4.1e-05\\
 1.76 & 8.0e-07 & 9.8e-06 & 1.2e-05 & 3.7e-05\\
 1.87 & 6.8e-07 & 8.3e-06 & 1.0e-05 & 3.2e-05\\
 1.98 & 5.6e-07 & 6.9e-06 & 8.6e-06 & 2.6e-05\\
 2.09 & 5.0e-07 & 6.2e-06 & 8.0e-06 & 2.5e-05\\
 2.20 & 4.6e-07 & 5.7e-06 & 7.6e-06 & 2.3e-05\\
 2.31 & 4.0e-07 & 4.9e-06 & 6.6e-06 & 2.0e-05\\
 2.42 & 3.5e-07 & 4.2e-06 & 5.9e-06 & 1.8e-05\\
 2.53 & 4.0e-07 & 4.8e-06 & 6.5e-06 & 2.0e-05\\
 2.64 & 3.9e-07 & 4.7e-06 & 6.4e-06 & 2.0e-05\\
 2.75 & 3.6e-07 & 4.3e-06 & 5.9e-06 & 1.8e-05\\
 2.86 & 3.4e-07 & 4.2e-06 & 6.1e-06 & 1.9e-05\\
 2.97 & 3.0e-07 & 3.6e-06 & 5.5e-06 & 1.7e-05\\
 3.08 & 2.7e-07 & 3.4e-06 & 5.2e-06 & 1.6e-05\\
 3.19 & 2.3e-07 & 2.8e-06 & 4.4e-06 & 1.3e-05\\
 3.30 & 1.8e-07 & 2.3e-06 & 3.7e-06 & 1.1e-05\\
 3.41 & 1.9e-07 & 2.3e-06 & 3.7e-06 & 1.1e-05\\
 3.52 & 1.9e-07 & 2.3e-06 & 3.9e-06 & 1.2e-05\\
 3.63 & 1.9e-07 & 2.3e-06 & 3.9e-06 & 1.2e-05\\
 3.74 & 1.7e-07 & 2.1e-06 & 3.6e-06 & 1.1e-05\\
 3.85 & 1.7e-07 & 2.0e-06 & 3.3e-06 & 1.0e-05\\
 3.96 & 1.7e-07 & 2.2e-06 & 3.1e-06 & 9.5e-06\\
 4.07 & 1.6e-07 & 1.9e-06 & 2.7e-06 & 8.3e-06\\
 4.18 & 1.6e-07 & 2.0e-06 & 2.6e-06 & 8.0e-06\\
 4.29 & 1.4e-07 & 1.7e-06 & 2.3e-06 & 7.0e-06\\
 4.40 & 1.3e-07 & 1.5e-06 & 2.2e-06 & 6.7e-06\\
 4.51 & 1.1e-07 & 1.3e-06 & 1.9e-06 & 6.0e-06\\
 4.62 & 1.0e-07 & 1.2e-06 & 1.8e-06 & 5.6e-06\\
 4.73 & 9.2e-08 & 1.1e-06 & 1.7e-06 & 5.1e-06\\
 4.84 & 8.2e-08 & 9.5e-07 & 1.5e-06 & 4.7e-06\\
 4.95 & 8.3e-08 & 9.8e-07 & 1.5e-06 & 4.5e-06\\
 5.06 & 7.9e-08 & 9.5e-07 & 1.4e-06 & 4.2e-06\\
 5.17 & 7.5e-08 & 9.1e-07 & 1.3e-06 & 3.9e-06\\
 5.28 & 6.7e-08 & 8.2e-07 & 1.1e-06 & 3.4e-06\\
 5.39 & 6.2e-08 & 7.5e-07 & 1.0e-06 & 3.2e-06\\
 5.50 & 5.5e-08 & 6.7e-07 & 9.7e-07 & 3.0e-06\\
 5.61 & 4.4e-08 & 5.3e-07 & 8.3e-07 & 2.5e-06\\
 5.72 & 4.7e-08 & 5.7e-07 & 8.8e-07 & 2.7e-06\\
 5.83 & 4.7e-08 & 5.7e-07 & 8.5e-07 & 2.6e-06\\
 5.94 & 4.6e-08 & 5.6e-07 & 8.4e-07 & 2.6e-06\\
 6.05 & 4.8e-08 & 5.7e-07 & 8.6e-07 & 2.7e-06\\
 6.16 & 4.3e-08 & 5.1e-07 & 8.4e-07 & 2.6e-06\\
 6.27 & 4.3e-08 & 5.2e-07 & 8.3e-07 & 2.6e-06\\
 6.38 & 4.0e-08 & 4.9e-07 & 8.1e-07 & 2.5e-06\\
 6.49 & 3.7e-08 & 4.6e-07 & 7.9e-07 & 2.4e-06\\
 6.60 & 3.4e-08 & 4.1e-07 & 7.2e-07 & 2.2e-06\\
 6.71 & 3.1e-08 & 3.8e-07 & 6.5e-07 & 2.0e-06\\
 6.82 & 3.0e-08 & 3.7e-07 & 6.0e-07 & 1.8e-06\\
 6.93 & 2.7e-08 & 3.3e-07 & 5.4e-07 & 1.7e-06\\
 7.04 & 2.7e-08 & 3.3e-07 & 5.2e-07 & 1.6e-06\\
 7.15 & 2.5e-08 & 3.0e-07 & 4.7e-07 & 1.4e-06\\
 7.26 & 2.4e-08 & 2.9e-07 & 4.5e-07 & 1.4e-06\\
 7.37 & 2.4e-08 & 2.9e-07 & 4.5e-07 & 1.4e-06\\
 7.48 & 2.3e-08 & 2.7e-07 & 4.3e-07 & 1.3e-06\\
 7.59 & 2.4e-08 & 2.7e-07 & 4.3e-07 & 1.3e-06\\
 7.70 & 2.5e-08 & 2.9e-07 & 4.5e-07 & 1.4e-06\\
 7.81 & 2.5e-08 & 2.8e-07 & 4.6e-07 & 1.4e-06\\
 7.92 & 2.7e-08 & 2.9e-07 & 4.8e-07 & 1.5e-06\\
 8.03 & 2.6e-08 & 2.8e-07 & 4.8e-07 & 1.5e-06\\
 8.14 & 2.5e-08 & 2.8e-07 & 4.8e-07 & 1.5e-06\\
 8.25 & 2.3e-08 & 2.6e-07 & 4.5e-07 & 1.4e-06\\
 8.36 & 2.0e-08 & 2.2e-07 & 4.1e-07 & 1.3e-06\\
 8.47 & 1.7e-08 & 2.0e-07 & 3.6e-07 & 1.1e-06\\
 8.58 & 1.5e-08 & 1.8e-07 & 3.1e-07 & 9.5e-07\\
 8.69 & 1.4e-08 & 1.7e-07 & 2.8e-07 & 8.6e-07\\
 8.80 & 1.4e-08 & 1.7e-07 & 2.7e-07 & 8.2e-07\\
 8.91 & 1.3e-08 & 1.6e-07 & 2.6e-07 & 7.9e-07\\
 9.02 & 1.4e-08 & 1.6e-07 & 2.7e-07 & 8.2e-07\\
 9.13 & 1.6e-08 & 1.8e-07 & 3.0e-07 & 9.1e-07\\
 9.24 & 1.6e-08 & 1.9e-07 & 3.1e-07 & 9.6e-07\\
 9.35 & 1.6e-08 & 1.8e-07 & 3.2e-07 & 9.7e-07\\
 9.46 & 1.6e-08 & 1.7e-07 & 3.2e-07 & 9.9e-07\\
 9.57 & 1.6e-08 & 1.7e-07 & 3.1e-07 & 9.7e-07\\
 9.68 & 1.4e-08 & 1.5e-07 & 3.0e-07 & 9.3e-07\\
 9.79 & 1.3e-08 & 1.4e-07 & 2.8e-07 & 8.7e-07\\
 9.90 & 1.2e-08 & 1.3e-07 & 2.6e-07 & 8.1e-07\\
10.01 & 1.1e-08 & 1.2e-07 & 2.4e-07 & 7.4e-07\\
10.12 & 1.0e-08 & 1.1e-07 & 2.2e-07 & 6.9e-07\\
10.23 & 1.0e-08 & 1.1e-07 & 2.1e-07 & 6.5e-07\\
10.34 & 9.5e-09 & 1.1e-07 & 2.0e-07 & 6.1e-07\\
10.45 & 9.3e-09 & 1.1e-07 & 1.9e-07 & 5.8e-07\\
10.56 & 9.0e-09 & 1.0e-07 & 1.8e-07 & 5.5e-07\\
10.67 & 8.2e-09 & 9.2e-08 & 1.7e-07 & 5.2e-07\\
10.78 & 8.6e-09 & 9.7e-08 & 1.7e-07 & 5.2e-07\\
10.89 & 8.4e-09 & 9.1e-08 & 1.6e-07 & 5.0e-07\\
11.00 & 9.1e-09 & 9.8e-08 & 1.7e-07 & 5.3e-07\\

\end{longtable}

\newpage

\subsection{F1140C}
\begin{longtable}{ccccc}
\caption{F1140C 1 $\sigma$ contrast curves as a function of the angular distance (Ang.Dist) for observational cases $k_A$, $k_B$, $k_C$, $k_D$.}\\
\hline\hline
Ang. Dist.[$\arcsec$]& $k_A$ & $k_B$ & $k_C$ & $k_D$  \\
\hline
\endfirsthead
\caption{continued.}\\
\hline\hline
Ang.Dist.[$\arcsec$]& $k_A$ & $k_B$ & $k_C$ & $k_D$ \\
\hline
\endhead
\hline
\endfoot
 0.11 & 9.9e-06 & 1.2e-04 & 1.7e-04 & 5.1e-04\\
 0.22 & 8.8e-06 & 1.0e-04 & 1.4e-04 & 4.2e-04\\
 0.33 & 8.6e-06 & 1.0e-04 & 1.3e-04 & 4.0e-04\\
 0.44 & 7.4e-06 & 8.8e-05 & 1.1e-04 & 3.3e-04\\
 0.55 & 5.0e-06 & 6.1e-05 & 7.4e-05 & 2.3e-04\\
 0.66 & 4.0e-06 & 4.9e-05 & 6.3e-05 & 2.0e-04\\
 0.77 & 3.2e-06 & 3.9e-05 & 5.3e-05 & 1.7e-04\\
 0.88 & 3.0e-06 & 3.8e-05 & 5.1e-05 & 1.6e-04\\
 0.99 & 2.2e-06 & 2.7e-05 & 3.8e-05 & 1.2e-04\\
 1.10 & 1.8e-06 & 2.2e-05 & 3.0e-05 & 9.4e-05\\
 1.21 & 1.7e-06 & 2.0e-05 & 2.7e-05 & 8.4e-05\\
 1.32 & 1.6e-06 & 2.0e-05 & 2.7e-05 & 8.3e-05\\
 1.43 & 1.5e-06 & 1.8e-05 & 2.5e-05 & 7.7e-05\\
 1.54 & 1.1e-06 & 1.3e-05 & 1.9e-05 & 5.7e-05\\
 1.65 & 9.7e-07 & 1.2e-05 & 1.6e-05 & 5.1e-05\\
 1.76 & 8.8e-07 & 1.1e-05 & 1.5e-05 & 4.5e-05\\
 1.87 & 7.4e-07 & 9.0e-06 & 1.2e-05 & 3.7e-05\\
 1.98 & 6.4e-07 & 7.8e-06 & 1.1e-05 & 3.4e-05\\
 2.09 & 5.7e-07 & 7.0e-06 & 9.7e-06 & 3.0e-05\\
 2.20 & 5.1e-07 & 6.3e-06 & 8.6e-06 & 2.7e-05\\
 2.31 & 4.4e-07 & 5.4e-06 & 7.6e-06 & 2.4e-05\\
 2.42 & 3.8e-07 & 4.7e-06 & 6.6e-06 & 2.1e-05\\
 2.53 & 3.8e-07 & 4.6e-06 & 6.4e-06 & 2.0e-05\\
 2.64 & 3.5e-07 & 4.2e-06 & 6.0e-06 & 1.9e-05\\
 2.75 & 3.4e-07 & 4.1e-06 & 6.0e-06 & 1.8e-05\\
 2.86 & 3.6e-07 & 4.3e-06 & 6.2e-06 & 1.9e-05\\
 2.97 & 3.2e-07 & 3.9e-06 & 5.7e-06 & 1.8e-05\\
 3.08 & 3.4e-07 & 4.2e-06 & 6.1e-06 & 1.9e-05\\
 3.19 & 3.0e-07 & 3.6e-06 & 5.4e-06 & 1.7e-05\\
 3.30 & 2.2e-07 & 2.7e-06 & 4.2e-06 & 1.3e-05\\
 3.41 & 2.1e-07 & 2.6e-06 & 4.0e-06 & 1.2e-05\\
 3.52 & 2.1e-07 & 2.5e-06 & 3.9e-06 & 1.2e-05\\
 3.63 & 2.0e-07 & 2.4e-06 & 3.8e-06 & 1.2e-05\\
 3.74 & 1.8e-07 & 2.3e-06 & 3.6e-06 & 1.1e-05\\
 3.85 & 1.8e-07 & 2.2e-06 & 3.5e-06 & 1.0e-05\\
 3.96 & 1.8e-07 & 2.1e-06 & 3.5e-06 & 1.1e-05\\
 4.07 & 1.6e-07 & 2.0e-06 & 3.2e-06 & 9.7e-06\\
 4.18 & 1.8e-07 & 2.3e-06 & 3.3e-06 & 1.0e-05\\
 4.29 & 1.6e-07 & 1.9e-06 & 2.8e-06 & 8.7e-06\\
 4.40 & 1.7e-07 & 2.1e-06 & 2.9e-06 & 9.0e-06\\
 4.51 & 1.5e-07 & 1.8e-06 & 2.6e-06 & 8.1e-06\\
 4.62 & 1.3e-07 & 1.5e-06 & 2.3e-06 & 7.1e-06\\
 4.73 & 1.1e-07 & 1.4e-06 & 2.1e-06 & 6.4e-06\\
 4.84 & 1.0e-07 & 1.2e-06 & 1.9e-06 & 5.8e-06\\
 4.95 & 9.7e-08 & 1.1e-06 & 1.8e-06 & 5.5e-06\\
 5.06 & 9.3e-08 & 1.1e-06 & 1.7e-06 & 5.3e-06\\
 5.17 & 9.0e-08 & 1.1e-06 & 1.6e-06 & 5.0e-06\\
 5.28 & 8.3e-08 & 9.8e-07 & 1.5e-06 & 4.6e-06\\
 5.39 & 8.0e-08 & 9.5e-07 & 1.4e-06 & 4.4e-06\\
 5.50 & 7.4e-08 & 9.0e-07 & 1.3e-06 & 4.0e-06\\
 5.61 & 6.4e-08 & 7.8e-07 & 1.1e-06 & 3.5e-06\\
 5.72 & 6.7e-08 & 8.1e-07 & 1.2e-06 & 3.6e-06\\
 5.83 & 5.9e-08 & 7.1e-07 & 1.0e-06 & 3.3e-06\\
 5.94 & 5.2e-08 & 6.3e-07 & 9.5e-07 & 3.0e-06\\
 6.05 & 4.8e-08 & 5.8e-07 & 8.8e-07 & 2.7e-06\\
 6.16 & 4.0e-08 & 4.8e-07 & 7.6e-07 & 2.4e-06\\
 6.27 & 4.7e-08 & 5.6e-07 & 8.5e-07 & 2.7e-06\\
 6.38 & 4.8e-08 & 5.7e-07 & 8.8e-07 & 2.7e-06\\
 6.49 & 4.5e-08 & 5.3e-07 & 8.6e-07 & 2.7e-06\\
 6.60 & 4.6e-08 & 5.5e-07 & 8.8e-07 & 2.7e-06\\
 6.71 & 4.3e-08 & 5.2e-07 & 8.7e-07 & 2.7e-06\\
 6.82 & 4.0e-08 & 4.8e-07 & 8.3e-07 & 2.6e-06\\
 6.93 & 3.7e-08 & 4.6e-07 & 8.1e-07 & 2.5e-06\\
 7.04 & 3.6e-08 & 4.4e-07 & 7.8e-07 & 2.4e-06\\
 7.15 & 3.3e-08 & 4.0e-07 & 7.1e-07 & 2.2e-06\\
 7.26 & 3.1e-08 & 3.7e-07 & 6.4e-07 & 2.0e-06\\
 7.37 & 2.9e-08 & 3.4e-07 & 5.5e-07 & 1.7e-06\\
 7.48 & 2.7e-08 & 3.2e-07 & 5.0e-07 & 1.5e-06\\
 7.59 & 2.5e-08 & 3.0e-07 & 4.6e-07 & 1.4e-06\\
 7.70 & 2.5e-08 & 3.0e-07 & 4.7e-07 & 1.5e-06\\
 7.81 & 2.2e-08 & 2.6e-07 & 4.3e-07 & 1.3e-06\\
 7.92 & 2.4e-08 & 2.8e-07 & 4.5e-07 & 1.4e-06\\
 8.03 & 2.2e-08 & 2.5e-07 & 4.2e-07 & 1.3e-06\\
 8.14 & 2.4e-08 & 2.8e-07 & 4.4e-07 & 1.4e-06\\
 8.25 & 2.5e-08 & 2.8e-07 & 4.4e-07 & 1.4e-06\\
 8.36 & 2.5e-08 & 2.7e-07 & 4.5e-07 & 1.4e-06\\
 8.47 & 2.6e-08 & 2.9e-07 & 4.9e-07 & 1.5e-06\\
 8.58 & 2.7e-08 & 2.9e-07 & 5.1e-07 & 1.6e-06\\
 8.69 & 2.6e-08 & 2.8e-07 & 5.0e-07 & 1.6e-06\\
 8.80 & 2.4e-08 & 2.7e-07 & 5.0e-07 & 1.5e-06\\
 8.91 & 2.1e-08 & 2.4e-07 & 4.5e-07 & 1.4e-06\\
 9.02 & 1.8e-08 & 2.1e-07 & 3.9e-07 & 1.2e-06\\
 9.13 & 1.6e-08 & 1.9e-07 & 3.2e-07 & 1.0e-06\\
 9.24 & 1.4e-08 & 1.7e-07 & 2.7e-07 & 8.5e-07\\
 9.35 & 1.2e-08 & 1.5e-07 & 2.4e-07 & 7.5e-07\\
 9.46 & 1.2e-08 & 1.4e-07 & 2.4e-07 & 7.3e-07\\
 9.57 & 1.3e-08 & 1.6e-07 & 2.5e-07 & 7.8e-07\\
 9.68 & 1.3e-08 & 1.5e-07 & 2.5e-07 & 7.9e-07\\
 9.79 & 1.5e-08 & 1.7e-07 & 2.8e-07 & 8.8e-07\\
 9.90 & 1.6e-08 & 1.8e-07 & 3.0e-07 & 9.4e-07\\
10.01 & 1.7e-08 & 1.9e-07 & 3.4e-07 & 1.1e-06\\
10.12 & 1.7e-08 & 1.8e-07 & 3.4e-07 & 1.1e-06\\
10.23 & 1.7e-08 & 1.8e-07 & 3.5e-07 & 1.1e-06\\
10.34 & 1.6e-08 & 1.7e-07 & 3.3e-07 & 1.0e-06\\
10.45 & 1.4e-08 & 1.5e-07 & 3.1e-07 & 9.7e-07\\
10.56 & 1.2e-08 & 1.3e-07 & 2.7e-07 & 8.3e-07\\
10.67 & 1.1e-08 & 1.2e-07 & 2.5e-07 & 7.9e-07\\
10.78 & 1.0e-08 & 1.2e-07 & 2.3e-07 & 7.3e-07\\
10.89 & 9.1e-09 & 1.0e-07 & 2.1e-07 & 6.7e-07\\
11.00 & 1.0e-08 & 1.2e-07 & 2.1e-07 & 6.5e-07\\

\end{longtable}

\newpage

\subsection{F1550C}
\begin{longtable}{ccccc}
\caption{F1140C 1 $\sigma$ contrast curves as a function of the angular distance (Ang.Dist) for observational cases $k_A$, $k_B$, $k_C$, $k_D$.}\\
\hline\hline
Ang. Dist.[$\arcsec$]& $k_A$ & $k_B$ & $k_C$ & $k_D$  \\
\hline
\endfirsthead
\caption{continued.}\\
\hline\hline
Ang.Dist.[$\arcsec$]& $k_A$ & $k_B$ & $k_C$ & $k_D$ \\
\hline
\endhead
\hline
\endfoot

 0.11 & 2.8e-06 & 3.2e-05 & 4.8e-05 & 1.5e-04\\
 0.22 & 2.7e-06 & 3.1e-05 & 4.7e-05 & 1.4e-04\\
 0.33 & 2.7e-06 & 3.0e-05 & 4.2e-05 & 1.3e-04\\
 0.44 & 2.6e-06 & 3.0e-05 & 4.0e-05 & 1.2e-04\\
 0.55 & 2.3e-06 & 2.7e-05 & 3.4e-05 & 1.1e-04\\
 0.66 & 1.9e-06 & 2.3e-05 & 2.8e-05 & 8.6e-05\\
 0.77 & 1.7e-06 & 2.1e-05 & 2.5e-05 & 7.8e-05\\
 0.88 & 1.3e-06 & 1.7e-05 & 2.2e-05 & 6.9e-05\\
 0.99 & 1.1e-06 & 1.3e-05 & 2.0e-05 & 6.3e-05\\
 1.10 & 7.6e-07 & 9.4e-06 & 1.6e-05 & 4.9e-05\\
 1.21 & 6.8e-07 & 8.6e-06 & 1.4e-05 & 4.5e-05\\
 1.32 & 6.3e-07 & 8.0e-06 & 1.3e-05 & 4.1e-05\\
 1.43 & 5.2e-07 & 6.5e-06 & 1.0e-05 & 3.2e-05\\
 1.54 & 5.0e-07 & 6.2e-06 & 8.9e-06 & 2.7e-05\\
 1.65 & 5.1e-07 & 6.2e-06 & 8.4e-06 & 2.6e-05\\
 1.76 & 5.0e-07 & 6.1e-06 & 8.2e-06 & 2.5e-05\\
 1.87 & 4.9e-07 & 5.9e-06 & 7.8e-06 & 2.4e-05\\
 1.98 & 4.4e-07 & 5.3e-06 & 7.0e-06 & 2.1e-05\\
 2.09 & 4.2e-07 & 5.0e-06 & 6.9e-06 & 2.1e-05\\
 2.20 & 3.7e-07 & 4.4e-06 & 6.2e-06 & 1.9e-05\\
 2.31 & 3.3e-07 & 4.0e-06 & 5.4e-06 & 1.6e-05\\
 2.42 & 2.9e-07 & 3.5e-06 & 4.4e-06 & 1.3e-05\\
 2.53 & 2.6e-07 & 3.2e-06 & 3.9e-06 & 1.2e-05\\
 2.64 & 2.3e-07 & 2.8e-06 & 3.6e-06 & 1.1e-05\\
 2.75 & 2.0e-07 & 2.5e-06 & 3.1e-06 & 9.7e-06\\
 2.86 & 1.8e-07 & 2.2e-06 & 2.7e-06 & 8.2e-06\\
 2.97 & 1.6e-07 & 2.0e-06 & 2.5e-06 & 7.6e-06\\
 3.08 & 1.5e-07 & 1.8e-06 & 2.4e-06 & 7.5e-06\\
 3.19 & 1.2e-07 & 1.5e-06 & 2.2e-06 & 6.8e-06\\
 3.30 & 1.2e-07 & 1.4e-06 & 2.2e-06 & 6.7e-06\\
 3.41 & 1.2e-07 & 1.5e-06 & 2.4e-06 & 7.3e-06\\
 3.52 & 1.2e-07 & 1.4e-06 & 2.3e-06 & 6.9e-06\\
 3.63 & 1.2e-07 & 1.4e-06 & 2.4e-06 & 7.3e-06\\
 3.74 & 1.1e-07 & 1.4e-06 & 2.3e-06 & 7.0e-06\\
 3.85 & 1.1e-07 & 1.4e-06 & 2.4e-06 & 7.4e-06\\
 3.96 & 1.0e-07 & 1.3e-06 & 2.2e-06 & 6.8e-06\\
 4.07 & 1.0e-07 & 1.2e-06 & 2.2e-06 & 6.8e-06\\
 4.18 & 8.9e-08 & 1.1e-06 & 2.0e-06 & 6.2e-06\\
 4.29 & 8.6e-08 & 1.1e-06 & 1.9e-06 & 5.9e-06\\
 4.40 & 7.9e-08 & 9.9e-07 & 1.7e-06 & 5.3e-06\\
 4.51 & 6.8e-08 & 8.5e-07 & 1.5e-06 & 4.5e-06\\
 4.62 & 6.1e-08 & 7.7e-07 & 1.4e-06 & 4.2e-06\\
 4.73 & 5.3e-08 & 6.6e-07 & 1.2e-06 & 3.8e-06\\
 4.84 & 5.0e-08 & 6.3e-07 & 1.2e-06 & 3.5e-06\\
 4.95 & 4.7e-08 & 6.0e-07 & 1.1e-06 & 3.3e-06\\
 5.06 & 4.5e-08 & 5.6e-07 & 1.1e-06 & 3.3e-06\\
 5.17 & 4.5e-08 & 5.7e-07 & 1.2e-06 & 3.5e-06\\
 5.28 & 4.5e-08 & 5.7e-07 & 1.2e-06 & 3.5e-06\\
 5.39 & 4.6e-08 & 5.9e-07 & 1.3e-06 & 3.7e-06\\
 5.50 & 4.3e-08 & 5.4e-07 & 1.1e-06 & 3.3e-06\\
 5.61 & 4.6e-08 & 5.9e-07 & 1.2e-06 & 3.6e-06\\
 5.72 & 4.7e-08 & 6.0e-07 & 1.2e-06 & 3.5e-06\\
 5.83 & 4.5e-08 & 5.7e-07 & 1.1e-06 & 3.3e-06\\
 5.94 & 4.5e-08 & 5.6e-07 & 1.0e-06 & 3.1e-06\\
 6.05 & 3.9e-08 & 4.9e-07 & 8.5e-07 & 2.6e-06\\
 6.16 & 3.9e-08 & 4.8e-07 & 8.1e-07 & 2.5e-06\\
 6.27 & 3.6e-08 & 4.4e-07 & 7.4e-07 & 2.3e-06\\
 6.38 & 3.3e-08 & 4.0e-07 & 6.6e-07 & 2.0e-06\\
 6.49 & 3.1e-08 & 3.7e-07 & 6.0e-07 & 1.8e-06\\
 6.60 & 2.9e-08 & 3.4e-07 & 5.5e-07 & 1.7e-06\\
 6.71 & 2.7e-08 & 3.1e-07 & 5.0e-07 & 1.5e-06\\
 6.82 & 2.6e-08 & 3.0e-07 & 4.9e-07 & 1.5e-06\\
 6.93 & 2.4e-08 & 2.8e-07 & 4.6e-07 & 1.4e-06\\
 7.04 & 2.3e-08 & 2.7e-07 & 4.4e-07 & 1.3e-06\\
 7.15 & 2.2e-08 & 2.6e-07 & 4.1e-07 & 1.2e-06\\
 7.26 & 2.0e-08 & 2.4e-07 & 3.7e-07 & 1.1e-06\\
 7.37 & 1.9e-08 & 2.3e-07 & 3.5e-07 & 1.1e-06\\
 7.48 & 1.7e-08 & 2.1e-07 & 3.2e-07 & 1.0e-06\\
 7.59 & 1.9e-08 & 2.3e-07 & 3.4e-07 & 1.1e-06\\
 7.70 & 1.6e-08 & 2.0e-07 & 2.9e-07 & 9.1e-07\\
 7.81 & 1.7e-08 & 2.1e-07 & 3.2e-07 & 9.8e-07\\
 7.92 & 1.5e-08 & 1.8e-07 & 2.9e-07 & 8.9e-07\\
 8.03 & 1.4e-08 & 1.8e-07 & 3.0e-07 & 9.2e-07\\
 8.14 & 1.3e-08 & 1.6e-07 & 2.8e-07 & 8.6e-07\\
 8.25 & 1.1e-08 & 1.4e-07 & 2.4e-07 & 7.5e-07\\
 8.36 & 1.1e-08 & 1.4e-07 & 2.4e-07 & 7.5e-07\\
 8.47 & 1.1e-08 & 1.3e-07 & 2.3e-07 & 7.1e-07\\
 8.58 & 1.1e-08 & 1.3e-07 & 2.2e-07 & 6.8e-07\\
 8.69 & 1.0e-08 & 1.2e-07 & 2.1e-07 & 6.4e-07\\
 8.80 & 1.0e-08 & 1.2e-07 & 2.0e-07 & 6.2e-07\\
 8.91 & 1.0e-08 & 1.2e-07 & 2.0e-07 & 6.2e-07\\
 9.02 & 9.8e-09 & 1.2e-07 & 2.0e-07 & 6.1e-07\\
 9.13 & 9.2e-09 & 1.1e-07 & 1.9e-07 & 5.9e-07\\
 9.24 & 8.6e-09 & 1.0e-07 & 1.8e-07 & 5.6e-07\\
 9.35 & 8.1e-09 & 9.9e-08 & 1.8e-07 & 5.5e-07\\
 9.46 & 7.9e-09 & 9.7e-08 & 1.8e-07 & 5.4e-07\\
 9.57 & 7.5e-09 & 9.0e-08 & 1.7e-07 & 5.3e-07\\
 9.68 & 7.3e-09 & 8.8e-08 & 1.7e-07 & 5.1e-07\\
 9.79 & 7.1e-09 & 8.5e-08 & 1.6e-07 & 4.8e-07\\
 9.90 & 7.0e-09 & 8.5e-08 & 1.5e-07 & 4.7e-07\\
10.01 & 6.7e-09 & 8.1e-08 & 1.5e-07 & 4.6e-07\\
10.12 & 6.4e-09 & 7.6e-08 & 1.4e-07 & 4.4e-07\\
10.23 & 6.1e-09 & 7.3e-08 & 1.3e-07 & 4.1e-07\\
10.34 & 5.8e-09 & 6.9e-08 & 1.3e-07 & 3.9e-07\\
10.45 & 5.7e-09 & 6.9e-08 & 1.3e-07 & 3.8e-07\\
10.56 & 5.2e-09 & 6.2e-08 & 1.2e-07 & 3.5e-07\\
10.67 & 5.3e-09 & 6.4e-08 & 1.2e-07 & 3.7e-07\\
10.78 & 5.0e-09 & 5.9e-08 & 1.1e-07 & 3.4e-07\\
10.89 & 5.5e-09 & 6.6e-08 & 1.2e-07 & 3.6e-07\\
11.00 & 5.6e-09 & 6.7e-08 & 1.2e-07 & 3.7e-07\\

\end{longtable}
%\textcolor{red}{table contrasts curves}
\end{appendix}

\end{document}